\documentclass[a4paper,10pt]{scrartcl}
\usepackage[scale={.8,.86}]{geometry}
\usepackage{fancyhdr,cite}
\usepackage[T1]{fontenc}
\usepackage[latin1]{inputenc}
\usepackage{amsmath,amssymb,dsfont,dcolumn,mathrsfs,multirow,slashed}
\usepackage{pstricks,psfrag,scalefnt}
\usepackage{axodraw,graphicx,wrapfig,subfig}
\usepackage{color}

\usepackage{caption}
\renewcommand{\captionfont}{\slshape}

\newcolumntype{d}{D{.}{.}{6}}

\DeclareMathOperator{\diag}{diag}
\DeclareMathOperator{\re}{Re}

\DeclareMathOperator{\sgn}{sgn}
\DeclareMathOperator{\tr}{tr}

\newcommand{\eq}[1]{Eq.~(\ref{#1})}
\newcommand{\lsim}{\stackrel{<}{_\sim}}

\newcommand{\bbs}{\ensuremath{B_s\!-\!\Bbar{}_s\,}}
\newcommand{\bbms}{\bbs\ mixing}
\newcommand{\bbd}{\ensuremath{B_d\!-\!\Bbar{}_d\,}}
\newcommand{\bbmd}{\bbd\ mixing}

\newcommand{\kk}{\ensuremath{K\!-\!\Kbar\,}}
\newcommand{\kkm}{\kk\ mixing}
\newcommand{\Bbar}{\,\overline{\!B}}
\newcommand{\Kbar}{\,\overline{\!K}}

\mathcode`,="013B
\mathcode`.="613A

\title{{\normalsize TTP09-38 \hfill arXiv:0910.2663}
  \\[-12pt]
  {\normalsize SFB/CPP-09-94 \hfill October 2009}
  \\[12pt]
  Lepton flavour violation in the MSSM} 

\author{Jennifer Girrbach$^1$, Susanne Mertens$^{1,2}$, Ulrich Nierste$^1$
  and S\"oren Wiesenfeldt$^{1,3}$
  \\[6pt]
  {\normalsize 
   {\slshape $^1$ \parbox[t]{0.8\textwidth}{Institut f\"ur Theoretische
      Teilchenphysik, Karlsruhe Institute of Technology,\\  
      Universit\"at Karlsruhe, 76128 Karlsruhe, Germany}
    }}
  \\
  {\normalsize {\slshape $^2$ \parbox[t]{0.8\textwidth}{Institut 
        f\"ur Experimentelle Kernphysik,
        Karlsruhe Institute of Technology,\\  
        Universit\"at Karlsruhe, 76128 Karlsruhe, Germany }}} 
  \\
  {\normalsize {\slshape $^3$  \parbox[t]{0.8\textwidth}{Helmholtz 
       Association,
      Anna-Louisa-Karsch-Stra{\ss}e 2, 10178 Berlin, Germany }} }}

\date{}

\begin{document}

\maketitle

\begin{center}
  {\large \bfseries Abstract}
\end{center}
\begin{abstract}
  \noindent
     We derive new constraints on the quantities $\delta_{XY}^{ij}$,
    $X,Y=L,R$, which parametrise the flavour-off-diagonal terms of the
    charged slepton mass matrix in the MSSM. Considering mass and
    anomalous magnetic moment of the electron we obtain the bound
    $|\delta^{13}_{LL} \delta^{13}_{RR}|\lsim 0.1$ 
    for $\tan\beta=50$, which involves the
    poorly constrained element $\delta^{13}_{RR}$.  We improve the
    predictions for the decays $\tau \to \mu \gamma$, $\tau \to e
    \gamma$ and $\mu \to e \gamma$ by including two-loop corrections
    which are enhanced if $\tan\beta$ is large. The finite
    renormalisation of the PMNS matrix from soft SUSY-breaking terms is
    derived and applied to the charged-Higgs-lepton vertex. We find that
    the experimental bound on $BR(\tau\to e \gamma)$ severely limits the
    size of the MSSM loop correction to the PMNS element $U_{e3}$, which
    is important for the proper interpretation of a future $U_{e3}$
    measurement. Subsequently we confront our new values for
    $\delta^{ij}_{LL}$ with a GUT analysis.  Further, we include the effects
    of dimension-5 Yukawa terms, which are needed to fix the
    Yukawa unification of the first two generations.  If universal
    supersymmetry breaking occurs above the GUT scale, we find the
    flavour structure of the dimension-5 Yukawa couplings tightly
    constrained by $\mu \to e \gamma$.
\end{abstract}

\section{Introduction}

Weak-scale supersymmetry (SUSY) is an attractive framework for physics
beyond the standard model (SM) of particle physics.  The SM fields are
promoted to superfields, with additional constituents of opposite spin.
Due to their identical couplings, they cancel the quadratic divergent
corrections to the Higgs mass.  Since none of the SUSY partners have
been observed in experiments, supersymmetry must be broken and the
masses of the SUSY partners are expected to be in the multi-GeV region.

A supersymmetric version of the standard model requires a second Higgs
doublet in order to cancel the Higgsino-related anomalies and to achieve
electroweak symmetry breaking.  At tree level, one of the Higgs
doublets, $H_u$, couples to the up-type particles, whereas the other
doublet, $H_d$, couples to the down type particles.  The Yukawa
couplings of the minimal supersymmetric standard model (MSSM) read
\begin{subequations}
  \label{equ:SuperpotenzialMR}
  \begin{align}\label{eq:1a}
    W_\text{MSSM} & = Y_{u}^{ij} u^c_{i} Q_{j} H_{u} + Y_{d}^{ij}
    d^c_{i} Q_{j} H_{d} + Y_{l}^{ij} e^c_{i} L_{j} H_{d} + \mu
    H_{d}H_{u} \,.
    \intertext{Neutrinos are massless in the MSSM; however, experiments
      and cosmological observations consistently point at small but
      non-vanishing masses in the sub-eV region.  We will therefore
      consider an extended MSSM with three right handed neutrinos, where
      the Yukawa couplings are given by}
    W & = W_\text{MSSM} + Y_{\nu}^{ij} \nu^c_{i} L_{j} H_{u} +
    \frac{1}{2} M_{R}^{ij} \nu^c_{i} \nu^c_{j} \,.
  \end{align}
\end{subequations}
Here, $Q$ and $L$ denote the chiral superfields of the quark and lepton
doublets and $u^c$, $d^c$, $e^c$ and $\nu^c$ the up and down-quark,
electron and neutrino singlets, respectively.  Each
  chiral superfield consists of a fermion and its scalar partner, the
  sfermion.  The Yukawa coupling matrices $Y_{u,d,l,\nu}$ are defined
with the right-left (RL) convention.  The field $\nu^c$ is sterile under
the SM group, so we allow for a Majorana mass term in addition to the
Dirac coupling.  The respective mass matrix is denoted by $M_R$ and the
scale of $M_R$ is undetermined but expected to be above the electroweak
scale, $M_\text{ew}$ (see Sec.~\ref{sec:RGE}).

The Higgs fields acquire the vacuum expectation values (vevs)
\begin{align}
  \langle H_u \rangle & = v_u \,, \mspace{60mu} \langle H_d \rangle =
  v_d \,.
\end{align}
where $\vert v_u \vert ^2 + \vert v_d \vert ^2 = v^2 = (174 \
\text{GeV})^2$.  The ratio of the two vevs is undetermined and defines
the parameter $\tan\beta$,
\begin{align}
  \frac{v_u}{v_d} =: \tan\beta \ .
\end{align}
While $\tan\beta$ is a free parameter of the theory, there exist lower
and upper bounds on its value.  Experimentally, Higgs searches at LEP
rule out the low-$\tan\beta$ region in simple SUSY models
\cite{Gianotti:2002vg}.  This result fits nicely with the theoretical
expectation that the top Yukawa coupling should not be larger than one.
The region of the MSSM parameter space with large values of
  $\tan\beta$ is of special importance for the flavour physics of quarks
  and leptons. We therefore have a brief critical look at the upper
  bounds on this parameter: Demanding a perturbative bottom Yukawa
  coupling $y_b$ na\"ively leads to an upper limit on $\tan\beta$
of about 50 inferred from the tree-level relation
\begin{align} 
  \label{yukawatanbeta}
  y_b = - \frac{m_b}{v \cdot \cos\beta} \approx - \frac{m_b}{v}
  \tan\beta \ .
\end{align}
Similarly, the MSSM provides a natural radiative breaking mechanism of the
electroweak symmetry as long as $y_b < y_t$ at a low scale
\cite{Drees:2004jm}.  At tree level, the ratio of the Yukawa couplings
is given by
\begin{align}
  \label{ratiotreelevel}
  1 > \left|\frac{y_b}{y_t}\right| = \frac{m_b}{m_t} \tan\beta \;.
\end{align}
Since $m_t(\mu)/m_b(\mu) \approx 60$ at the electroweak scale,
$\tan\beta$ should not exceed this value.

Both arguments, however, do only hold at tree level.  In particular,
down quarks as well as charged leptons couple to $H_u$ via loops.  As a
result, if we take $\tan\beta$-enhanced contributions into account, an
explicit mass renormalisation changes the relation of Yukawa coupling
and mass \cite{Hall:1993gn,Carena:1993ag,Hempfling:1993kv}.  The
$\tan\beta$ enhancement of the bottom coupling in
Eq.~(\ref{yukawatanbeta}) can be compensated; similarly, the ratio of
the Yukawa couplings is changed due to an explicit bottom quark mass
renormalisation.  We will find that values of $\tan\beta$ up to 100 both
provide small enough Yukawa couplings and do not destroy natural
electroweak symmetry breaking.

Large values for $\tan\beta$ are interesting for two reasons.  One, in
various grand-unified theories (GUTs), top and bottom Yukawa couplings
are unified at a high scale.  In this case, it is natural to expect
$\tan\beta=m_t/m_b$, as shown above.  Two, many supersymmetric loop
processes are $\tan\beta$-enhanced due to chirality-flipping loop
processes with supersymmetric particles in the loop.  This enhancement
can compensate the loop suppression and therefore large values of
$\tan\beta$ lead to significant SUSY corrections.

In this paper, we will study the lepton sector in the (extended) MSSM.
Since the neutrinos are massive, the leptonic mixing matrix,
$U_\text{PMNS}$, is no longer trivial and leads to lepton flavour
violation (LFV).  In its standard parametrisation, it reads
\begin{align}
  \label{PMNSMATRIX}
  U_\text{PMNS} & =  
  \begin{pmatrix}
    1 & 0 & 0 \cr 0 & c_{23} & s_{23} \cr 0 & -s_{23} & c_{23}
  \end{pmatrix}
  \begin{pmatrix}
    c_{13} & 0 & s_{13}e^{i\delta} \cr 0 & 1 & 0 \cr -s_{13} e^{i\delta}
    & 0 & c_{13}
  \end{pmatrix}
  \begin{pmatrix}
    c_{12} & s_{12} & 0 \cr -s_{12} & c_{12} & 0 \cr 0 & 0 & 1
  \end{pmatrix}
  \begin{pmatrix}
    e^{i\frac{\alpha_1}{2}} & 0 & 0 \cr 0 & e^{i\frac{\alpha_2}{2}} & 0
    \cr 0 & 0 & 1
  \end{pmatrix} 
  ,
\end{align}
with $s_{ij} = \sin\theta_{ij}$ and $c_{ij} = \cos\theta_{ij}$.  The two
phases $\alpha_{1,2}$ appear if neutrinos are Majorana particles.  They
are only measurable in processes which uncover the Majorana nature of
neutrinos, such as neutrinoless double beta decay.  

The PMNS matrix allows for flavour transitions in the lepton sector, in
particular neutrino oscillations, through which its parameters are well
constrained.  Compared with the mixing angles of the quark mixing
matrix, $V_\text{CKM}$, two mixing angles, namely the atmospheric and
solar mixing angles, $\theta_{23} = \theta_\text{atm}$ and $\theta_{12}
= \theta_\text{sol}$, are surprisingly large, whereas the third mixing
angle is small.  The current experimental status at $1\sigma$ level is
as follows \cite{GonzalezGarcia:2007ib}:\footnote{Recently, a hint for
  non-zero $\theta_{13}$, $\sin^2{\theta_{13}}=0.016 \pm 0.010\
  (1\sigma)$, was claimed in Ref.~\cite{Fogli:2008jx}.}
\begin{align}
  \theta_{12} & = 34.5 \pm 1.4 \;, & \Delta m_{21}^{2} & =
  7.67^{+0.22}_{-0.21} \cdot 10^{-5}\ \textrm{eV}^{2}, \nonumber
  \\
  \theta_{23} & = 42.3^{+5.1}_{-3.3} \;, & \Delta m_{31}^{2} & =
  \begin{cases}
    -2.37\pm 0.15 \cdot 10^{-3}\ \textrm{eV}^{2} & \textrm{inverted
      hierarchy},
    \\
    +2.46\pm 0.15 \cdot 10^{-3}\ \textrm{eV}^{2} & \textrm{normal
      hierarchy},
  \end{cases}
  \nonumber
  \\
  \theta_{13} & = 0.0^{+7.9}_{-0.0} \;.
  \label{nu-best-fit}
\end{align}
These values are determined by the atmospheric and solar mass splitting
$\Delta m_\text{atm}^{2}= \Delta m_{13}^{2}$, $\Delta m_\text{sol}^{2}=
\Delta m_{21}^{2}$, leaving the absolute mass scale open.  
The pattern of mixing angles is close to tri-bimaximal, corresponding to
$\theta_{23} = 45^{\circ}$, $\theta_{12} \simeq 35^{\circ}$, and
$\theta_{13} = 0^{\circ}$ \cite{Harrison:2002er}.  Due to the smallness
of $\theta_{13}$, the CP phase $\delta$ is unconstrained.
Tri-bimaximal mixing can be motivated by symmetries (see
  Ref.~\cite{King:2009db} and references therein), which constrain
  fundamental quantities like Yukawa couplings or soft SUSY-breaking
  terms. Measurable quantities like $U_{e3}$ usually do not point
  directly to fundamental parameters, but are sensitive to corrections
  from all sectors of the theory. The analysis of such corrections is
  therefore worthwhile.  A large portion of this paper is devoted to the
  influence of supersymmetric loops and higher-dimensional Yukawa terms
  on observables in the lepton sector of the MSSM.

In a supersymmetric framework, additional lepton flavour violation can
be induced by off-diagonal entries in the slepton mass matrix, which
parametrise the lepton-slepton misalignment in a model independent way.
However, a generic structure of the soft masses is already excluded
because too large decay rates for $l_{j}\rightarrow l_{i} \gamma$ would
arise.  To avoid this flavour problem, the SUSY breaking mechanism is
often assumed to be flavour blind, yielding universal soft masses at a
high scale.  Then the PMNS matrix is the only source of flavour
violation in the lepton sector, as is the CKM matrix for the quarks;
this ansatz is called minimal flavour violation.  The soft terms do not
cause additional flavour violation and the various mass and coupling
matrices are flavour-diagonal at some scale in the basis of fermion mass
eigenstates, e.g.,
\begin{align}
  m_{\tilde{L}}^{2} = m_{\tilde{e}}^{2} & = m_{0}^{2}\, \mathds{1} \,, &
  m_{H_{u}}^{2} = m_{H_{d}}^{2} & = m_{0}^{2} \;, & A_l & = A_{0} Y_l
  \;.
\end{align}
Here, $m_{\tilde{L},\tilde{e}}^2$ denote the soft mass matrices of the
sleptons (see Eq.~(\ref{Sleptonmassenmatrix})), $m_H^2$ the analogous
soft masses of the Higgs doublets, and $A_l$ is the trilinear coupling
matrix of the leptons.

Even if the soft terms are universal at the high scale, renormalisation
group equations (RGE) can induce non vanishing off-diagonal entries in the
slepton mass matrix at the electroweak scale.  Lepton flavour violation
can be parametrised by non-vanishing $\delta_{XY}^{ij}$ at the
electroweak scale in a model-independent way, where $\delta_{XY}^{ij}$
is defined as the ratio of the flavor-violating elements of the slepton
mass matrix (\ref{Sleptonmassenmatrix}) and an average slepton mass (see
Eq.~(\ref{eq:deltas})),
\begin{align}
  \label{eq:introdelta}
  \delta_{XY}^{ij} = \frac{\Delta m_{XY}^{ij}}{\sqrt{m_{i_{X}}^{2}
      m_{j_{Y}}^{2}}}, \mspace{60mu} X,\,Y = L,\,R, \qquad i,\,j =
  1,\,2,\,3 \quad (i\not= j) \ .  
\end{align}
The flavour-off-diagonal elements $\Delta m_{XY}^{ij}$ are defined in a
weak basis in which the lepton Yukawa matrix $Y_l$ in \eq{eq:1a} is
diagonal.  According to the chiralities of the sfermion involved, there
are four different types, $\delta_{LL}$, $\delta_{RR}$, $\delta_{LR}$,
and $\delta_{RL}$.  The tolerated deviation from alignment can be
quantified by upper bounds on $\delta_{XY}^{ij}$, as discussed above
and are already extensively studied in the literature
  (see for example \cite{Gabbiani:1996hi,Masina:2002mv,Paradisi:2005fk}
  and references therein).

Being generically small, the sfermion propagator can be expanded in
terms of these off-diagonal elements, corresponding to the mass
insertion approximation (MIA)
\cite{Hall:1985dx,Gabbiani:1988rb}.  Instead of
diagonalising the full slepton mass matrix and dealing with mass
eigenstates and rotation matrices at the vertices, in
  MIA one faces flavour-diagonal couplings and LFV appears as a mass
insertion in the slepton propagator.  This approach is valid as long as
$\left|\delta_{XY}^{ij}\right| \ll 1$ and makes it possible to identify certain
contributions easily.  For a numerical analysis an exact diagonalisation
of all mass matrices is, of course, possible.  In
  Ref. \cite{Paradisi:2005fk} a systematic comparison between the full
  computation and the MIA both in the slepton and chargino/neutralino
  sector clarifies the applicability of these approximations. 

\smallskip

This paper provides a comprehensive analysis of the lepton sector in the
MSSM, focusing on the phenomenological constraints on the
  parameters $\delta_{XY}^{ij}$ in \eq{eq:introdelta}. In Sec.~2 we
  briefly review the supersymmetric threshold corrections to $y_b$ in
  \eq{yukawatanbeta} and relax the usually quoted upper bounds on
  $\tan\beta$. In Sec.~3 we derive new constraints on the
  $\delta_{XY}^{ij}$'s by studying loop corrections to the electron
  mass, finite renormalisations of the PMNS matrix and the magnetic
  moment of the electron at large $\tan\beta$. As a byproduct we
identify all
  $\tan\beta$-enhanced corrections to the charged-Higgs coupling to
  leptons. We then improve the MSSM predictions for the decay rates of
  $l_j\to l_i\gamma$ by including $\tan\beta$-enhanced two-loop
  corrections. In Sec.~4 we embed the MSSM into SO(10) GUT scenarios
  and study RGE effects.  Even for flavour-universal soft breaking terms
  at the GUT scale $M_\text{GUT}$ sizable flavour violation can be
  generated between $M_\text{GUT}$ and the mass scales of the
  right-handed neutrinos \cite{Borzumati:1986qx}.  Comparing the results
  to the upper bounds on $|\delta_{XY}^{ij}|$ found in Sec.~3 enables
  us to draw general conclusions on GUT scenarios. We
  include corrections to the Yukawa couplings from dimension-5 terms and
  constrain their possible flavour misalignment with the dimension-4
  Yukawa matrices. In Sec.~5 we summarise our results. Our notations
  and conventions are listed in three appendices.
\section{Upper bound for
  \boldmath{$\tan\beta$}}
\label{massrenormalization}

The tree level mass of a particle receives corrections due to virtual
processes.  Supersymmetric loop corrections are small unless a large
value for $\tan\beta$ compensates for the loop suppression.  We will
therefore start with a discussion of $\tan\beta$-enhanced loops and
derive a relation between Yukawa coupling and mass, coming from the
resummation of $\tan\beta$-enhanced corrections to the mass to all
orders.

Corrections to the mass with more than one loop and more than one
coupling to Higgs fields do not produce further factors of $\tan\beta$.
Nevertheless, $\tan\beta$-enhanced loops can become important in an
explicit mass renormalisation, where $\tan\beta$-enhanced contributions
to all orders are taken into account, because counterterms are
themselves $\tan\beta$-enhanced.

\begin{figure}
  \centering
  \scalebox{0.98}{
    \begin{picture}(180,65)(0,0)
      \SetColor{Black}
      \ArrowLine(10,30)(55,30) \Text(30,20)[]{$l_{L}$}
      \ArrowLine(95,30)(140,30) \Text(120,20)[]{$l_{R}$}
      \DashArrowArc(75,30)(20,180,360){2}
      \Text(75,0)[]{$\tilde{\nu}_{l}$} 
      \ArrowArc(75,30)(20,0,180) 
      \PhotonArc(75,30)(20,0,180){3}{4.5}
      \Text(75,62)[]{$\tilde{\chi}_{j}^{\pm}$}
    \end{picture}
    \begin{picture}(140,65)(160,0)
      \ArrowLine(160,30)(205,30)\Text(180,20)[]{$l_{L}$}
      \ArrowLine(245,30)(290,30)\Text(270,20)[]{$l_{R}$}
      \DashArrowArc(225,30)(20,180,360){2}
      \Text(225,0)[]{$\tilde{l}_{i}$} 
      \ArrowArc(225,30)(20,0,180) 
      \PhotonArc(225,30)(20,0,180){3}{4.5} 
      \Text(225,62)[]{$\tilde{\chi}_{j}^{0}$}
    \end{picture}
  }%
  \caption{Contribution to the self-energy $\Sigma$ arising from
    chargino-sneutrino and neutralino-slepton loops}
  \label{Selbstenergie} 
\end{figure}
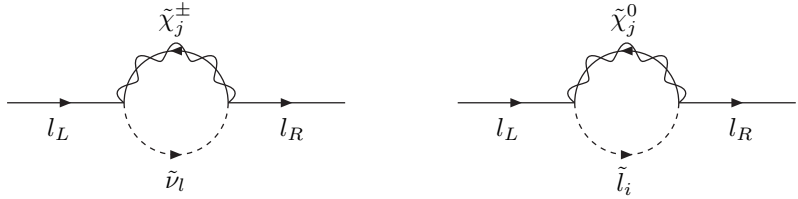

Down-quarks and charged leptons receive $\tan\beta$-enhanced corrections
to their masses, due to loops with $H_u$.  As a coupling to $H_u$ does
not exist at tree level, $\tan\beta$-enhanced loops are finite; there is
no counterterm to this loop induced coupling.  Moreover,
$\tan\beta$-enhanced contributions to self-energies do not decouple.
The coupling of $H_u$ to the charged slepton is proportional to $\mu$
with $\mu=\mathcal{O}(M_{\text{SUSY}})$.  On the other hand, the
integration over the loop momentum gives a factor $1/M_{\text{SUSY}}$.
Thus the dependence on the SUSY mass scale cancels out.  In the large
$\tan\beta$ regime, neutralino-slepton and chargino-sneutrino loops can
significantly change the relation between the Yukawa couplings and
masses \cite{Carena:1999py,Marchetti:2008hw,Hofer:2009xb},
\begin{align}
  \label{Massrenfinal}
  m^{(0)}_l = m^\text{phys}_l + \sum_{n=1}^{\infty} \left(-\Delta\right)^{n}
  m^\text{phys}_l = \frac{m^\text{phys}_l} {1+\Delta_l} \
  ,
\end{align}
where the corrections $\Delta_l$ are related to the self-energy $\Sigma_l$
as $\Delta_l=-\Sigma_l/m_l$.  This relation includes all $\tan\beta$-enhanced
contributions and can be determined by only calculating two diagrams,
according to chargino-sneutrino and neutralino-slepton loops
(Figure~\ref{Selbstenergie}).  As a result, the relation between Yukawa
coupling and physical mass is given by
\begin{align}
  \label{equ:YukKorrektur} 
  y_l = -\frac{m^{(0)}_l}{v_{d}} =
  -\frac{m^\text{phys}_l}{v_{d}\left(1+\Delta_l\right)} \ .
\end{align}

The individual contributions from the two diagrams to the self-energy
$\Sigma = \Sigma^{\tilde{\chi}^{\pm}} + \Sigma^{\tilde{\chi}^{0}}$ are
\begin{align}
  \Sigma_{l_{L}-l_{R}}^{\tilde{\chi}^{\pm}} & = \frac{1}{16\pi^{2}}
  \sum_{j=1,2} M_{\tilde{\chi}_{j}^{\pm}}\, P_{L}\,
  \Gamma_{l}^{\tilde{\chi}_{j}^{\pm}\tilde{\nu}_{l}*}
  \Gamma_{l}^{\tilde{\chi}_{j}^{\pm}\tilde{\nu}_{l}} P_{L}\,
  B_{0}\left(M_{\tilde{\chi}_{j}^{\pm}}^{2},\,
    m_{\tilde{\nu}_{l}}^{2}\right) , \nonumber
  \\
  \label{Selbstexaktneu}
  \Sigma_{l_{L}-l_{R}}^{\tilde{\chi}^{0}} & = \frac{1}{16\pi^{2}}
  \sum_{i=1,2} \sum_{j=1}^{4} M_{\tilde{\chi}_{j}^{0}}\, P_{L}\,
  \Gamma_{l}^{\tilde{\chi}_{j}^{0}\tilde{l}_{i}*}
  \Gamma_{l}^{\tilde{\chi}_{j}^{0}\tilde{l}_{i}} P_{L}\,
  B_{0}\left(M_{\tilde{\chi}_{j}^{0}}^{2},\,
    m_{\tilde{l}_{i}}^{2}\right) .
\end{align}
The loop integrals, couplings and rotation matrices are defined in the
appendices \ref{appendix:loopintegrals} and
\ref{appendix:massesmixingfeynmanrules}; $M_{\tilde{\chi}^{\pm}}$ and
$M_{\tilde{\chi}^{0}}$ denote the chargino and neutralino masses,
respectively.  The $\tan\beta$-enhanced transitions require a chirality
flip and are given by
\begin{align}
  \label{Selbsttanbeta}
  \Delta_{l} = -\frac{\Sigma_{l}}{m_{l}} = \epsilon_{l}\tan\beta & =
  \frac{\alpha_{1}}{4\pi} M_1\, \mu\, \tan\beta \left[\frac{1}{2}
    f_1\left(M_{1}^{2},\, \mu^{2},\, m_{\tilde{l}_{L}}^{2}\right) -
    f_1\left(M_{1}^{2},\, \mu^{2},\, m_{\tilde{l}_{R}}^{2}\right) +
    f_1\left(M_{1}^{2},\, m_{\tilde{l}_{L}}^{2},\,
      m_{\tilde{l}_{R}}^{2}\right) \right] \nonumber
  \\
  & \quad -\frac{\alpha_{2}}{4\pi} M_{2}\, \mu\, \tan\beta \left[
    \frac{1}{2} f_1\left(M_{2}^{2},\, \mu^{2},\,
      m_{\tilde{l}_{L}}^{2}\right) + f_1\left(M_{2}^{2},\, \mu^{2},\,
      m_{\tilde{\nu}_{l}}^{2}\right) \right] ,
\end{align}
with the loop-function $f_1(x,y,z)$ defined in Eq.~(\ref{i-def}).

\begin{figure}
  \centering
  \includegraphics[width=0.5\linewidth]{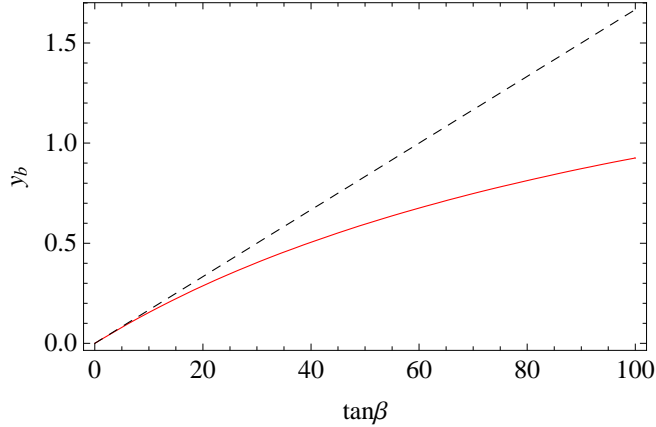}
  \caption{Bottom quark coupling as a function of $\tan\beta$ for
    $\epsilon_b = 0$ (black-dashed) and $\epsilon_b = 0.008$
    (red-solid).}
  \label{yukawabtanbetaplot1}
\end{figure}

The improved relation between the Yukawa coupling and the physical mass
(\ref{equ:YukKorrektur}) relaxes the upper bound for $\tan\beta$, as
indicated in the Introduction.  Equation (\ref{yukawatanbeta}) changes
to \cite{Carena:1999py}
\begin{align}
  y_b & = - \frac{m_b}{v}\frac{\tan\beta}{1+\epsilon_b\tan\beta} \,.
\end{align}
For down-quarks, the SUSY contributions are dominated by gluino loops
such that
\begin{align}
  \epsilon_b & \simeq \frac{2\alpha_s}{3\pi} m_{\tilde g}\, \mu\,
  f_1\left(m_{\tilde{b}_1},\, m_{\tilde{b}_2},\, m_{\tilde g}\right) .
\label{epsb}
\end{align}
A typical value is $\epsilon_b \approx 0.008$,
leading to $y_b = \mathcal{O}(1)$ for $\tan\beta \approx 100$ (see
Fig.~\ref{yukawabtanbetaplot1}).  Similarly, the bound for natural
electroweak symmetry breaking shifts to $\tan\beta \lesssim 100$.

In the following, we will use this relation (\ref{equ:YukKorrektur}) to
constrain lepton flavour violating parameters.

\section{Constraints on the flavour-violating parameters}
\label{se:lfv}

Various processes can be used to constrain lepton flavour violating
(LFV) parameters.  Remarkably, we can also use lepton flavor conserving
(LFC) observables, due to double lepton flavour violation (LFV).  Two
LFV vertices lead to lepton flavor conservation (LFC) and so contribute
to the LFC self-energies.  In a similar manner, we will consider
multiple flavour changes contributing to the magnetic moment of the
electron.  In addition, we consider LFV processes, in particular
radiative decays.

As mentioned in the Introduction, we introduce the dimensionless
parameters $\delta_{XY}^{ij}$ via
\begin{align}
  \label{eq:deltas}
  \Delta m_{XY}^{ij} = \delta_{XY}^{ij}
  \left(\overline{m}_{XY}^{ij}\right)^{2} = \delta_{XY}^{ij}
  \sqrt{m_{i_{X}}^{2} m_{j_{Y}}^{2}} \ , \quad X,\,Y = R,\, L \ , \quad
  i \not= j \ .
\end{align}
Note that $\Delta m_{XY}^{ij}$ has mass-dimension two.  Both
$m_{i_{X}}^2$ and $\Delta m_{XY}^{ij}$ are the diagonal and off-diagonal
entries of the slepton mass matrix (\ref{Sleptonmassenmatrix}), so
$\overline{m}_{XY}^{ij}$ is an average slepton mass.

{The effects discussed in this section stem from chirally enhanced
self-energies, which involve an extra factor of $\tan\beta$ compared
to the tree-level result, analogous to $\epsilon_b$ in \eq{epsb}.
Such effects have been widely studied before in the quark sector, yet
most authors have performed their studies in the decoupling limit
$M_{\rm SUSY}\gg v$, where $M_{\rm SUSY}$ denotes the mass scale of
the soft SUSY breaking parameters. (For a guide through the literature
see Ref.~\cite{Hofer:2009xb}.) If one relaxes the condition $M_{\rm
SUSY}\gg v$, novel effects (namely those which vanish like some power
of $v/M_{\rm SUSY}$) can be analysed. Analytical results covering the
case $M_{\rm SUSY}\sim v$ have been derived in
Refs.~\cite{Carena:1999py,Crivellin:2008mq,Hofer:2009xb,Crivellin:2009ar},
numerical approaches were pursued in
Refs.~\cite{Buras:2002vd,Ellis:2007kb}. Superparticle contributions to
physical processes vanish for $M_{\rm SUSY}\to \infty$, typically as
$v^2/M_{\rm SUSY}^2$, and can only be addressed with the methods of
the latter papers. However, by combining the decoupling supersymmetric
loop with resummation formulae valid for $M_{\rm SUSY}\gg v$ one can
correctly reproduce the resummed all-order result to leading
non-vanishing order in $v/M_{\rm SUSY}$. This approach has been used
in an analysis of ``flavoured'' electric dipole moments in
Ref.~\cite{Hisano:2008hn}.}
 
{The possibility to constrain $\delta_{XY}^{ij}$ through LFC
processes has been pointed out in Ref.~\cite{Masiero}, which addresses
leptonic Kaon decays. Here we analyse the constraints from two LFC
observables which have not been considered before: In
Sec.~\ref{se:lfcse} we apply a naturalness argument to the electron
mass, demanding that the supersymmetric loop corrections are smaller
than the measured value. In Sec.~\ref{se:ge} we study the anomalous
magnetic moment of the electron.}

\subsection{Flavour-conserving self-energies}
\label{se:lfcse}

The masses of the SM fermions are protected from radiative corrections
by the chiral symmetry $\Psi\rightarrow e^{i\alpha \gamma_{5}}\Psi$.
According to Weinberg, Susskind and 't Hooft, a theory with small
parameters is natural if the symmetry is enhanced when these parameters
vanish.  The smallness of the parameters is then protected against large
radiative corrections by the concerned symmetry.  This naturalness
principle makes the smallness of the electron mass natural.  Radiative
corrections are proportional to the electron mass itself $\delta
m_{e}\propto m_{e} \ln\left(\Lambda/m_{e}\right)$ and vanish for
$m_{e}=0$.  If such a small parameter is composed of some different
terms and one does not want any form of fine-tuning, one should require
that all contributions should be roughly of the same order of magnitude;
no accidental cancellation between different terms should occur.  Hence,
the counterterm of the electron mass should be less than the measured
electron mass,
\begin{align}
  \left|\frac{\delta m_{e}}{m_{e}^\text{phys} }\right| = \left|
    \frac{m_{e}^{(0)}-m_{e}^\text{phys}}{m_{e}^\text{phys}}
  \right| < 1 \ .
\end{align}
This naturalness argument for the light fermion masses
  was already discussed in the pioneering study of
  Ref. \cite{Gabbiani:1996hi}.  Since the authors of this analysis want
  to provide a model-independent analysis on classes of SUSY theories they
  consider only photinos and do not include flavour violation in the
  corrections to light fermion masses. Their derived upper bound for
  $\delta_{LR}^{11}$ depends on the overall SUSY mass scale and actually
  becomes stronger for larger SUSY masses.  The case of radiatively
  generated fermion masses via soft trilinear terms is studied in
  \cite{Borzumati:1999sp} and an updated version including two
  flavour-violating self-energies can be found in
  \cite{Crivellin:2010gw}.  Here we concentrate on the
  chirality-conserving flavour-violating mass insertions and use this
  argument to restrict the product $\delta_{LL}^{13}\delta_{RR}^{13}$
  which is then independent of the SUSY scale.   Considering double
lepton flavour violation, we demand that the radiative corrections
should not exceed the tree-level contribution.  For the electron mass,
the dominant diagrams involve couplings to the third generation.  As a
result, we can constrain the product
$\left|\delta_{LL}^{13}\delta_{RR}^{13}\right|$.  Note that
$\left|\delta_{RR}^{13}\right|$ has so far been unconstrained from
radiative decay $l_{j}\rightarrow l_{i}\gamma$ as we will discuss
shortly in the following Section~\ref{se:lfvse}. This
  cancellation of the RR sensitivity is analysed in
  \cite{Masina:2002mv,Paradisi:2005fk} with the conclusion that even
  a better experimental sensitivity on $BR(l_j\to l_i\gamma)$ can not help
  to set strong constraints in the RR sector. However, with double mass
  insertions and the bound from $\mu\to e\gamma$ it is possible to
  derive bounds on products like $\delta_{LL}^{23}\delta_{RR}^{31}$. 

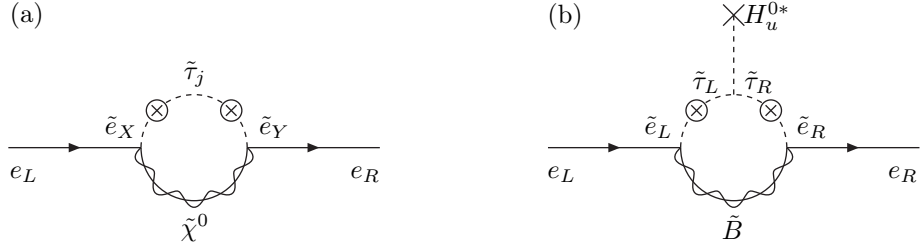
\begin{figure}
  \centering
  \scalebox{1.0}{%
    \begin{picture}(200,100)(-170,10)
      \SetColor{Black}
      \ArrowLine(-170,50)(-120,50) \ArrowLine(-80,50)(-30,50)
      \Text(-170,40)[l]{$e_{L}$} \Text(-30,40)[r]{$e_{R}$}
      \DashCArc(-100,50)(20,0,180){2}
      \Text(-100,78)[]{$\tilde{\tau}_{j}$}
      \BCirc(-86,64){4}\BCirc(-114,64){4}
      \Line(-88,62)(-84,66)\Line(-88,66)(-84,62)
      \Line(-116,62)(-112,66)\Line(-116,66)(-112,62) 
      \CArc(-100,50)(20,180,0)
      \Text(-128,58)[]{$\tilde{e}_{X}$}
      \Text(-70,58)[]{$\tilde{e}_{Y}$}
      \PhotonArc(-100,50)(20,180,0){2.5}{4.5}
      \Text(-100,20)[]{$\tilde{\chi}^{0}$}
      \Text(-170,100)[l]{(a)}
    \end{picture}
    \begin{picture}(130,100)(30,10)
      \ArrowLine(30,50)(80,50) \ArrowLine(120,50)(170,50)
      \Text(30,40)[l]{$e_{L}$} \Text(170,40)[r]{$e_{R}$}
      \DashLine(100,100)(100,70){2}
      \Line(96,96)(104,104) \Line(96,104)(104,96)
      \Text(113,100)[]{$H_{u}^{0*}$}
      \DashCArc(100,50)(20,0,180){2}
      \Text(90,75)[]{$\tilde{\tau}_{L}$}
      \Text(110,75)[]{$\tilde{\tau}_{R}$}
      \BCirc(114,64){4}\BCirc(86,64){4}
      \Line(112,62)(116,66)\Line(112,66)(116,62)
      \Line(84,62)(88,66)\Line(84,66)(88,62) 
      \CArc(100,50)(20,180,0)
      \Text(72,58)[]{$\tilde{e}_{L}$}
      \Text(130,58)[]{$\tilde{e}_{R}$}
      \PhotonArc(100,50)(20,180,0){2.5}{4.5}
      \Text(100,20)[]{$\tilde{B}$}
      \Text(30,100)[l]{(b)}
    \end{picture}
  }%
  \caption{(a) LFC self-energy through double LFV and (b) dominant
    double LFV contribution to the electron mass renormalisation.}
  \label{doppelteLVFdominant}
\end{figure}

The diagram in Figure~\ref{doppelteLVFdominant}(a) achieves an
$m_{\tau}\tan\beta$ enhancement only if there is a helicity flip in the
stau propagator.  Since $\alpha_{1}/\left(4\pi\right) \gg
Y^{2}_{e}/\left(16\pi^{2}\right)$, the higgsino contribution is
negligible.  A chargino loop can also be neglected, because only
left-left (LL) insertions for the sneutrinos can be performed and the
helicity flip is associated with an electron Yukawa coupling.  The
left-right (LR) insertions are either not associated with an
$\tan\beta$-enhanced contribution or suppressed by $v/M_\text{SUSY}$,
compared to right-right (RR) and LL-insertions.  Thus the dominant
diagram involves a Bino and the scalar tau-Higgs coupling, as shown in
Figure~\ref{doppelteLVFdominant}(b).
For simplicity, we choose all parameters real and obtain
\begin{align}
  \label{Sigma_{e}^{FV}}
  \Sigma_{e}^\text{FV} & \simeq \frac{\alpha_{1}}{4\pi} M_{1}\, \mu\,
  \frac{ m_{\tau}^\text{phys}\tan\beta}{1 + \Delta_{\tau}} \Delta m_{LL}^{13} \Delta
  m_{RR}^{13}\, F_{0}\left(M_{1}^{2},\, m_{\tilde{e}_{L}}^{2},\,
    m_{\tilde{e}_{R}}^{2},\, m_{\tilde{\tau}_{L}}^{2},\,
    m_{\tilde{\tau}_{R}}^{2}\right) \nonumber
  \\[3pt]
  & \simeq -\frac{\alpha_{1}}{4\pi} M_{1}\, \mu\, 
  \frac{m_{\tau}^\text{phys}\tan\beta}{1 + \Delta_{\tau}} \Delta m_{LL}^{13} \Delta m_{RR}^{13}\,
  f_1''\left(M_{1}^{2},\, m_{\tilde{L}}^{2},\, m_{\tilde{R}}^{2}\right)
  .
\end{align}
For equal SUSY masses this simplifies to $\Sigma_{e}^\text{FV} =
-\frac{\alpha_1}{48 \pi}\frac{m_\tau\tan\beta}{1 +
\Delta_{\tau}}\delta_{LL}^{13}\delta_{RR}^{13}$.
This term is proportional to $m_{\tau}$, in contrast to the LFC self
energy, which is proportional to $m_{e}$.  Thus the counterterm receives
an additional constant term $\Sigma_{e}^\text{FV}$,
\begin{center}
  \begin{picture}(370,30)(10,0)
    \SetColor{Black}
    \ArrowLine(10,20)(60,20)
    \Text(35,0)[]{$-i m^\text{phys}_l$}
    \Text(70,20)[]{$+$}
    \ArrowLine(80,20)(110,20) \ArrowLine(110,20)(140,20)
    \BCirc(110,20){10} 
    \Text(110,0)[]{$i\Sigma^{(1)}=-m^\text{phys}_l\Delta_l$} 
    \Text(150,20)[]{$+$}
    \ArrowLine(160,20)(190,20) \ArrowLine(190,20)(220,20)
    \BCirc(190,20){10} \Text(190,0)[]{$i\Sigma^\text{FV}_l$} 
    \Text(230,20)[]{$+$}
    \ArrowLine(240,20)(270,20) \ArrowLine(270,20)(300,20)
    \Text(270,0)[]{$-i \delta m^{(1)}_l$} \BCirc(270,20){5}
    \Line(267,23)(273,17) \Line(267,17)(273,23) 
    \Text(310,22)[]{$\stackrel{!}{=}$}
    \ArrowLine(320,20)(370,20) \Text(345,0)[]{$-i m^\text{phys}_l$} 
  \end{picture}
\end{center}
\medskip

\noindent
Substituting $m^\text{phys}_l \to m^\text{phys}_l + \delta m^{(1)}_l$, one gets the second
order contributions since the only real diagrams of order $n$ are
one-loop diagrams in which a counterterm of order $n-1$ is inserted.

We will use the on-shell renormalisation scheme, where the mass and the
wave-function counterterms cancel all loop contributions to the
self-energy such that the pole of the lepton propagator is equal to the
physical mass of the lepton. Then we obtain the relation
\begin{align}
  m^{(0)}_l & = m^\text{phys}_l + \sum_{n=1}^{\infty} \delta m^{(n)}_l =
  \frac{m^\text{phys}_l}{1+\Delta_l} + \frac{\Sigma^\text{FV}_l}{1+\Delta_l} \ .
\end{align}

\begin{table}
  \centering
  \begin{tabular}{|c|rrrrr|rr|}
    \hline
    SPS & 1a & 1b & 2 & 3 & 4 & A & B\\
    \hline
    $m_{0}$ & $100 $&$200$ & $1450$& $90$ & $400$ & $500$ & $500$ \\
    $A_{0}$ & $-100$& $0$ & $0$&  $0$&$ 0$ & $0$& $0$\\
    $m_{1/2}$ &$ 250$& $400$ &$300$ &$400$  &$300$ & $500$& $500$\\
    $\tan\beta$ & $10$& $30$ &$10$ &  $10$&$ 50$&$ 40$&$ 10$\\
    $sgn(\mu)$ & $ + 1$&$ + 1$ &$  + 1$& $ + 1$ &$ + 1$ &$ + 1$&$ + 1$\\
    \hline
    $\mu(M_{Z})$ & $352$ & $507$ & $422$ & $516$ & $388$ & $614$& $629$\\
    \hline
  \end{tabular}
  \caption{Snowmass Points and Slopes \cite{Allanach:2002nj} and two
    additional scenarios (masses in GeV)}
  \label{SPS}
\end{table}

\begin{figure}
  \centering
  \psfrag{DeltaLL}{\hspace{0.2cm}$\delta_{LL}^{13}$}
  \psfrag{DeltaRR}{\hspace{0.2cm}$\delta_{RR}^{13}$}
  \psfrag{mel}{\scalefont{0.8}$m_{e}$}
  \psfrag{0.1 mel}{\scalefont{0.8}$0.1m_{e}$}
  \psfrag{0.5 mel}{\scalefont{0.8}$0.5m_{e}$}
  \includegraphics[width=7cm]{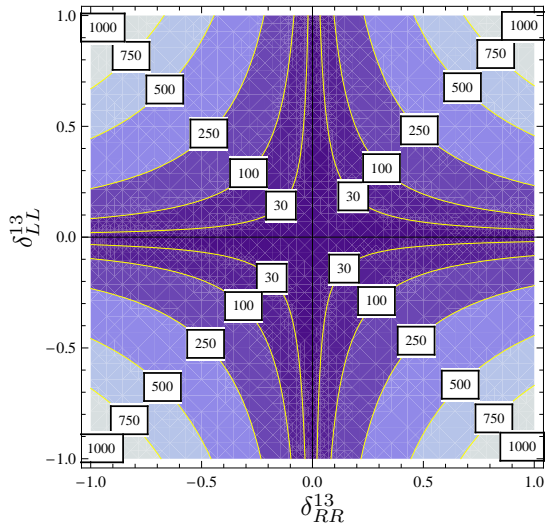}
  \caption{Percentage deviation of the electron mass through SUSY loop
    correction in dependence of real $\delta_{RR}^{13}$ and
    $\delta_{LL}^{13}$ for SPS4.}
  \label{LFVelRRLL}
\end{figure}

For the numerical analysis, let us consider the mSUGRA scenario SPS4;
its parameter values are given in Table~\ref{SPS}.  For this model, the
constraint reads (Figure~\ref{LFVelRRLL})
\begin{align}
\label{eq:bdme}
  \left|\delta_{RR}^{13}\delta_{LL}^{13}\right| & <
  \begin{cases}
    0.097, & \text{if}\ \delta_{RR}^{13} \delta_{LL}^{13} > 0
    \\[2pt]
    0.083, & \text{if}\ \delta_{RR}^{13} \delta_{LL}^{13} < 0
  \end{cases}
\end{align}

As discussed above, the constraint is independent of the overall SUSY
mass scale.  It does, however, depend on the ratio of the various
masses, in particular $x=\mu/m_{R}$.  $m_{R,L}$ denotes the average
right and left-handed slepton mass matrix, respectively. To be sure that
this result is no special feature of SPS4, we consider other scenarios
in Table~\ref{LFVRes}, which differ by ratios of the SUSY breaking
masses. The sparticle mass spectrum at the electroweak scale for SPS4
corresponds most likely to scenario 2 with $x \approx 0.9-1$.  The
bounds are very strong for larger values of $x$ and weaken for a small
$\mu$-parameter. The upper bounds do not change considerably for values
of $x$ larger than one and therefore are relatively stable and
independent of the parameter space. Electroweak symmetry breaking yields
a relation between the $\mu$-Parameter, the mass of the $Z$ boson and the
two Higgs fields such that in absence of any fine tuning $\mu^2$ should be
within an order of magnitude of $M_{Z}^2$ (also known as the $\mu$
problem).

\begin{table}
  \centering
  \begin{tabular}{|c|c||c|c|c|c|c|}
    \hline
    \multicolumn{2}{|c||}{scenario} & $x=0.3$ & $x=1$ & $x=1.5$ & $x=3.0$
    & for 
    \\
    \hline
    \multirow{2}{*}{1} & 
    \multirow{2}{*}{$M_{1} = M_{2} = m_{L} = m_{R}$} & 
    0.261 & 0.073 & 0.050 & 0.026 &
    $\delta_{RR}^{13}\delta_{LL}^{13}>0$ \\
    & & 0.234 & 0.059 & 0.040 & 0.023 & 
    $\delta_{RR}^{13}\delta_{LL}^{13}<0$ 
    \\
    \hline
    \multirow{2}{*}{2} & 
    \multirow{2}{*}{$3 M_{1} = M_{2} = m_{L} = m_{R}$} & 
    0.301 & 0.083 & 0.057 & 0.029 &
    $\delta_{RR}^{13}\delta_{LL}^{13}>0$ \\
    & & 0.269 & 0.067 & 0.045 & 0.024 &
    $\delta_{RR}^{13}\delta_{LL}^{13}<0$
    \\
    \hline
    \multirow{2}{*}{3} & 
    \multirow{2}{*}{$M_{1} = M_{2} =  3 m_{L} = m_{R}$} &
    0.292 & 0.082 & 0.057 & 0.031 &
    $\delta_{RR}^{13}\delta_{LL}^{13}>0$ \\
    & & 0.235 & 0.067 & 0.042 & 0.027 &
    $\delta_{RR}^{13}\delta_{LL}^{13}<0$
    \\
    \hline
    \multirow{2}{*}{4} & 
    \multirow{2}{*}{$M_{1} = M_{2} = \frac{m_{L}}{3} =  m_{R}$} &
    0.734 & 0.210 & 0.142 & 0.071 &
    $\delta_{RR}^{13}\delta_{LL}^{13}>0$ \\
    & & 0.702 & 0.190 & 0.127 & 0.064 &
    $\delta_{RR}^{13}\delta_{LL}^{13}<0$
    \\
    \hline
    \multirow{2}{*}{5} & 
    \multirow{2}{*}{$3 M_{1} = M_{2} = m_{L} = 3 m_{R}$}  & 
    0.731 & 0.205 & 0.137 & 0.067 & 
    $\delta_{RR}^{13}\delta_{LL}^{13}>0$ \\
    & & 0.693 & 0.179 & 0.116 & 0.054 &
    $\delta_{RR}^{13}\delta_{LL}^{13}<0$
    \\
    \hline
  \end{tabular}
  \caption{Different mass scenarios and the corresponding
    upper bounds for $\left|\delta_{RR}^{13}\delta_{LL}^{13}\right|$.
    $m_{R,L}$ denotes the average right and left-handed slepton mass,
    respectively, $M_{1}$ and $M_2$ the bino and wino masses.
    In all scenarios, $\tan\beta=50$ and $\sgn(\mu)=+1$.
    The upper line is valid for $\delta_{RR}^{13}\delta_{LL}^{13}>0$,
    the lower for $\delta_{RR}^{13}\delta_{LL}^{13}<0$.}
  \label{LFVRes}
\end{table}

We will see in the following section that the  dominating RR terms in
the flavor violating self-energy $\tau \rightarrow e$ cancel in part of
the parameter space. That is why so far no upper bound on  $\delta_{RR}$
could be derived \cite{Masina:2002mv,Paradisi:2005fk}.
  In these regions, we can use the constraint on
$\left|\delta_{RR}^{13}\delta_{LL}^{13}\right|$ as a restriction on
 $\delta_{RR}^{13}$.

\subsection{Lepton-flavour violating self-energies}
\label{se:lfvse}

Lepton flavour violating self-energies can be $\tan\beta$-enhanced and,
moreover, they also have a non-decoupling behaviour.  They occur in the
renormalisation of the PMNS matrix and lead to a correction of the
radiative decays $l_{j}\rightarrow l_{i}\gamma$.

\begin{figure}
   \centering
   \scalebox{0.75}{%
     \begin{picture}(600,100)(0,0)
       \SetColor{Black}
       \ArrowLine(10,50)(50,50) \ArrowLine(100,50)(140,50)
       \Text(20,40)[]{$\tau_{R}$} \Text(130,40)[]{$e_{L}$} 
       \DashLine(40,85)(75,50){2} \Line(40,80)(40,90)
       \Line(35,85)(45,85) 
       \CCirc(75,50){25}{White}{White}
       \DashArrowArc(75,50)(25,0,180){2} \BCirc(75,75){5}
       \Line(72,72)(78,78) \Line(72,78)(78,72) 
       \Text(45,67)[]{$\tilde{\tau}_{R}$}
       \Text(60,80)[]{$\tilde{\tau}_{L}$}
       \Text(100,75)[]{$\tilde{e}_{L}$} 
       \CArc(75,50)(25,180,360) \PhotonArc(75,50)(25,180,360){3}{5.5}
       \Text(75,20)[]{$\tilde{B}$}
       
       \ArrowLine(160,50)(200,50) \ArrowLine(250,50)(290,50)
       \Text(170,40)[]{$\tau_{R}$} \Text(280,40)[]{$e_{L}$}
       \CCirc(225,50){25}{White}{White}
       \DashArrowArc(225,50)(25,0,180){2} \BCirc(225,75){5}
       \Line(222,72)(228,78) \Line(222,78)(228,72)
       \Text(198,75)[]{$\tilde{\tau}_{L}$}
       \Text(250,75)[]{$\tilde{e}_{L}$} 
       \CArc(225,50)(25,180,360) \PhotonArc(225,50)(25,270,360){3}{3.5}
       \Text(200,30)[]{$\tilde{H}$} \Text(252,30)[]{$\tilde{B}$}
       \DashLine(225,25)(225,5){2} \Line(220,10)(230,0)
       \Line(220,0)(230,10) \Text(240,5)[]{$v_{u}$}
       
       \ArrowLine(310,50)(350,50) \ArrowLine(400,50)(440,50)
       \Text(330,40)[]{$\tau_{R}$} \Text(430,40)[]{$e_{L}$}
       \CCirc(375,50){25}{White}{White}
       \DashArrowArcn(375,50)(25,180,0){2}\BCirc(375,75){5}
       \Line(372,72)(378,78) \Line(372,78)(378,72)
       \Text(350,70)[]{$\tilde{\tau}_{L}$}
       \Text(403,70)[]{$\tilde{e}_{L}$} 
       \CArc(375,50)(25,180,360) \PhotonArc(375,50)(25,270,360){3}{3.5}
       \Text(350,30)[]{$\tilde{H}$} \Text(402,30)[]{$\tilde{W}$}
       \DashLine(375,25)(375,5){2} \Line(370,10)(380,0)
       \Line(370,0)(380,10) \Text(390,5)[]{$v_{u}$} 

      \ArrowLine(460,50)(500,50) \ArrowLine(550,50)(590,50)
       \Text(470,40)[]{$\tau_{R}$} \Text(580,40)[]{$e_{L}$}
       \CCirc(525,50){25}{White}{White}
       \DashArrowArcn(525,50)(25,180,0){2}\BCirc(525,75){5}
       \Line(522,72)(528,78) \Line(522,78)(528,72)
       \Text(500,70)[]{$\tilde{\nu}_{\tau}$}
       \Text(553,70)[]{$\tilde{\nu}_{e}$} 
       \CArc(525,50)(25,180,360) \PhotonArc(525,50)(25,270,360){3}{3.5}
       \Text(500,30)[]{$\tilde{H}$} \Text(552,30)[]{$\tilde{W}$}
       \DashLine(525,25)(525,5){2} \Line(520,10)(530,0)
       \Line(520,0)(530,10) \Text(540,5)[]{$v_{u}$}
     \end{picture}
   }%
   \caption{Dominant diagrams for the neutralino-slepton and chargino-sneutrino loop for the $\tau_R\rightarrow e_L$-transition.}\label{tauRmuLDiag2}
 \end{figure}
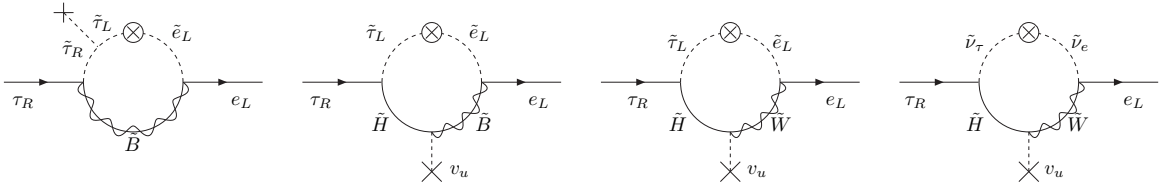

We consider all diagrams with one LFV MI in the slepton propagator and
start with $\tau_{R} \to e_{L}$; the dominant diagrams are shown in
Figure~\ref{tauRmuLDiag2}.  In fact, we can do an approximate
diagonalisation of the neutralino mass matrix
(\ref{Neutralinomassenmatrix}) and use the MI approximation in the
slepton propagator.  Furthermore, we will choose $\mu$ to be real.

The dominant diagrams are proportional to the mass of the tau and
sensitive to the LL element (see diagram 1 to 3 in Fig. (\ref{tauRmuLDiag2})), while the RR dependence is suppressed,
\begin{align}
  \label{tauRmuLFormel2}
  \Sigma_{\tau_{R}-e_{L}}^{\tilde{\chi}^{0}} & \simeq
  \frac{m_{\tau}^\text{phys}}{1 + \Delta_{\tau}}\, \mu\, \tan\beta\; m_{\tilde{e}_{L}} m_{\tilde{\tau}_{L}} \delta_{LL}^{31} \left\{
    -\frac{\alpha_{1}}{4\pi} M_{1}\, f_{2}\left(M_{1}^{2},\,
      m_{\tilde{e}_{L}}^{2},\, m_{\tilde{\tau}_{R}}^{2},\,
      m_{\tilde{\tau}_{L}}^{2}\right) \right. \nonumber
  \\[2pt]
  & \mspace{200mu} + \left. \left( - \frac{1}{2} \frac{\alpha_{1}}{4\pi}
      M_1\, + \frac{1}{2} \frac{\alpha_{2}}{4\pi} M_{2} \right)
    f_{2}\left(M_{1}^{2},\, \mu^{2},\, m_{\tilde{\tau}_{L}}^{2},\,
      m_{\tilde{e}_{L}}^{2}\right) \right\} .
\end{align}
This self-energy will contribute to the renormalisation of the PMNS
matrix (Section~\ref{se:pmns}).  As in the previous section, this
contribution potentially leads to an upper bound on $\delta_{LL}^{13}$ when a naturalness argument is applied to the PMNS element $U_{e3}$.
Note that in Eq.~(\ref{tauRmuLFormel2}) the LFC mass renormalisation is
already taken into account.

The dominant contributions for the opposite helicity transition,
$\tau_{L} \to e_{R}$, are analogous to the first and second diagrams in
Fig.~\ref{tauRmuLDiag2},
\begin{align}
  \label{tauLmuRFormel2}
  \Sigma_{\tau_{L}-e_{R}}^{\tilde{\chi}^{0}} & \simeq
  \frac{\alpha_{1}}{4\pi}\, \frac{m_{\tau}^\text{phys}}{1 + \Delta_{\tau}}\, M_{1}\, \mu\,
  \tan\beta\; m_{\tilde{e}_{R}} m_{\tilde{\tau}_{R}}\delta_{RR}^{31} \left( f_{2}\left(M_{1}^{2},\,
      \mu^{2},\, m_{\tilde{\tau}_{R}}^{2},\,
      m_{\tilde{e}_{R}}^{2}\right) - f_{2}\left(M_{1}^{2},\,
      m_{\tilde{e}_{R}}^{2},\, m_{\tilde{\tau}_{R}}^{2},\,
      m_{\tilde{\tau}_{L}}^{2}\right) \right) .
\end{align}
They are sensitive to the RR element; however, the relative minus sign
due to the different hypercharges potentially leads to cancellations.
In this approximation, the RR sensitivity vanishes completely for $\mu =
m_{\tilde{\tau}_{L}}$ and hence no upper bounds on $\delta_{RR}^{ij}$
has been derived, as mentioned in the previous section.
In Ref.~\cite{Paradisi:2005fk} the processes $\mu\to
  e\gamma$ and $\mu \to e$ conversion in nuclei are combined. The
  corresponding one-loop amplitudes suffer from similar 
  cancellations, albeit in different regions of the parameter space leading to
  the constraint $\delta_{RR}^{12}\leq 0.2$.

\smallskip

Let us now turn to the chargino-sneutrino loops.  As the left-handed
charged sleptons and sneutrinos form a doublet, they have the same SUSY
breaking soft mass and therefore the same off-diagonal elements.  The
neutrino is always left-handed so that the chargino loop can only be
sensitive to the LL element and a chirality flip of the charged lepton
is needed.  The higgsino component of the chargino couples to the
right-handed lepton and the wino part to the left-handed lepton.  Thus,
the self-energy is proportional to the mass of the right-handed lepton,
\begin{align}
  \Sigma_{l_{iL}-l_{jR}}^{\tilde{\chi}^{\pm}} & = \frac{g_{2}}{16
    \pi^{2}}Y_{j}\sum_{n=1}^{3}\sum_{k=1,2} Z_{-}^{2k} Z_{+}^{1k}
  Z_{\nu}^{jn} Z_{\nu}^{in*} m_{\tilde{\chi}_{k}}
  B_{0}\left(m_{\tilde{\nu}_{n}}^{2},
    m_{\tilde{\chi}_{k}^{\pm}}^{2}\right) \nonumber
  \\
  & \simeq \frac{\alpha_{2}}{4 \pi}\, \frac{m_{j}^\text{phys}}{1 + \Delta_{l_j}}\, M_{2}\, \mu\,
  \tan\beta\; m_{\tilde{\nu}_{i}} m_{\tilde{\nu}_{j}}\,
  \delta_{LL}^{ij}\, f_{2}\left(m_{\tilde{\nu}_{i}}^{2},
    m_{\tilde{\nu}_{j}}^{2}, M_{2}^{2}, \mu^{2}\right) \ .
  \label{charnichtMFV}
\end{align}
In the second line, we used the MI approximation in the chargino
propagator and for the LFV (see last diagram in Fig. (\ref{tauRmuLDiag2})).

\subsection{PMNS matrix renormalisation}
\label{se:pmns}

Up to now, we have only an upper bound for the matrix element $U_{e3}$
and thus for the mixing angle $\theta_{13}$; the best-fit value is at or
close to zero (cf.~Eq.~(\ref{nu-best-fit})).
It might well be that it vanishes at tree level due to a particular
symmetry and obtains a non-zero value due to corrections. 
So we can ask the question if threshold corrections to the PMNS matrix could spoil the prediction $\theta_{13} = 0^{\circ}$ at the weak scale. 
What does it mean for the physics at the high scale if experiment will tell us that $\theta_{13}$ does not vanish?
 As before, we
demand the absence of fine-tuning for these corrections and therefore
require that the SUSY loop contributions do not exceed the value of
$U_{e3}$,
\begin{align}
  \label{equ:finetuningUe3}
  \left|\delta U_{e3}\right| \leq\left| U_{e3}^\text{phys}\right| .
\end{align}
Then we can in principle use the smallness of $U_{e3}$ to constrain
$\delta_{LL}^{13}$.

In an effective field theory approach, the renormalisation of the PMNS
matrix is done via rotation matrices that diagonalise the mass matrix,
which receives contribution from both the tree-level coupling of the
fermions to $H_d$ and the loop-induced coupling to $H_u$ (see
Refs.~\cite{Hall:1985dx,Hamzaoui:1998nu,PhysRevLett.84.228,Buras:2002vd}
for the quark sector).  Lepton flavour violating self-energies induce
off-diagonal entries in the mass matrix.  In order to deal with physical
fields, one has to rotate them in flavour space to achieve a diagonal
mass matrix.

\begin{wrapfigure}{r}{0.27\linewidth}
  \begin{picture}(100,80)(80,25)
    \ArrowLine(150,50)(110,90) \Text(130,60)[]{$\nu_{k}$}
    \ArrowLine(164.14,64.14)(150,50) \Text(162,52)[]{$l_{l}$}
    \ArrowLine(190,90)(176.86,76.84) \Text(192,80)[]{$l_{j}$}
    \ArrowArc(170,70)(10,45,225)
    \DashArrowArc(170,70)(10,225,45){2.5} \Text(180,55)[]{$\Sigma$}
    \Photon(150,20)(150,50){2}{5} \Text(165,35)[]{$W_{\mu}$}
  \end{picture}
\end{wrapfigure}

As a drawback, the effective field theory method is only valid if the
masses of the supersymmetric particles in the loop are much larger than
$v=174$ GeV. For sleptons and neutralinos this assumption is doubtful,
so that we resort to the diagrammatic method of
Refs.~\cite{Carena:1999py,Marchetti:2008hw,Hofer:2009xb} which does not
rely on any hierarchy between $M_{\rm SUSY}$ and $v$.  In our
diagrammatic approach, we consider chargino and neutralino loops in the
external lepton propagator and resum all $\tan\beta$-enhanced
corrections explicitly.  Once again the on-shell scheme is used.  The
loop corrections are finite and the counterterms are defined such that
they exactly cancel the loop diagrams:
\begin{align}
  U^{(0)} = U^\text{phys}+\sum_{n}\delta U^{(n)} = U^\text{phys} +
  \delta U \ .
\end{align}
The first-order correction is displayed in the adjoining figure.
The counterterm reads
\begin{align}
  \label{equ:deltaUerste}
  \delta U_{jk}^{(1)} & = \sum_{l\neq j} U_{lk}^\text{phys}
  \frac{(\slashed{p}_{j}+m_{l})}{p_{j}^{2}-m_{l}^{2}}
  \left(\Sigma_{l_{j}-l_{l}}\right)^{*} \nonumber
  \\[3pt]
  & \simeq
  \begin{cases}
    \sum_{l\neq j}U_{lk}^\text{phys}\frac{1}{m_{j}}
    \left(\Sigma_{l_{jR}-l_{lL}}\right)^{*} , & j>l \ , \ \textrm{i.e.,
      heavy particle as external leg;}
    \\
    -\sum_{l\neq j}U_{lk}^\text{phys}\frac{1}{m_{l}}
    \left(\Sigma_{l_{jL}-l_{lR}}\right)^{*} , & j<l \ , \ \textrm{i.e.,
      heavy particle as internal propagator.}
  \end{cases}
\end{align}
As for the mass renormalisation there are no genuine
$\tan\beta$-enhanced two-loop diagrams.  The corrections in second order
come from one-loop diagrams in which a counterterm of first order is
inserted, corresponding to the substitution $U_{lk}^\text{phys}\to
U_{lk}^\text{phys} + \delta U_{lk}^{(1)}$.  In contrast to the
resummation of the mass counterterms, these counterterms are not
directly proportional to the PMNS-element under consideration.
The sum of the counterterms has to cancel the corrections up to that
order, so at the $n^\text{th}$ order, one gets
\begin{align}
  \label{equ:deltaUnte}
  \sum_{m=1}^{n}\delta U_{jk}^{(m)} & =
  \begin{cases}
    \sum_{l\neq j}\left(U_{lk}^\text{phys}+\sum_{m}^{n-1}\delta
      U_{lk}^{(m)}\right)\frac{1}{m_{j}}
    \left(\Sigma_{l_{jR}-l_{lL}}\right)^{*} \ , & j>l
    \\
    -\sum_{l\neq j}\left(U_{lk}^\text{phys}+\sum_{m}^{n-1}\delta
      U_{lk}^{(m)}\right)\frac{1}{m_{l}}
    \left(\Sigma_{l_{jL}-l_{lR}}\right)^{*} \ , & j<l
  \end{cases}
  .
\end{align}
Now we can take the limit $n \to \infty$ and obtain a linear system of
equations for the $U^{(0)}$ elements ($ k = 1,\,2,\,3$):
\begin{subequations}\label{equ:PMNSsystem}
 \begin{align}
  U_{ek}^{(0)} + \frac{1}{m_{\mu}}\Sigma_{\mu_{R}-e_{L}}U_{\mu k}^{(0)} + \frac{1}{m_{\tau}}\Sigma_{\tau_{R}-e_{L}}U_{\tau k}^{(0)}  = & U_{ek}^{\text{phys}},\\
 U_{\mu k}^{(0)} - \frac{1}{m_{\mu}}\Sigma_{e_{L}-\mu_{R}} U_{e k}^{(0)} + \frac{1}{m_{\tau}}\Sigma_{\tau_{R}-\mu_{L}}U_{\tau k}^{(0)}  = & U_{\mu k}^{\text{phys}},\\
 U_{\tau k}^{(0)} - \frac{1}{m_{\tau}}\Sigma_{e_{L}-\tau_{R}}U_{e k}^{(0)} - \frac{1}{m_{\tau}}\Sigma_{\mu_{L}-\tau_{R}}U_{\mu k}^{(0)}  = & U_{\tau k}^{\text{phys}}.
 \end{align}
\end{subequations}

\begin{figure}
 \psfrag{Grenze}{\hspace{0.0cm}\scalefont{0.7}$\text{bound}$}
 \psfrag{Prozent}{\hspace{-0.5cm}\scalefont{0.8}$\text{Corrections in \%}$}

  \includegraphics[width=.49\linewidth]{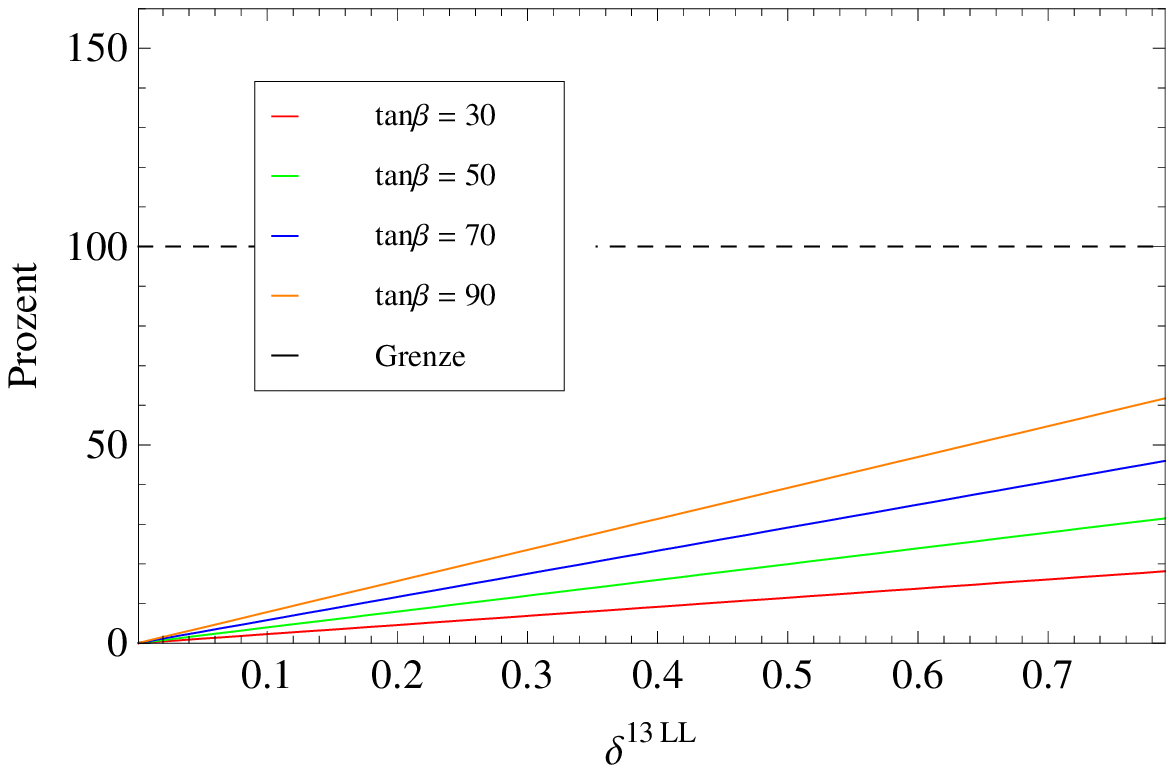}
  \hspace{.01\linewidth}
  \includegraphics[width=.49\linewidth]{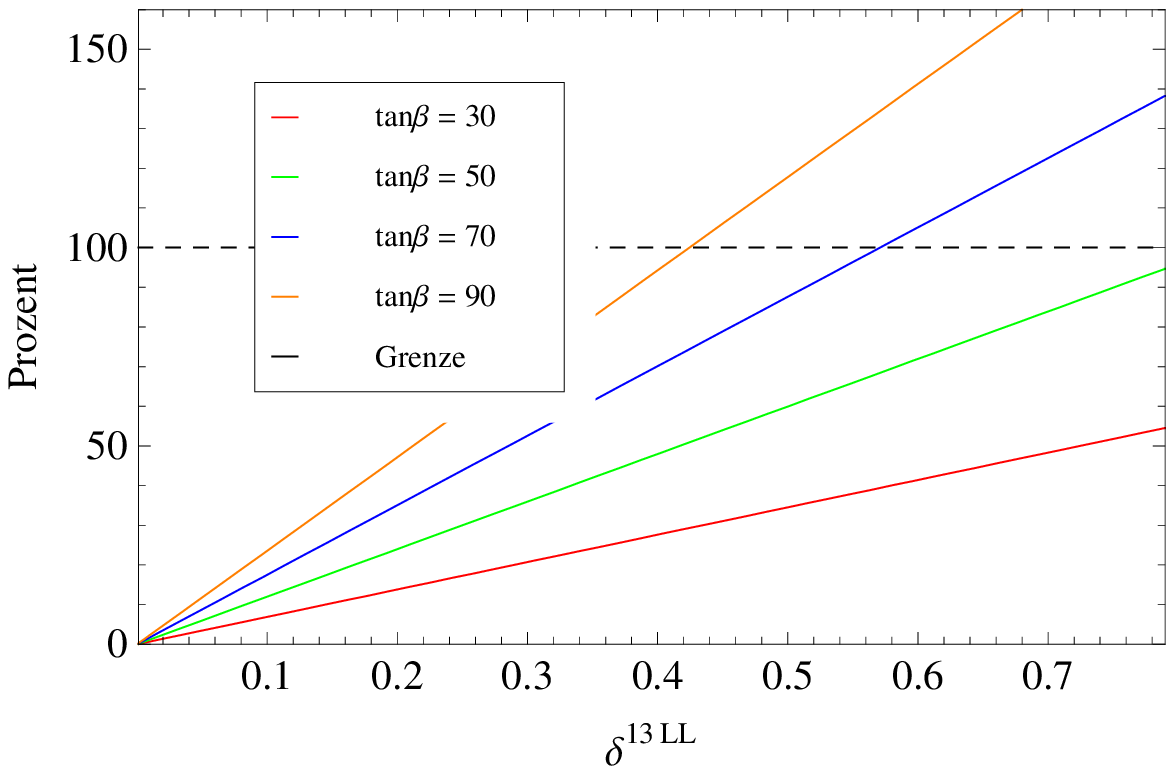}
  \caption{$\left|\delta U^{13}\right|/U_{13}^\text{phys}$ in percent as
    a function of $\delta^{13}_{LL}$ and $\tan\beta$ for $\theta_{13} =
    3^{\circ}$ (left) and $\theta_{13} = 1^{\circ}$ (right).
    }
  \label{PMNS1}
\end{figure}

In the  MSSM, we have $\Sigma = \Sigma^{\tilde{\chi}^{0}} +
\Sigma^{\tilde{\chi}^{\pm}}$.  As shown above, $\Sigma_{\tau_{R}-e_{L}}$
is sensitive to $\delta_{LL}^{13}$ and so is $\delta U_{e3}$.   We aim to avoid accidental
cancellations
and set all off-diagonal elements to zero except for $\delta_{LL}^{13}$\footnote{In
\cite{Crivellin:2010gw} we study the case with nonvanishing $\delta_{LR}^{13}$.}. In this
case we can
explicitly solve the linear system of equations
\begin{align}
  U_{e3}^{(0)} = \frac{U_{e3}^\text{phys} -
    \frac{1}{m_{\tau}}\Sigma_{\tau_{R}-e_{L}}U_{\tau 3}^\text{phys} }{1
    + \left|\frac{1}{m_{\tau}}\Sigma_{\tau_{R}-e_{L}}\right|^{2}}.
\end{align}
By means of Eq.~(\ref{equ:finetuningUe3}), we can in principle derive upper bounds
for $\delta_{LL}^{13}$.  As shown in Figs.~\ref{PMNS1}, they strongly
depend on $\tan\beta$ and the assumed value for $U_{e3}^\text{phys}$.

After three years of running, the DOUBLE CHOOZ experiment will be
sensitive to $\theta_{13}=3^{\circ}$, which corresponds to $U_{e3} =
0.05$. A future neutrino factory may probe $\theta_{13}$
    down to $\theta_{13}=0.6^{\circ}$ \cite{Bandyopadhyay:2007kx}.  In
    general, even with future experimental facilities, we can conclude
    that the corrections from SUSY loops to the small element $U_{e3}$
    stay unobservably small. This means at the same time that if some
    experiment measures $\theta_{13}\neq 0$, this will not be compatible
    with tri-bimaximal mixing at the high scale and moderate sparticle
    masses, since SUSY threshold corrections cannot account for such an
    effect: Even for large $\tan\beta$ the already existing constraints
    on $\delta_{LL}^{13}$ from $\tau\rightarrow e \gamma$ are stronger
    assuming reasonable SUSY masses. However, since $\tau\rightarrow e
    \gamma$ decouples, our method leads to a sharper bound for very large
    SUSY masses, especially with $\theta_{13}=1^{\circ}$ and large
    $\tan\beta$. 

\subsection{Counterterms in the flavour basis and charged Higgs
  couplings}

Neutrinos are both produced and detected as flavour eigenstates.  In
order to have flavour diagonal $W$ couplings, however, it is necessary
to introduce counterterms, $\delta V_{ij}$, which cancel the LFV loops.
By doing this you perform a renormalisation of the unit matrix.  
In an effective field theory approach this is achieved via a wave function renormalisation by
rotating the lepton fields leading to a diagonal mass matrix and
physical fields.  This rotation of the fields induce LFV in the charged
Higgs coupling to lepton and neutrino and the same is true for the
counterterms in the diagrammatic approach.

The first-order correction is displayed in the figure above; the
flavour-diagonal vertices do not get any counterterms, since the
external loops are already included in the mass renormalisation.  We
obtain
\begin{align}
  \delta V & =
  \begin{pmatrix}
    0 & -\frac{1}{m_{\mu}^\text{phys} }\Sigma_{e_{L}-\mu_{R}} &
    -\frac{1}{m_{\tau}^\text{phys} }\Sigma_{e_{L}-\tau_{R}} 
    \\[2pt]
    \frac{1}{m_{\mu}^\text{phys} }\Sigma_{\mu_{R}-e_{L}} & 0 &
    -\frac{1}{m_{\tau}^\text{phys} }\Sigma_{\mu_{L}-\tau_{R}}
    \\[2pt]
    \frac{1}{m_{\tau}^\text{phys} }\Sigma_{\tau_{R}-e_{L}} &
    \frac{1}{m_{\tau}^\text{phys} }\Sigma_{\tau_{R}-\mu_{R}}&0
  \end{pmatrix}
  .
\end{align}

You can translate this to the mass eigenstate basis used in Eq. (\ref{equ:PMNSsystem}) via $\delta U_{ik} = \delta V_{ij}^* U_{jk}^{(0)}$. 
These counterterms induce LFV in the charged Higgs coupling to lepton
and neutrino, due to the different helicity structure of the Higgs and
$W$ coupling and the different lepton masses.  The $H^{+} e \nu_{\tau}$
vertex can be of particular importance, since it is possible to pick up
terms with a tau Yukawa coupling.  As discussed before, this coupling is
enhanced in the large $\tan\beta$ regime and can partly compensate the
loop suppression factor.

The chargino contributions from the counterterm and the LFV loop cancel
in the charged Higgs coupling as the chargino loop is exactly
proportional to the mass of the right handed lepton.  Therefore only the
neutralino contributions remain.

The charged Higgs coupling to electrons reads
\begin{subequations}
  \label{effHiggsenu}
  \begin{align}
    i\Gamma^{H^{+}}_{e\nu_{\tau}} & = \frac{ig_{2}}{\sqrt{2}M_{W}}
    \tan\beta \left( m_{e}^{(0)}\, \delta V_{13} +
      m_{\tau}^{(0)} \frac{\Sigma_{e_{R}
        - \tau_{L}}}{m_{\tau}^\text{phys}}  \right)  =  \frac{ig_{2}}{\sqrt{2}M_{W}}
    \tan\beta \left( -\frac{m_{e}^\text{phys}}{m_{\tau}^\text{phys}}\frac{\Sigma_{e_{L}
        - \tau_{R}}}{1 + \Delta_e} +
      \frac{\Sigma_{e_{R}
        - \tau_{L}}}{1 + \Delta_{\tau}}  \right),
    \label{effHiggsenutau}
    \\
    i\Gamma^{H^{+}}_{e\nu_{\mu}} & = \frac{ig_{2}}{\sqrt{2}M_{W}}
    \tan\beta \left( m_{e}^{(0)} \delta V_{12} +
      m_{\mu}^{(0)} \frac{\Sigma_{e_{R} -
        \mu_{L}}}{m_{\mu}^\text{phys}}  \right)  = \frac{ig_{2}}{\sqrt{2}M_{W}}
    \tan\beta \left( -\frac{m_{e}^\text{phys}}{m_{\mu}^\text{phys}}\frac{\Sigma_{e_{L} -
        \mu_{R}}}{1 +\Delta_e} +
      \frac{\Sigma_{e_{R} -\mu_{L}}}{1 + \Delta_{\mu}}  \right),
    \\
    i\Gamma^{H^{+}}_{e\nu_{e}} & = \frac{ig_{2}}{\sqrt{2}M_{W}}
    \frac{m_{e}^\text{phys}}{1+\Delta_{e}} \tan\beta \left( 1 +
      \frac{m_{\tau}^\text{phys}}{m_{e}^\text{phys}}
      \frac{\tan\beta}{1+\Delta_{\tau}} \Delta^{e}_{LR} \right) ,
    \label{effHiggsenue}
  \end{align}
\end{subequations}
where $\Sigma^\text{FV}_{e} = \frac{m_{\tau}^\text{phys}}{1 + \Delta_{\tau}}\, \tan\beta\;
\Delta_{LR}^{e}$.  
We see that the counterterms are suppressed with the electron mass.  As
the lepton mass cancels out in $\Sigma_{e_R-\ell_{iL}}/m_{\ell_i}$, the
LFV loop contributions with $\nu_\tau$ and $\nu_\mu$ differ by a factor
$m_{\tau}/m_{\mu}$.

Similarly, we obtain for the coupling to muons,
{\allowdisplaybreaks
  \begin{subequations}
    \begin{align}
      i\Gamma^{H^{+}}_{\mu\nu_{\tau}} & = \frac{ig_{2}}{\sqrt{2}M_{W}}
      \tan\beta \left( m_{\mu}^{(0)} \delta V_{23} +
        m_{\tau}^{(0)} \frac{\Sigma_{\mu_{R}-\tau_{L}}}{m_{\tau}^\text{phys}}
         \right) = \frac{ig_{2}}{\sqrt{2}M_{W}}
      \tan\beta \left( -\frac{m_{\mu}^\text{phys}}{m_{\tau}^\text{phys}}\frac{\Sigma_{\mu_{L}-\tau_{R}}}{1 + \Delta_{\mu}}  + \frac{\Sigma_{\mu_{R}-\tau_{L}}}{1 + \Delta_{\tau}}
         \right) ,
      \\
      i\Gamma^{H^{+}}_{\mu\nu_{\mu}} & = \frac{ig_{2}}{\sqrt{2}M_{W}}
      \frac{m_{\mu}^\text{phys}}{1+\Delta_{\mu}} \tan\beta \left( 1 +
        \frac{m_{\tau}^\text{phys}}{m_{\mu}^\text{phys}}
        \frac{\tan\beta}{1+\Delta_{\tau}} \Delta^{\mu}_{LR} \right) .
      \\
      i\Gamma^{H^{+}}_{\mu\nu_{e}} & = \frac{ig_{2}}{\sqrt{2}M_{W}}
      \tan\beta \left( m_{\mu}^{(0)} \delta V_{21} -
        m_{e}^{(0)} \frac{\Sigma_{\mu_{R}-e_{L}}}{m_{e}^\text{phys}}
         \right) = \frac{ig_{2}}{\sqrt{2}M_{W}}
      \tan\beta \left(  \frac{\Sigma_{\mu_{L}-e_{R}}}{1 + \Delta_{\mu}} -\frac{m_e^\text{phys}}{m_{\mu}^\text{phys}}
         \frac{\Sigma_{\mu_{R}-e_{L}}}{1 + \Delta_e}
         \right) ,
    \end{align}
  \end{subequations}
}%
While the first term is similar to the couplings to electrons, the
counterterm dominates over the loop contribution if there is an electron
neutrino in the final state.

Finally, for the $\tau$ coupling one finds
\begin{subequations}
\label{effHiggstaunu}
  \begin{align}
    i\Gamma^{H^{+}}_{\tau\nu_{\tau}} & = \frac{ig_{2}}{\sqrt{2}M_{W}}
    \frac{m_{\tau}^\text{phys}}{1+\Delta_{\tau}} \tan\beta \ ,
    \\
    i\Gamma^{H^{+}}_{\tau\nu_{\mu}} & = \frac{ig_{2}}{\sqrt{2}M_{W}}
    \tan\beta \left( m_{\tau}^{(0)} \delta V_{32} -
      m_{\mu}^{(0)} \frac{\Sigma_{\tau_{L}-\mu_{R}}}{m_{\tau}^\text{phys}}
       \right)  = \frac{ig_{2}}{\sqrt{2}M_{W}}
    \tan\beta \left(\frac{\Sigma_{\tau_{R}-\mu_{L}}}{1 + \Delta_{\tau}} -
      \frac{m_{\mu}^\text{phys}}{m_{\tau}^\text{phys}} \frac{\Sigma_{\tau_{L}-\mu_{R}}}{1 + \Delta_{\mu}}
       \right),
    \\
    i\Gamma^{H^{+}}_{\tau\nu_{e}} & = \frac{ig_{2}}{\sqrt{2}M_{W}}
    \tan\beta \left( m_{\tau}^{(0)} \delta V_{31} -
      m_{e}^{(0)} \frac{\Sigma_{\tau_{L}-e_{R}}}{m_{\tau}^\text{phys}}
       \right) = \frac{ig_{2}}{\sqrt{2}M_{W}}
    \tan\beta \left( \frac{\Sigma_{\tau_{R}-e_{L}}}{1 + \Delta_{\tau}} -
      \frac{m_{e}^\text{phys}}{m_{\tau}^\text{phys}} \frac{\Sigma_{\tau_{L}-e_{R}}}{1 + \Delta_e}
       \right) .
  \end{align}
\end{subequations}
{The results of Eqs.~(\ref{effHiggsenu})-(\ref{effHiggstaunu}) are
  given in Eqs.~(92-95) of Ref.~\cite{Hisano:2008hn} for the
  decoupling limit $M_{\rm SUSY}\gg v$. In Appendix~C of
  Ref.~\cite{Hisano:2008hn} an iterative procedure (analogous to the    
  one in Ref.~\cite{Buras:2002vd}) has been outlined which achieves
  the all-order resummation of the $\tan\beta$-enhanced higher-order 
  corrections.  Eqs.~(\ref{effHiggsenu})-(\ref{effHiggstaunu}) comprise
  analytical formulae for the limits to which this iterative procedure
  converges.}  

The $\tan\beta$-enhanced lepton flavour violating Higgs couplings can
become important in the leptonic decay of charged Kaons, $K\rightarrow
l\nu$, where they potentially induce lepton non-universality.  Then the
current experimental data and our fine-tuning argument together
constrain the various terms in Eqs.~(\ref{effHiggsenu}), as they
contribute to the electron self-energy as well.  In particular, if the
second term in Eq.~(\ref{effHiggsenue}) had a significant effect in the
ratio $R_K = \Gamma(K\rightarrow e \nu)/\Gamma(K\rightarrow \mu \nu)$,
as was assumed in Ref.~\cite{Masiero}, $\Delta^{e}_{LR}$ would give a
large contribution to the electron mass \cite{JG}.  (The value
\mbox{$\Delta_{RL}^{11} \stackrel{\scriptscriptstyle\wedge}{=}
  \Delta_{LR}^{e}= 10^{-4}$} \cite{Masiero} corresponds to
$\delta_{RR}^{13}\delta_{LL}^{13}\approx 2$ in the SPS4 scenario and
thus gives a more than 2000\% correction to the electron mass.)  While
in the improved analysis \cite{Masiero:2008cb} the contribution of
$\Sigma_e^{FV}$to the electron mass was not considered, their scanned
values of $\delta_{LL,RR}^{13}$ are in agreement with the fine-tuning
argument.  The scan respects
$\left|\delta_{LL}^{13}\delta_{RR}^{13}\right|\leq 0.25$, in marginal
agreement with our results of Sects.~3.1 and 3.5.
The NA62 experiment at CERN aims to reduce the error of $R_K$ from
$1.3\%$ to $0.3\%$.  This prospective error is used in
Ref.~\cite{Ellis:2008st} to derive large, phenomenologically interesting
values for $\delta_{LL}^{13}$ and $\delta_{RR}^{13}$.

\subsection{Anomalous Magnetic Moment of the Electron}
\label{se:ge}

The anomalous magnetic moment of the electron plays a central role in
quantum electrodynamics.  The precise measurements provide the best
source of the fine structure constant $\alpha_\text{em}$ if one assumes
the validity of QED \cite{Hanneke:2008tm}.  Conversely, one can use a
value of $\alpha_\text{em}$ from a (less precise) measurement and insert
it into the theory prediction for $a_e$ to probe new physics in the
latter quantity. The most recent calculation yields
\cite{Aoyama:2007mn}
\begin{align}
  \label{eq:ae-theory}
  a_e & = 1\; 159\; 652\; 182.79 \left(7.71\right) \times 10^{-12} \;,
\end{align}
where the largest uncertainty comes from the second-best measurement of
$\alpha_\text{em}$ which is $ \alpha_\text{em}^{-1} = 137.03599884(91)$ from a Rubidium atom experiment \cite{Clade:2006zz}.

Supersymmetric contributions to the magnetic moment are usually small,
due to the smallness of the electron Yukawa coupling and the SUSY mass
suppression.  However, multiple flavour changes, resulting in a LFC
loop, insert the $\tau$ Yukawa coupling, which strongly enhances the
amplitude.  As a result, supersymmetric contributions can be as large as
$\mathcal{O}(10^{-12})$, comparable to the weak or hadronic
contributions \cite{Aoyama:2007mn}.  The amplitude can exceed a
$3\sigma$ deviation of the theoretical mean value, which enables us to
constrain the LFV parameters $\delta^{13}_{LL}$ and $\delta^{13}_{RR}$.
In Ref.~\cite{Masina:2002mv} the 
magnetic and electric dipole moments $a_i$ and $d_i$ of the charged
lepton $\ell_i$ were calculated in the MSSM, considering  
flavour-conserving and flavour-violating contributions within 
the mass insertion approximation. The authors found that the na\"ive mass scaling 
can be overcome with double mass insertions. However, in their phenomenological analysis
to constrain the flavour-violating parameters $\delta_{XY}^{ij}$, they
only used $a_\mu$ and the experimental bounds on $d_\mu$ and $d_e$ but
did not consider $a_e$. Our consideration of $a_e$  adds a novel aspect to the
phenomenological study of LFV parameters in the MSSM and complements the
analysis of Ref.~\cite{Masina:2002mv} in this respect.

\smallskip

The supersymmetric contributions to the anomalous magnetic moment $a_e$ are
generated by chargino and neutralino loops, where the photon couples
to any charged particle in the loop.  The full analytic result can be
found in Ref.~\cite{Stockinger:2006zn}.  Here, we will neglect the terms
which are both proportional to the electron mass and not (potentially)
$\tan\beta$-enhanced and are therefore left with
\begin{subequations}
  \label{Ergebnisg-2}
  \begin{align}
    \label{Ergebnisg-21}
    a^{\chi^0}_{e} & = -\frac{m_{e}}{16\pi^2}
    \sum_{A=1}^{4}\sum_{X=1}^{6} \frac{m_{\chi^0_A}}{3m^2_{\tilde{l}_X}}
    \re \left[ N^{L}_{1AX} N^{R\ast}_{1AX}\right] F_2^N \left(x_{AX} \right)
    , & x_{AX} & =
    \frac{m_{\tilde{\chi}^{0}_{A}}^{2}}{m_{\tilde{l}_{X}}^{2}} ,
    \\
    \label{Ergebnisg-22}
    a^{\chi^{\pm}}_{e} & = \frac{m_{e}}{16\pi^2} \sum_{A = 1,2}
    \sum_{X=1}^{3} \frac{2m_{\chi^{\pm}_A}}{3m^2_{\tilde{\nu}_{X}}} \re
    \left[ C^{L}_{1AX} C^{R\ast}_{1AX}\right] F_2^C \left(x_{AX}\right) , &
    x_{AX} & =
    \frac{m_{\tilde{\chi}^{\pm}_{A}}^{2}}{m_{\tilde{\nu}_{X}}^{2}} .
  \end{align}
\end{subequations}
The loop functions are listed in Eq.~(\ref{Schleifenfunktionen}) and
the couplings read \cite{Rosiek}
\begin{subequations}
  \label{equ:Kopplungeni}
  \begin{align}
    N_{iAX}^{L} & = -\sqrt{2} g_{1} \left(Z_{L}^{i+3,X}\right)^{*}
    Z_{N}^{1A} + Y_{l_{i}} \left(Z_{L}^{i,X}\right)^{*} Z_{N}^{3A} =
    \left( \Gamma_{l_{iR}}^{\tilde{\chi}^{0}_{A}\tilde{l}_{X}}
    \right)^{*} ,
    \\
    N_{iAX}^{R} & = \frac{(Z_{L}^{i,X})^{*}}{\sqrt{2}} \left(g_{1}
      \left(Z_{N}^{1A}\right)^{*} + g_{2} \left(Z_{N}^{2A}\right)^{*}
    \right) + Y_{l_{i}} \left(Z_{N}^{3A}\right)^{*}
    \left(Z_{L}^{i+3,X}\right)^{*} = \left(
      \Gamma_{l_{iL}}^{\tilde{\chi}^{0}_{A}\tilde{l}_{X}} \right)^{*} ,
    \\
    C_{iAX}^{L} & = -Y_{l_{i}} Z_{-}^{2A} Z_{\nu}^{i,X} = \left(
      \Gamma_{l_{iR}}^{\tilde{\chi}^{\pm}_{A}\tilde{\nu}_{X}}
    \right)^{*} ,
    \\
    C_{iAX}^{R} & = -g_{2} \left(Z_{+}^{1A}\right)^{*} Z_{\nu}^{i,X} =
    \left( \Gamma_{l_{iL}}^{\tilde{\chi}^{\pm}_{A}\tilde{\nu}_{X}}
    \right)^{*} .
  \end{align}
\end{subequations}
The mixing matrices are defined in
Appendix~\ref{appendix:massesmixingfeynmanrules}.  Note that they are
$6\times6$ matrices, in order to allow for flavour changes in the loop.

The dependence on $\tan\beta$ in Eqs.~(\ref{Ergebnisg-2}) is hidden in
the mixing matrices.  In principle, $\tan\beta$ comes from a chirality
flip on the selectron line and in the chargino case from the combination
of vacuum expectation value $v_u$ and the Yukawa coupling, $y_{e} v_u =
m_{e} \tan\beta$.
We can, however, simplify the expressions significantly as follows: We
assume a universal SUSY mass, real parameters and the same signs for
$M_1$ and $M_2$ \cite{Moroi:1995yh}, then expand $a_{e}$ in powers of
$M_W/M_{\text{SUSY}}$ or $1/\tan\beta$.  Then we obtain
\begin{align}
  \label{Ergebnisaproxg-2}
  \begin{split}
    a^{\chi^0}_{e} & = \sgn\left(\mu M_2\right)
    \frac{g^2_1-g^2_2}{192\pi^2} \frac{m^2_{e}}{M^2_{\text{SUSY}}}
    \tan\beta \left[ 1 + \mathcal{O}\left( \frac{1}{\tan\beta},\,
        \frac{M_W}{M_{\text{SUSY}}}\right)\right] ,
    \\
    a^{\chi^\pm}_{e} & = \sgn\left(\mu M_2\right) \frac{g^2_2}{32\pi^2}
    \frac{m^2_{e}}{M^2_{\text{SUSY}}} \tan\beta \left[ 1 +
      \mathcal{O}\left( \frac{1}{\tan\beta},\,
        \frac{M_W}{M_{\text{SUSY}}}\right)\right] .
  \end{split}
\end{align}
The result is again finite.  The $1/M_{\text{SUSY}}^2$ dependence
reflects the decoupling behaviour of supersymmetry.  Furthermore, we
note that a large value for $\tan\beta$ can dilute the
$1/M_{\text{SUSY}}^2$ suppression.
The numerical results are computed with the exact formula in
Eqs.~(\ref{Ergebnisg-2}).

So far, the Yukawa couplings are unrenormalised; the inclusion of the
mass renormalisation amounts to a loop contribution to $a_e$ which
approximately grows as $\tan^2\beta$ \cite{Marchetti:2008hw}.
Diagonalising the mixing matrices perturbatively, one finds a linear
dependence on the Yukawa coupling of the remaining second terms of
Eqs.~(\ref{Ergebnisg-2}).  In this way we find an easy expression, which
takes the corrections into account by a global factor,
\begin{align}
  \label{Korrekturg-2}
  a_{e}^{\text{SUSY},\,1L} + a_{e}^{\text{SUSY}, \Delta_{e}} =
  a_{e}^{\text{SUSY},\,1L}\left( \frac{1}{1+\Delta_{e}}\right) ,
\end{align}
where $a_{e}^{\text{SUSY},\,1L} = a^{\chi^0}_{e} + a^{\chi^\pm}_{e}$, as
discussed in Ref.~\cite{Marchetti:2008hw}.

\smallskip

For the numerical analysis, we only allow $\delta^{13}_{LL}$ and
$\delta^{13}_{RR}$ to be non-zero such that they are the only source of
flavour violation.  The theoretical uncertainty in
Eq.~(\ref{eq:ae-theory}) is taken as $1\sigma$ deviation and we require
that the SUSY contribution to $a_e$ is less than $3\sigma$.

\begin{figure}
  \subfloat[\label{deltaLLelektron}]{%
    \includegraphics[width=.45\linewidth]{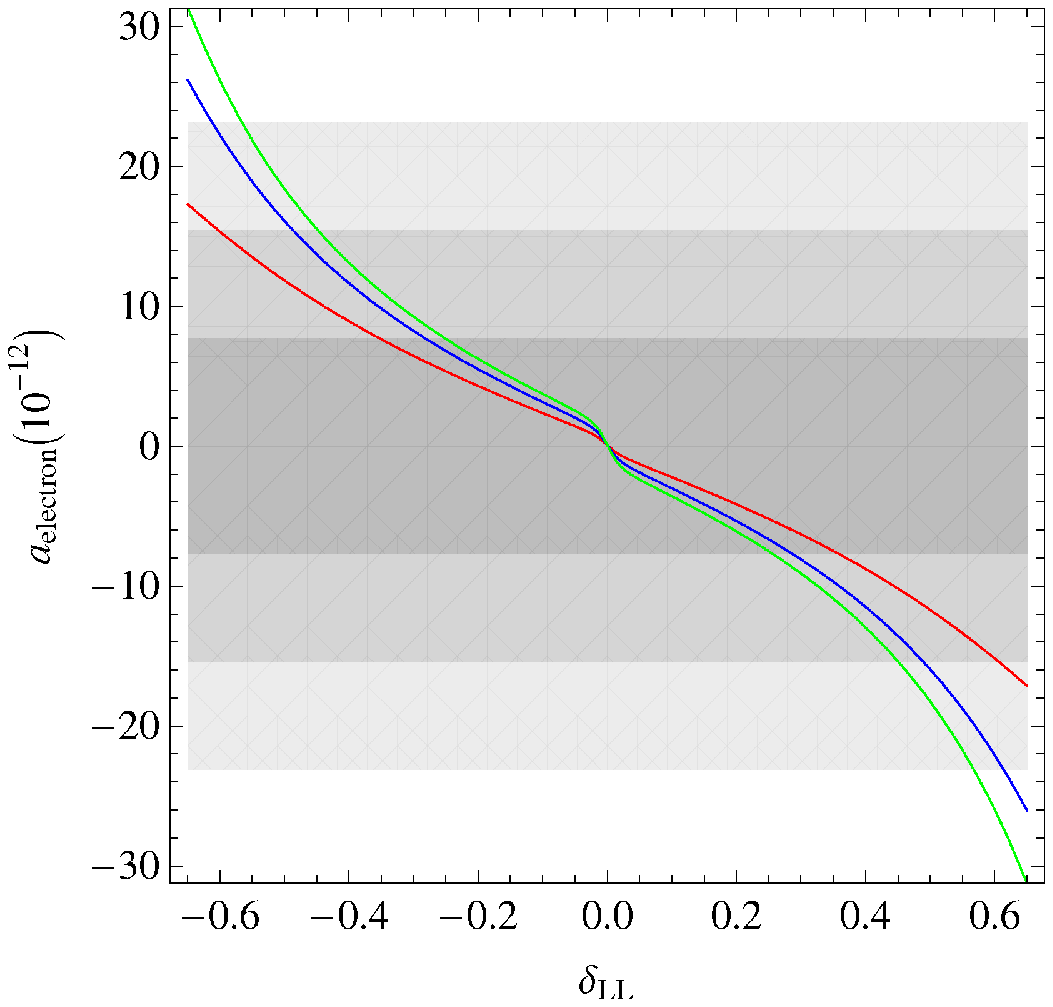}
  }%
  \hfill
  \subfloat[\label{deltaLLelektronkorr}]{%
    \includegraphics[width=.45\linewidth]{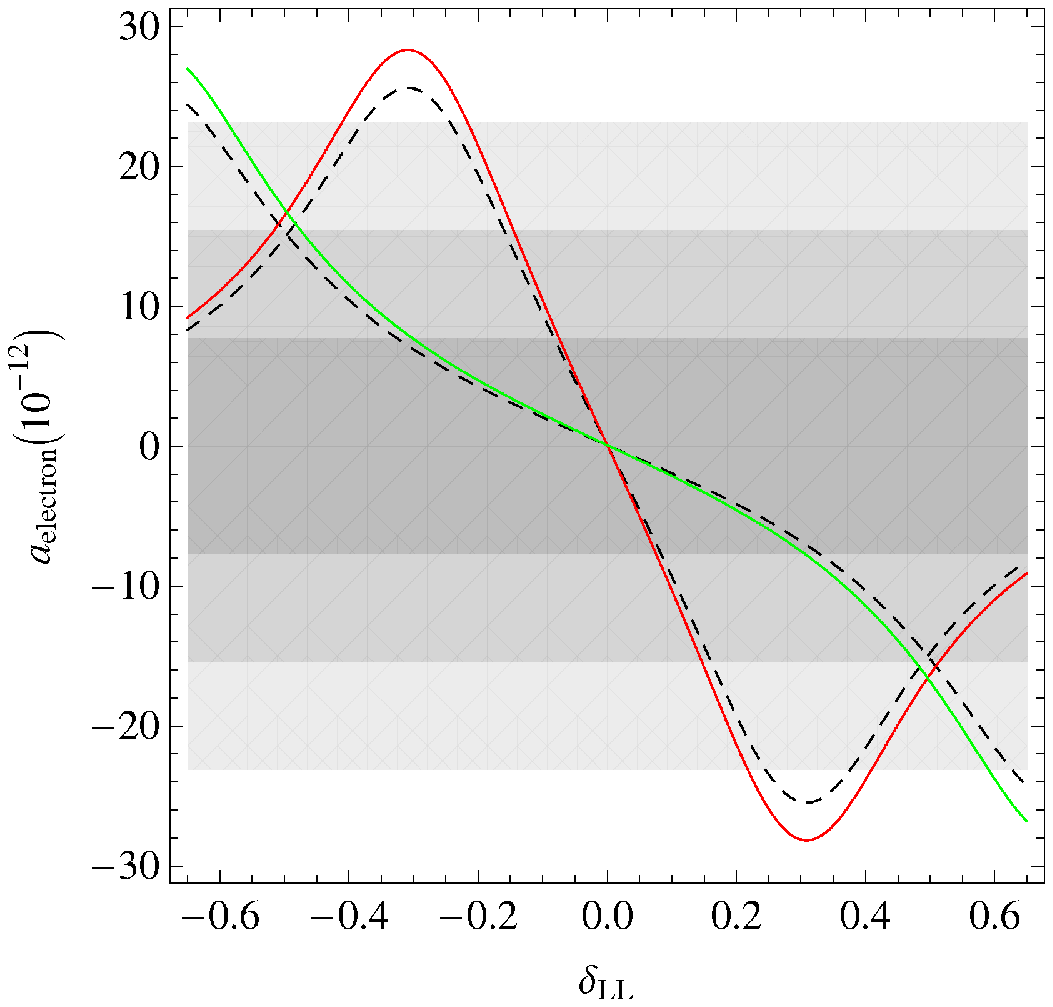}
  }%
  \caption{Supersymmetric contributions to $a_e$ as a function of
    $\delta^{13}_{LL}$ for $\delta^{13}_{RR}$ for (a) scenario 5 (from
    steep to level: $\delta^{13}_{RR} = 0.6$ (green); $0.4$ (blue);
    $0.2$ (red)); (b) scenario 2 ($\delta^{13}_{RR} = 0.6$ (green);
    $0.2$ (red)) of Tab.~\ref{LFVRes} with $M_\text{SUSY} = 500$,
    $\tan\beta = 50$, and $\mu = M_\text{SUSY}$.  The light, medium, and
    dark grey regions correspond to the theoretical $1\sigma$,
    $2\sigma$, and $3\sigma$ regions, respectively.  In (b), the dashed
    curve shows the result without the mass correction.}
  \label{fig:deltaLLelektron}
\end{figure}

We show the results for our scenarios 2 and 5 (see Table~\ref{LFVRes})
in Fig.~\ref{fig:deltaLLelektron}.  As $\delta^{13}_{RR}$ increases,
the bound on $\delta^{13}_{LL}$ becomes stronger and vice versa.  The bound
strongly depends on the SUSY mass.  Since $a_e$ decouples for large SUSY
masses, the bounds become very loose for $M_\text{SUSY} \gtrsim
500$~GeV.  On the other hand, small SUSY masses lead to complex slepton
masses, resulting in a lower bound on the SUSY mass.  For this reason,
the upper bounds on $\delta^{13}_{LL}$ and $\delta^{13}_{RR}$ are
limited by the SUSY mass constraints.  We find
$\left|\delta^{13}_{LL} \cdot \delta^{13}_{RR}\right| < 0.1$ for 
$M_\text{SUSY} \lsim 500$~GeV, coinciding with our non-decoupling bound 
in \eq{eq:bdme}.

\subsection{The radiative decay \boldmath{$l_j \rightarrow l_i \gamma$}}
\label{subsec:ljnachligamma}

Since their SM branching ratios are tiny, supersymmetric contributions
to lepton flavour violating decays $l_i \rightarrow l_j \gamma$ can be
sizable and vastly dominate over the SM values.  As indicated above,
these decays currently give the best constraints on the left-left (LL)
and left-right (LR) lepton flavour violating parameters.
At one-loop level and within MIA, $l_i \rightarrow l_j
  \gamma$ has for example extensively been studied in
  Ref.~\cite{Masina:2002mv}, constraining e.g.\ the mSUGRA parameters $M_1$ and
  $m_R$.
In this section, we compute the supersymmetric contributions to $l_i
\rightarrow l_j \gamma$, including both the mass renormalisation and the
two-loop contributions coming from flavour-violating loops.  The current
upper bounds for the branching ratios are listed in
Table~\ref{tab:expBRSchranken}.

\begin{table}[b]
  \centering
  \begin{tabular}{|c|c|}
    \hline
    \multicolumn{2}{|c|}{experimental  upper bounds}\\
    \hline
    \hline
    {\slshape BR}$\left(\mu\rightarrow e\gamma\right)$ & $1.2\cdot
    10^{-11}$\\ 
    \hline
    {\slshape BR}$\left(\tau\rightarrow e\gamma\right)$ & $1.1\cdot
    10^{-7}$\\ 
    \hline
    {\slshape BR}$\left(\tau\rightarrow \mu\gamma\right)$ & $6.8\cdot
    10^{-8}$\\ 
    \hline
  \end{tabular}
  \caption{Current upper bounds for {\slshape BR}$\left(l_{j}\rightarrow
      l_{i}\gamma\right)$, $j>i$ \cite{Ciuchini:2007ha}.}
  \label{tab:expBRSchranken} 
\end{table}

\smallskip

Let us briefly summarise the formalism.  Three SUSY diagrams contribute
to the amplitude of $l_j \rightarrow l_i \gamma$, corresponding to the
coupling of the photon to $l_j$, $l_i$, and the charged particle in the
loop.  The off shell amplitude can be written as \cite{Hisano:1995cp}
\begin{align}
  i\mathcal{M} = ie\epsilon^{\mu*} \overline{u}_i(p-q) \left[
    q^2\gamma_{\mu}(A_1^LP_L + A_1^R P_R) + m_{l_j} i\sigma_{\mu\nu}
    q^{\nu}(A_2^LP_L + A_2^R P_R)\right] u_j(p) \ ,
\end{align}
where $\epsilon^*$ is the photon polarisation vector.  If the photon is
on shell, the first part of the off-shell amplitude vanishes.  
 
The coefficients $A$ contain chargino and neutralino contributions,
\begin{align}
  A^{L,R} = A^{(\tilde{\chi}^0)L,R} + A^{(\tilde{\chi}^\pm)L,R} , \quad
  i=1,2 \ ,
\end{align}
so $A^L$ is given by the sum of \cite{Hisano:1995nq}
\begin{align}
  A_2^{(\tilde{\chi}^0)L} & = \frac{1}{32 \pi^{2}} \sum_{A=1}^{4}
  \sum_{X=1}^{6} \frac{1}{m_{\tilde{l}_{X}}^{2}} \left[ N_{iAX}^{L}
    N_{jAX}^{L*} \frac{1}{12}F_1^N(x_{AX}) + N_{iAX}^{L} N_{jAX}^{R*}
    \frac{m_{\tilde{\chi}^{0}_{A}}}{3m_{l_{j}}} F_2^N(x_{AX}) \right] ,
  \\
  A_2^{(\tilde{\chi}^{\pm})L} & = - \frac{1}{32 \pi^{2}} \sum_{A=1}^{2}
  \sum_{X=1}^{3} \frac{1}{m_{\tilde{\nu}_{X}}^{2}} \left[ C_{iAX}^{L}
    C_{jAX}^{L*}\frac{1}{12} F_1^C(x_{AX}) + C_{iAX}^{L} C_{jAX}^{R*}
    \frac{2m_{\tilde{\chi}^{\pm}_{A}}}{3m_{l_{j}}} F_2^C(x_{AX}) \right] ,
\end{align}
with the couplings given in Eqs.~(\ref{equ:Kopplungeni}).  We get $A^R$
by simply interchanging $L\leftrightarrow R$.

Finally, the decay rate is given by
\begin{align}
  \Gamma(l_j \rightarrow l_i\gamma) = \frac{e^2}{16 \pi} m^5_{l_j}
  \left( \left|A^L_2\right|^2 + \left|A^R_2\right|^2 \right) .
\end{align}

\smallskip

\begin{figure}
  \centering
  \includegraphics[width = 10cm]{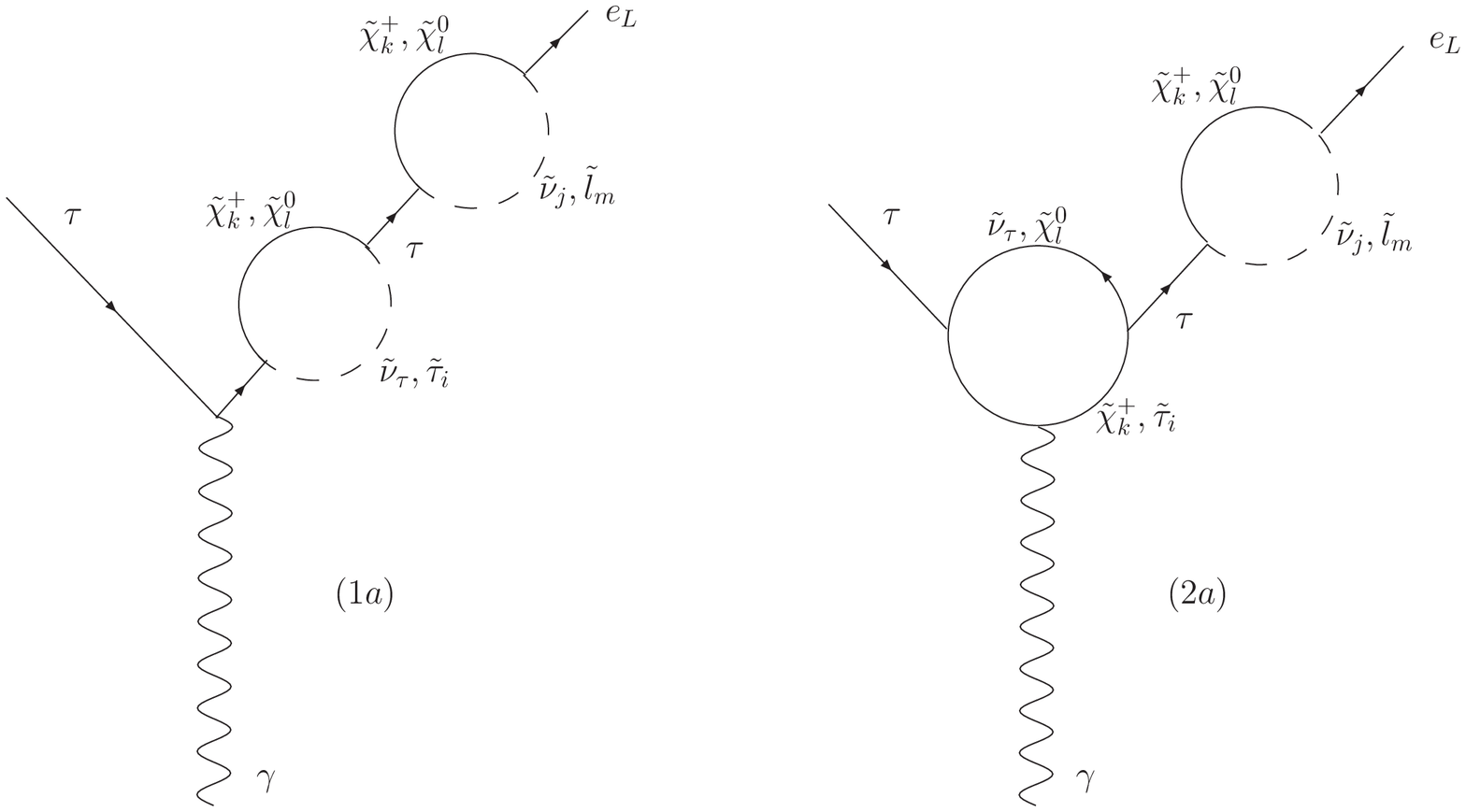}
  \\[18pt]
  \includegraphics[width = 13cm]{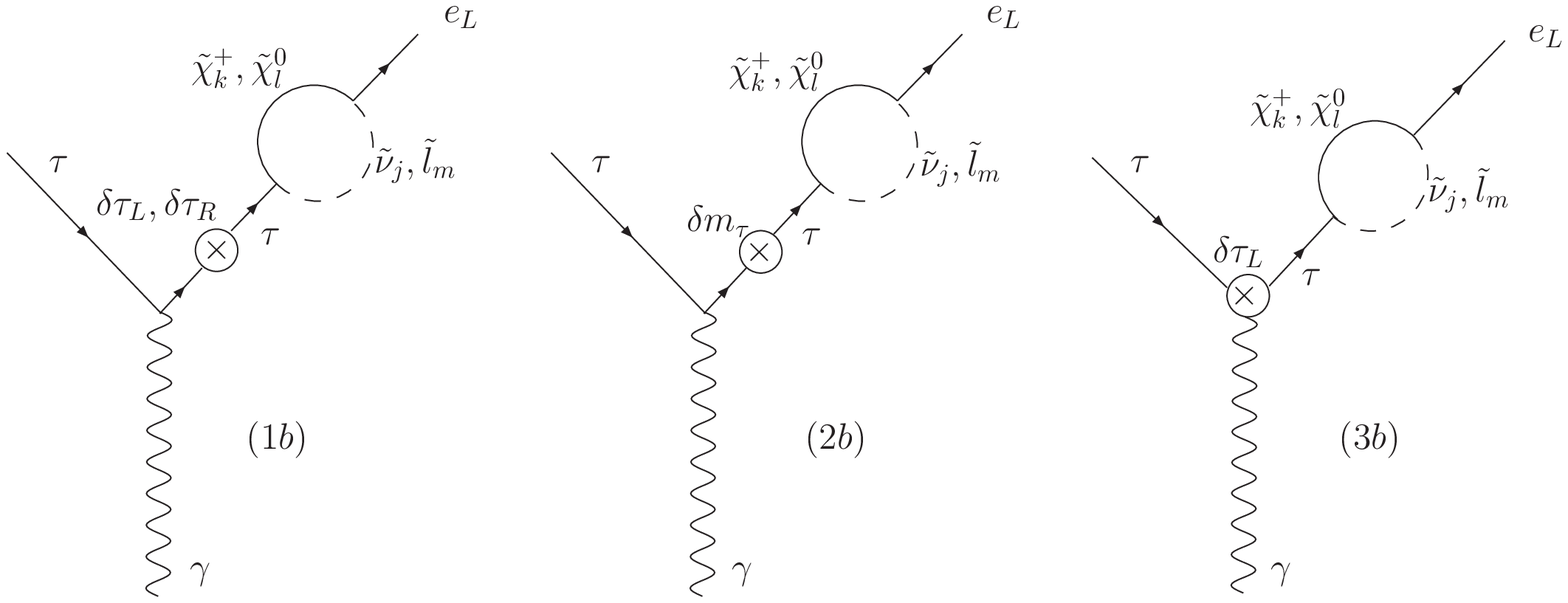}
  \caption{Two-loop contributions to $\tau \rightarrow e\gamma$ from (a)
    chargino and neutralino loops; (b) from the counterterm of the
    $\tau$ propagator.}
  \label{2.Korrektur}
\end{figure}

Both the flavor-conserving transition $l_i \rightarrow l_i \gamma$ and
the flavour-changing self-energies are $\tan\beta$-enhanced.  For this
reason, we do not only consider the effect of the mass renormalisation
but also include the two-loop contributions.  Because of the double
$\tan\beta$ enhancement they can compete with the first non-vanishing
contribution.  As for the corresponding counterterms, mass counterterms
have to be inserted.  In addition, wave-function renormalisation
counterterms play a role as the above-quoted result for $l_j \rightarrow
l_i \gamma$ presumes an expansion in the external momenta of the lepton.
Therefore, to be consistent, the counterterm has to be given in higher
order of the external momentum.  However, only the mass counterterm will
be $\tan\beta$-enhanced because of the chirality flip involved.
Corresponding diagrams are shown in Figs.~\ref{2.Korrektur}.

The wave-function and mass counterterms are given by:
\begin{align}
  \label{delta-LLRR-mtau}
  l_L^0 & = \left(1+\frac{1}{2}\delta l_L\right) l_L \,, &
  l_R^0 & = \left(1+\frac{1}{2}\delta l_R\right) l_R \,, &
  m_{l}^0 & = m_{l} + \delta m_{l} \,,
\end{align} 
where the fields and masses with a superscript 0 are the unrenormalised
fields.
In order to identify the counterterms, one first considers the kinetic
and the mass term of the Lagrangian.  The one-loop self-energy of the
lepton can be divided into a scalar and a vector-type part, where the
latter can further be divided in a left-left and a right-right
transition,
\begin{align}
  i\Sigma_{l}(p) = i\Sigma_{l_L-l_R}^S(p) + i \slashed{p}\, \Sigma_{l_L-l_L}(p)
  P_L + i \slashed{p}\, \Sigma_{l_R-l_R}(p) P_R \ .
\end{align}
Now we demand that the additional terms in the mass Lagrangian cancel
the scalar-part of the one-loop self-energy whereas the additional terms
in the wave-function Lagrangian cancel the vector-type part.  Therefore
the counterterms have to fulfil the following conditions:
\begin{alignat}{2}
  \label{Wellenfcounterterm}
  \delta l_L & = -\Sigma_{l_L-l_L}(p^2=m_l^2), &\quad \delta l_R & =
  -\Sigma_{l_R-l_R}(p^2=m_l^2) \ ,
  \\
  \label{Massencounterterm}
  \delta m_{l} & = \Sigma_{l_L-l_R}(p^2=m_l^2) - \frac{m_{l}}{2} (\delta
  l_L + \delta l_R) \ .  \quad
\end{alignat}
To give an explicit expression for the counterterms, we expand the 
self-energies up to the quadratic order in the external momentum and then
compute the two parts of the one-loop self-energy.  In the series, the
even and odd orders contribute to the scalar and the vector type part,
respectively.
The chargino contribution to the counterterm is then given by
\begin{align}
  \delta m_{l} & = \frac{1}{16\pi^2} \sum_k
  \Gamma_{l,\tilde{\nu},\tilde{\chi}_k}
  \Gamma^*_{l,\tilde{\nu},\tilde{\chi}_k}\left( - B_0 + m^2_{l}
    m_{\tilde{\chi}_k}
    C_0(m_{\tilde{\chi}}^2,\, m_{\tilde{\nu}}^2,\, m_{\tilde{\nu}}^2)
    \vphantom{\frac{4}{d}} - \frac{4}{d} m^2_{l} m_{\tilde{\chi}_k}
    D_2(m_{\tilde{\chi}}^2,\, m_{\tilde{\nu}}^2,
    m_{\tilde{\nu}}^2,\, m_{\tilde{\nu}}^2)\right) \nonumber
  \\
  \delta l_L & = \frac{1}{16\pi^2} \sum_k
  \Gamma^R_{l,\tilde{\nu},\tilde{\chi}_k}
  \Gamma^{R*}_{l,\tilde{\nu},\tilde{\chi}_k} \frac{2}{d}
  C_2(m_{\tilde{\chi}}^2,\, m_{\tilde{\nu}}^2,\, m_{\tilde{\nu}}^2) \nonumber
  \\
  \delta l_R & = \frac{1}{16\pi^2} \sum_k
  \Gamma^L_{l,\tilde{\nu},\tilde{\chi}_k}
  \Gamma^{L*}_{l,\tilde{\nu},\tilde{\chi}_k} \frac{2}{d}
  C_2(m_{\tilde{\chi}}^2,\, m_{\tilde{\nu}}^2,\, m_{\tilde{\nu}}^2) \ ,
\end{align}
where $d = 4-2 \epsilon$.
The wave-function counterterms also induce an additional lepton photon
vertex, $\delta l_L \, \bar l_L \gamma^\mu l_L A_\mu$.

Now we can compute the various diagrams (cf.~Fig.~\ref{2.Korrektur}).
Up to second order in the momentum $p$ all contributions indeed cancel
each other.  For the chargino two-loop contribution to $\tau \rightarrow
e \gamma$, we obtain
\begin{align}
  \mathcal{M}^\text{2-loop} & = \overline{u}_{e} P_R
  \Sigma_{\tau_R- e_L} m_{\tau} i \sigma^{\mu\nu} q_{\nu}u_\tau
  \frac{1}{32\pi^2} \left[ \sum_{A=1}^{4} \sum_{X=1}^{6}
    \frac{1}{m^2_{\tilde{l}_X}} \mathscr{N}_{AX} + \sum_{A=1,2}\sum_{X =
      1}^{3} \frac{1}{m^2_{\tilde{\nu}_{X}}} \mathscr{C}_{AX} \right] ,
  \intertext{where}
  \mathscr{N}_{AX} & = -\left( \left|N_{3AX}^{L}\right|^2 +
    \left|N_{3AX}^{R}\right|^2 \right) \frac{1}{12} F_1^N
  \left(x_{AX}\right) - \frac{m_{\tilde{\chi}^0_A}}{3m_\tau} \re \left[
    N_{3AX}^{L} N_{3AX}^{R\ast}\right] F_2^N \left(x_{AX}\right) , \nonumber
  \\
  \mathscr{C}_{AX} & = \left( \left|C_{3AX}^{L}\right|^2 +
    \left|C_{3AX}^{R}\right|^2 \right) \frac{1}{12} F_1^C
  \left(x_{AX}\right) + \frac{2m_{\tilde{\chi}^{\pm}_A}}{3m_{\tau}} \re
  \left[ C_{3AX}^{L} C_{3AX}^{R\ast}\right] F_2^C \left(x_{AX}\right)
  \nonumber
\end{align}
and the couplings $N_{3AX}^{L,R}$ and $C_{3AX}^{L,R}$ are defined in
Eqs.~(\ref{equ:Kopplungeni}).

For the numerical analysis, we first consider the mSUGRA scenarios
listed in Table~\ref{SPS} as well as the scenarios of
Table~\ref{LFVRes}.  The $\mu$ parameter at $M_\text{ew}$ is determined
with Isajet \cite{Kraml, Belanger:2005jk, Allanach:2003jw}.
Note that the different bounds for $\delta_{LL}^{ij}$ in the literature
can differ due to their dependence on the chosen point in the SUSY
parameter space, see, e.g.,
Refs.~\cite{Hisano:1995nq,Hisano:1995cp,Masiero:2002jn,Calibbi:2006nq}.
\begin{table}
  \centering
  \begin{tabular}{c|ddddddd}
    \hline
    \rule[-3pt]{0pt}{13pt}
    SPS & \multicolumn{1}{c}{1a} & \multicolumn{1}{c}{1b} &
    \multicolumn{1}{c}{2} & \multicolumn{1}{c}{3} &
    \multicolumn{1}{c}{4} & \multicolumn{1}{c}{A} &
    \multicolumn{1}{c}{B} \\
    \hline
    \hline
    \rule[-6pt]{0pt}{16pt}
    $\left|\delta_{LL}^{12}\right|\leq$ & 0.000221 & 0.00019 & 0.00179 &
    0.00047 & 0.000096 & 0.00028 & 0.00116 \\ 
    \hline
    \rule[-6pt]{0pt}{16pt}
    $\left|\delta_{LL}^{13}\right|\leq$ & 0.048 & 0.041 & 0.381 & 0.104
    & 0.0217 & 0.063 & 0.260 \\ 
    \hline
    \rule[-6pt]{0pt}{16pt}
    $\left|\delta_{LL}^{23}\right|\leq$ & 0.038 & 0.032 & 0.299 & 0.082
    & 0.017 & 0.049 & 0.204 \\ 
    \hline
  \end{tabular}
  \caption{Upper bounds on $\left|\delta_{LL}^{12}\right|$,
    $\left|\delta_{LL}^{13}\right|$ and $\left|\delta_{LL}^{23}\right|$
    for the mSUGRA scenarios of Table~\ref{SPS} from {\slshape
      BR}$\left(l_{j}\rightarrow l_{i}\gamma\right)$ 
    including mass renormalisation and two loop
    contributions.}
  \label{tab:deltaLLSchranken} 
\end{table}
Table~\ref{tab:deltaLLSchranken} summarises the bounds on
$\left|\delta_{LL}^{12}\right|$, $\left|\delta_{LL}^{13}\right|$ and
$\left|\delta_{LL}^{23}\right|$ for these scenarios; they include both
$\tan\beta$-enhanced corrections to $l_{j}\rightarrow l_{i}\gamma$,
namely the mass renormalisation and two loops contributions.
Interestingly, the two corrections tend to cancel each other: As
illustrated for SPS4 in Table~\ref{tab:SPS4boundscorrections}, the mass
renormalisation tightens the bound, whereas the two loops effects
increases them again.  Thus, the effect is generally smaller than 1\,\%,
particularly for the small $\tan\beta$ scenarios.  For large $\tan\beta$
(as is the case in SPS4), however, the deviation can reach 6\,\%.
Without the inclusion of our two corrections we recover
  the results already found in
  Refs.~\cite{Masina:2002mv,Paradisi:2005fk} after taking into account
  that the experimental upper bounds have changed a bit.

\begin{table}
  \centering
  \begin{tabular}{c|lll}
    \hline
    SPS4 & \multicolumn{1}{c}{tree level} & \multicolumn{1}{c}{+ mass
      renormalisation} & \multicolumn{1}{c}{+  two loops effects} 
    \\
    \hline
    \rule[-6pt]{0pt}{16pt}
    $\left|\delta_{LL}^{12}\right|\leq$ & 0.000101189 & 0.000094695
    $(-6.4\%)$ & 0.000095998 $(-5.1\%)$ \\
    \hline
    \rule[-6pt]{0pt}{16pt}
    $\left|\delta_{LL}^{13}\right|\leq$ & 0.021472 &
    0.020053\phantom{000} $(-6.6\%)$ & 0.021666\phantom{000} $(+0.9\%)$
    \\
    \hline
    \rule[-6pt]{0pt}{16pt}
    $\left|\delta_{LL}^{23}\right|\leq$ & 0.016778 &
    0.015671\phantom{000} $(-6.6\%)$ & 0.016925\phantom{000} $(+0.9\%)$
    \\
    \hline
  \end{tabular}
  \caption{Upper bounds on $|\delta_{LL}^{12}|$, $|\delta_{LL}^{13}|$
    and $|\delta_{LL}^{23}|$ for SPS4  without any corrections, with
    mass renormalisation and taking into account both mass
    renormalisation and two loops contribution. In parenthesis:
    deviation compared to the tree level bound in
    percent.}\label{tab:SPS4boundscorrections} 
\end{table}

\begin{figure}
  \centering

  \includegraphics[width=.6\linewidth]{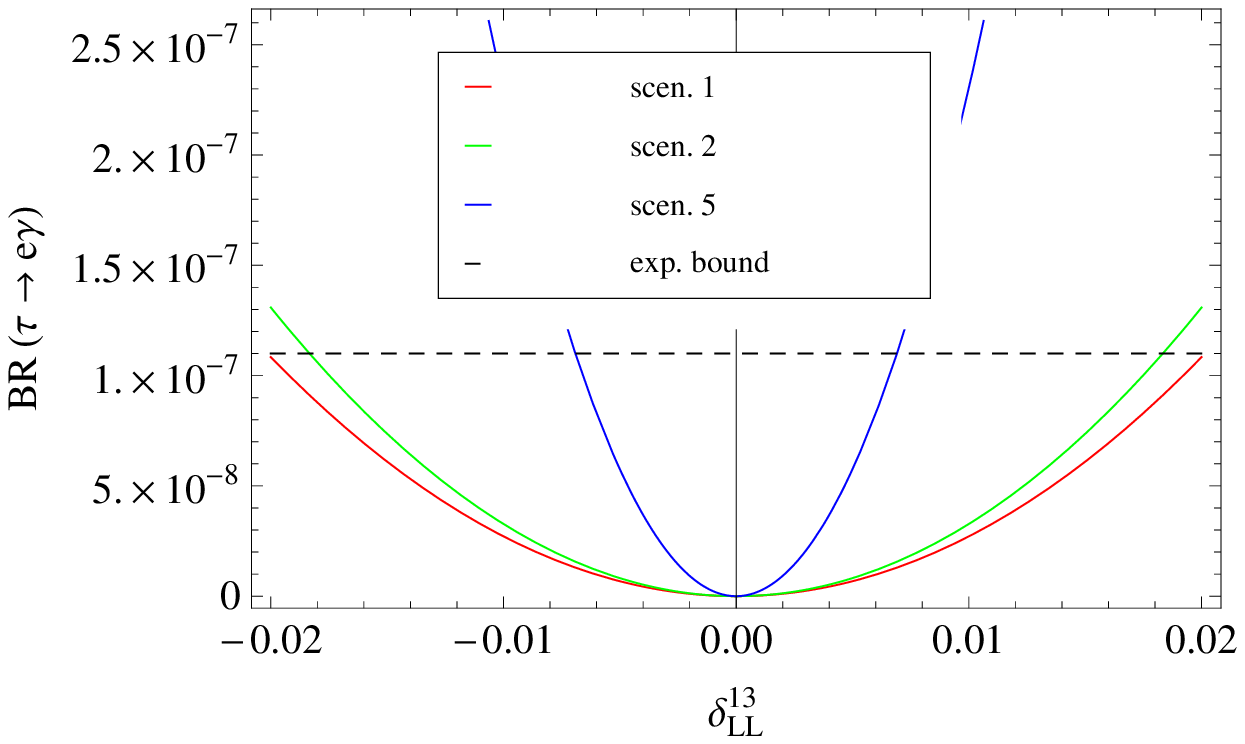}
  \includegraphics[width=.6\linewidth]{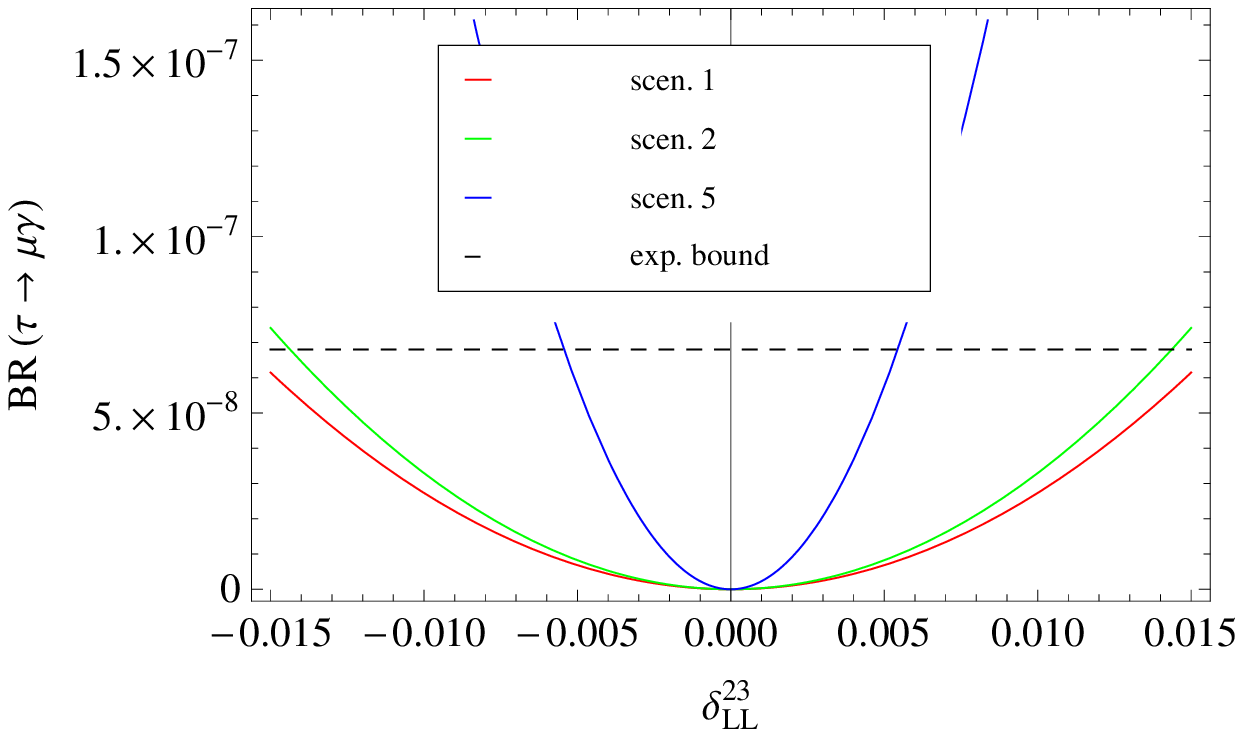}
  \includegraphics[width=.6\linewidth]{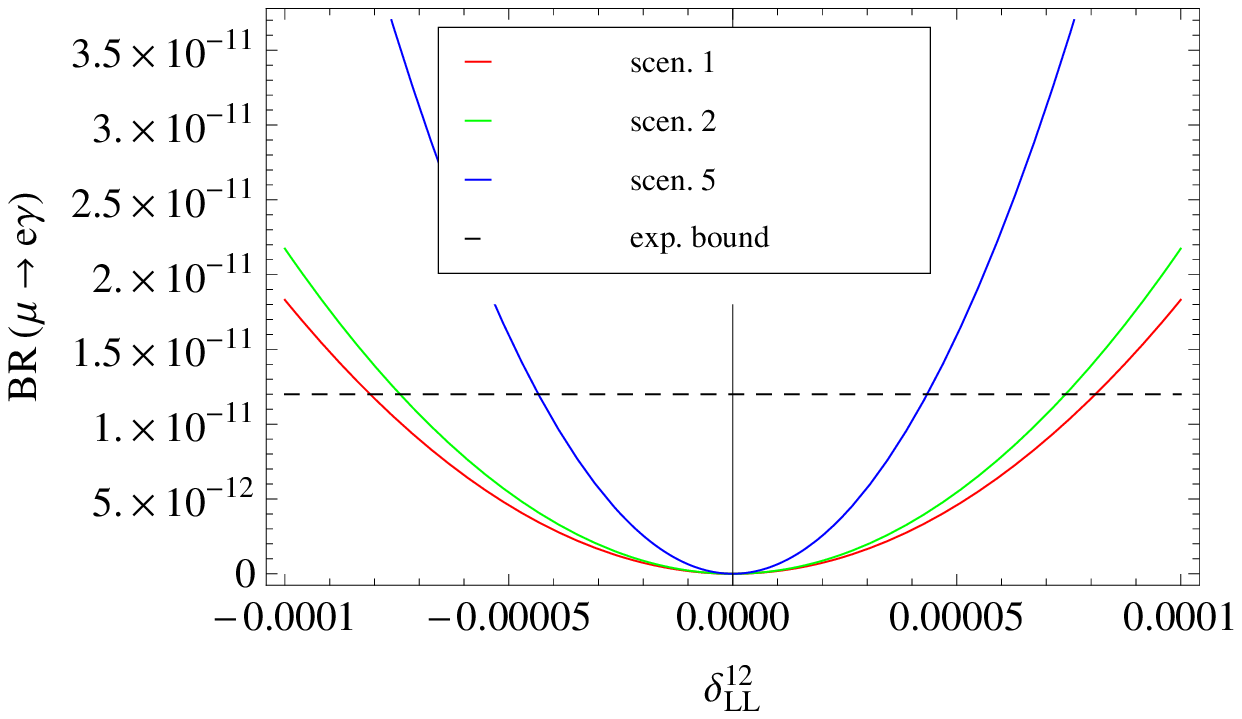}
  \caption{The branching ratio as a function of $\delta^{13}_{LL}$,
    $\delta^{23}_{LL}$ and $\delta^{12}_{LL}$ with corrections in the
    scenarios 1, 2, 5 (bottom to top) 
    at $M_{\text{SUSY}} = 300$ GeV and $\tan\beta = 50$.
}
  \label{tauegammakorrektur}
\end{figure}

\begin{figure}
  \centering
 \psfrag{Szenario}{\hspace{0.0cm}\scalefont{0.9}$\text{scen.}$}

  \includegraphics[width=.6\linewidth]{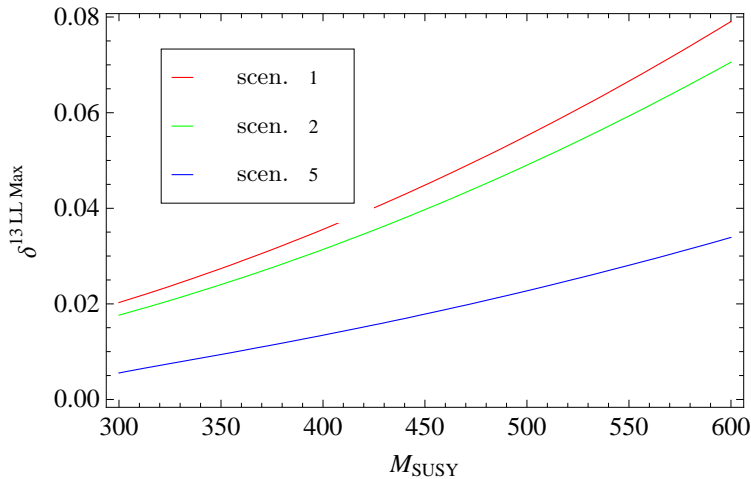}
  \caption{$\delta^{13(max)}_{LL}$ as a function of $M_{\text{SUSY}}$ in
    the scenarios 1, 2 and 5 (top to bottom).}
  \label{tauegammacontour}
\end{figure}

Let us therefore study the scenarios of Table~\ref{LFVRes} with
$M_{\text{SUSY}} = 300$ GeV and $\tan\beta = 50$.  The branching ratios
of $l_j \rightarrow l_i \gamma$ in the scenarios 1, 2 and 5 of table
\ref{LFVRes} are shown in Figure~\ref{tauegammakorrektur}.  We see that
scenario 5 gives the strongest constraint on $\delta^{13}_{LL}$.  In
addition, the corrections discussed here have the biggest effect in this
scenario.  The upper bound on $\delta^{13}_{LL}$ again depends on the
SUSY mass.  The branching ratio decouples for large SUSY masses so that
the upper bounds weakens for increasing $M_\text{SUSY}$
(Fig.~\ref{tauegammacontour}).

 As already noted in Sec.~\ref{se:pmns} the corrections
  from supersymmetric loops cannot reasonably push $U_{e3}$ into the
  reach of the DOUBLE CHOOZ experiment without violating the bound from
  $\tau \to e \gamma$: E.g.\ for sparticle masses of 500 GeV we find
$|\delta U_{e3}|<10^{-3}$ corresponding to a correction to the mixing
angle $\theta_{13}$ of at most 0.06 degrees.  That is, if the DOUBLE
CHOOZ experiment measures $U_{e3}\neq 0$, one will not be able to
ascribe this result to the SUSY breaking sector. Stated positively,
$U_{e3}\gtrsim 10^{-3}$ will imply that at low energies the flavour
symmetries imposed on the Yukawa sector to motivate tri-bimaximal mixing
are violated. If the same consideration is made for the most optimistic
reach $\theta_{13}\leq 0.6^\circ$ of a future neutrino factory (the
quoted bound corresponds to the best value of the CP phase in the PMNS
matrix), the threshold corrections become relevant only for sparticle
masses well above 1500$\,$GeV.  While the considered effects in both
$\tau\rightarrow e \gamma$ and $U_{e3}$ involve the product
$\delta^{33}_{RL}\delta^{13}_{LL}$, the qualitative result is equally
valid, if the needed flavour and chirality violations are triggered by
$\delta^{31}_{RL}$ or other combinations of the $\delta^{ij}_{XY}$'s.

{While we have only considered loops with a single flavour-changing
  $\delta^{ij}_{XY}$ in our discussion of $l_j \rightarrow l_i \gamma$
  decays, contributions proportional to
  $\delta^{23}_{RR}\delta^{31}_{LL}$ can be relevant for $\mu\to e
  \gamma$. For recent analyses including this effect we refer to
  Refs.~\cite{Hisano:2009ae,Altmannshofer:2009ne}. If both $\tan\beta$
  is large and the charged-Higgs-boson mass is small, further two-loop
  effects involving a virtual Higgs boson can be relevant
  \cite{Paradisi:2005tk,Paradisi:2006jp}. These effects are suppressed
  by one power of $\tan\beta$ with respect to the two-loop corrections
  included by us, but do not vanish for $M_{\rm SUSY}\gg M_{H^+},v$.}
In Ref.~\cite{Hisano:2008hn} effective
  lepton-slepton-gaugino vertices reproducing the chirally enhanced
  corrections in the leading order of $v^2/M_{\rm SUSY}^2$ have been
  derived and applied to electric dipole moments, cf.\ the overview  
  on previous work at the beginning of Sec.~\ref{se:lfv}.

\section{Renormalisation group equation with SUSY seesaw mechanism}
\label{sec:RGE}

In the previous section, we derived bounds on the off-diagonal elements
of the slepton mass matrix, parametrised by $\delta_{XY}^{ij}$.  These
are a priori free parameters in the MSSM; they are set once we know how
supersymmetry is broken.  We saw, however, that these elements are
well-constrained and this result generally applies to the soft terms.
Therefore one usually assumes universality of the supersymmetry breaking
terms at a high scale, e.g.\ $M_\text{GUT} = 2\cdot 10^{16}$ GeV where
the SM gauge couplings converge.  Then the renormalisation group
equation (RGE) running induces non-vanishing $\delta_{XY}^{ij}$ at the
electroweak scale.  Clearly, the size of $\delta_{XY}^{ij}$ is
model-dependent.

In this section, we will consider two GUT scenarios based on the gauge
group SO(10), which generically includes right-handed neutrinos.  The
breaking of SO(10) around $M_\text{GUT}$ generates heavy Majorana masses
for these right-handed neutrinos.  After the electroweak symmetry
breaking, the left-handed neutrinos receive small Majorana masses via
the seesaw mechanism.

\subsection{Neutrino Yukawa Couplings and Grand Unification}
\label{subsec:neutrinoYuk}

The seesaw mechanism naturally explains tiny neutrino masses.  As
already discussed in the Introduction, the right-handed neutrinos are
singlets under the standard model group.  Then we can write down an
explicit mass term, $(M_R)_{ij} \nu^c_{i} \nu^c_{j}$ (see
Eqs.~(\ref{equ:SuperpotenzialMR})).  Now if the entries $(M_{R})_{ij}$
are much larger than the electroweak scale, we can integrate out the
heavy neutrino fields at their mass scales.  Below the scale of the
lightest state the Yukawa couplings are then given by
\begin{align}
  W_\text{eff} = W_\text{MSSM} + \frac{1}{2} \left(Y_{\nu} L
    H_{u}\right)^\top M_{R}^{-1} \left(Y_{\nu} L H_{u}\right) .
\end{align}
After electroweak symmetry breaking, $W_\text{eff}$ leads to the
following effective mass matrix for the light neutrinos:
\begin{align}
  \label{equ:MnueffRGE}
  \mathcal{M}_{\nu} = -Y_{\nu}^\top\, M_{R}^{-1}\, Y_{\nu}\, v_{u}^{2}
  \equiv -\kappa\, v_{u}^{2} \;.
\end{align}
Since the light neutrinos cannot be heavier than 1~eV and the mass scale
of the atmospheric oscillations is of order 0.1~eV, the Majorana mass scale is
around $10^{14}$~GeV.

In the MSSM, it is convenient to choose both the Yukawa coupling matrix
of the charged leptons and the Majorana mass matrix of the right-handed
neutrinos diagonal.
In this basis, $\mathcal{M}_{\nu}$ is diagonalised by the PMNS matrix,
\begin{align}
  \label{equ:Mnueffdiag}
  U_\text{PMNS}^\top\, \mathcal{M}_{\nu}\, U_\text{PMNS} = \diag
  \left(m_{\nu_{1}},\,_{\nu_{2}},\,_{\nu_{3}}\right) \equiv
  -\mathcal{D}_{\kappa} v_{u}^{2} \;.
\end{align}
By means of Eqs.~(\ref{equ:MnueffRGE}) and (\ref{equ:Mnueffdiag}),
$Y_{\nu}$ can be expressed as \cite{Casas:2001sr}
\begin{align}
  \label{equ:Ynuallg}
  Y_{\nu} = \mathcal{D}_{\sqrt{M}}\, R\, \mathcal{D}_{\sqrt{\kappa}}\,
  U_\text{PMNS}^{\dagger} \;, \qquad \mathcal{D}_{M} \equiv \diag
  \left(M_{R_{1}},\, M_{R_{2}},\, M_{R_{3}}\right) ,
\end{align}
with an arbitrary orthogonal matrix $R$.  Thus $Y_\nu$ depends both on
the measurable parameters, contained in the diagonal matrix
$\mathcal{D}_{\kappa}$ and $U_\text{PMNS}$, and the model-dependent
parameters, namely three Majorana masses and three mixing parameters.
In the MSSM, these are completely free parameters.

The seesaw mechanism is automatic in grand-unified models with broken
$\text{U(1)}_{B-L}$ symmetry \cite{Minkowski:1977sc}.  ($B$ and $L$
denote baryon and lepton number, respectively.)  In SO(10), the SM
fermions of each generation are unified in one matter representation,
together with the singlet neutrino \cite{Georgi:1974my,Fritzsch:1974nn}.
No further fermionic multiplets are needed.  An additional Higgs field
acquires a vev, breaking the SO(10) subgroup $\text{SU(2)}_R \times
\text{U(1)}_{B-L}$ to hypercharge, $\text{U(1)}_Y$.  At the same time,
Majorana masses for the SM singlets are generated.  As indicated in
Eqs.~(\ref{equ:SuperpotenzialMR}), the up-quarks and neutrinos couple to
the same Higgs fields $H_u$ so that the Yukawa matrices $Y_{\nu}$ and
$Y_{u}$ are related.  The actual form of this relation is
model-dependent; however, there are two extreme cases
\cite{Masiero:2002jn,Masiero:2004js}.
\begin{enumerate}
\item \emph{Minimal (CKM) case}: The mixing in $Y_{\nu}$ is small and
  \begin{align}
    \label{equ:YnuCKM}
    Y_{\nu} = Y_{u} = V_\text{CKM}^\top\, \mathcal{D}_{u} \,
    V_\text{CKM}
  \end{align}
  holds at the GUT scale.  This case refers to minimal SO(10) scenarios
  with small mixing angles for the Dirac mass matrices.  The large
  leptonic mixing angles are a consequence of the interplay of $Y_\nu$
  and $M_R$ in the seesaw mechanism.
  
  For normal-hierarchical neutrino masses, i.e.~$m_{\nu_{1}} \ll
  m_{\nu_{2}} \simeq \sqrt{\Delta m_{21}^{2}} \ll m_{\nu_{3}} \simeq
  \sqrt{\Delta m_{31}^{2}}$ and the MNS matrix being close to its
  tri-bimaximal form, the masses of the right handed neutrinos are given
  by
  \begin{align}
    \label{equ:MRCKMFall}
    M_{R_{1}} \approx \frac{1}{\displaystyle \frac{m_{\nu_2}}{3m_u^2} +
      \frac{m_{\nu_3}}{2m_c^2}} \,, \qquad 
    M_{R_{2}} \approx 2\frac{m_c^{2}}{m_{\nu_3}} +
    3\frac{m_u^{2}}{m_{\nu_2}} \,, \qquad 
    M_{R_{3}} \approx \frac{m_t^{2}}{6m_{\nu_1}} \,.
  \end{align}

\item \emph{Maximal (PMNS) case}: Large mixing in $Y_{\nu}$ is achieved
  in models with
  \begin{align}
    \label{equ:YnuYu}
    Y_{\nu} = \mathcal{D}_{u}\, U_\text{PMNS}^{\dagger} \,.
  \end{align}
  This scenario is the analogon to the quark case: the mixing matrix
  arises in the Dirac couplings, with the Majorana matrix being
  diagonal.   Note that $Y_\nu$ is not symmetric any more and this
  relation is indeed realised in models with lopsided mass matrices.
  In this case, the masses for the right handed neutrinos are simply
  \begin{align}
    \label{equ:MRmnsFall}
    M_{R_{1}} = \frac{m_{u}^{2}}{m_{\nu_{1}}},\quad M_{R_{2}} =
    \frac{m_{c}^{2}}{m_{\nu_{2}}},\quad M_{R_{3}} =
    \frac{m_{t}^{2}}{m_{\nu_{3}}}.
  \end{align}
\end{enumerate}

In terms of the parametrisation (\ref{equ:Ynuallg}), the second case
corresponds to $R=\mathds{1}$.  Then the mixing in $Y_{\nu}$ is
determined by the PMNS matrix.  By contrast, small CKM-like mixing in
$Y_{\nu}$ (case 1) requires a non-trivial structure of $R$.

Clearly, these two cases are special; however, they provide two
well-motivated but distinct scenarios.
A more detailed introduction to these two cases are given in
\cite{Masiero:2002jn,Masiero:2004js}.  Note that the authors use the LR
convention for the neutrino Yukawa coupling so that their equations
differ from ours by the substitution $Y_{\nu}\leftrightarrow
Y_{\nu}^\top$.

\subsection{Renormalisation-group Analysis}
\label{subsec:solveRGE}

We list the renormalisation group equations (RGE) of the MSSM
\cite{Martin:1993zk,Kazakov:2000us,Bertolini:1990if} and the MSSM with
right-handed neutrinos \cite{Casas:2001sr,Hisano:1998fj,Antusch:2005gp}
in Appendix~\ref{appendix:RGE}.  The right-handed neutrinos are singlets
under the SM gauge group so that they do not change the RGE for gauge
couplings and gaugino masses.

\begin{figure}
  \centering
  \begin{small}
    \begin{picture}(400,190)(0,-20)
      \SetColor{Black}
      \Line(0,0)(150,150)\Line(40,90)(70,120)
      \Line(70,115)(70,120)\Line(65,120)(70,120)
      \Text(70,170)[]{\textbf{Gauge-, Yukawa couplings}}
      \CCirc(0,0){2}{Black}{Black}\Text(-5,10)[]{$M_{Z}$}	
      \CCirc(150,150){2}{Black}{Black}\Text(150,160)[]{$M_\text{GUT}$}
      \CCirc(75,75){2}{Black}{Black}\Text(70,85)[]{$M_{R_{1}}$}
      \CCirc(102.5,102.5){2}{Black}{Black}\Text(97.5,112.5)[]{$M_{R_{2}}$}
      \CCirc(130,130){2}{Black}{Black}\Text(125,140)[]{$M_{R_{3}}$}
      \Text(65,35)[]{$Y_{l}, Y_{u}, Y_{d}$}
      \Text(65,45)[]{$\alpha_{i},\kappa$}
      \Text(5,0)[l]{$\Delta m_\text{sol}^{2},\,\Delta
        m_\text{atm}^{2},\,U_{PMNS}$} 
      \Text(5,-10)[l]{$ M_{l},\,M_{u},\,M_{d},\,\tan\beta,\,\alpha_{i}$}
      \Text(5,-20)[l]{$ m_{\nu_{1}},\,M_{R_{i}}$}
      \Text(95,85)[l]{$\stackrel{(2)}{Y_{\nu}},\stackrel{(2)}{\kappa},
        \stackrel{(2)}{M_{R}}$}
      \Text(125,115)[l]{$\stackrel{(3)}{Y_{\nu}},\stackrel{(3)}{\kappa},
        \stackrel{(3)}{M_{R}}$}
      \Text(150,135)[]{$ Y_{\nu},M_{R}$}
      \Line(150,150)(300,0)
      \Text(200,170)[]{\textbf{A-terms, $M_{i}$}}
      \CCirc(225,75){2}{Black}{Black}
      \CCirc(197.5,102.5){2}{Black}{Black}
      \CCirc(170,130){2}{Black}{Black}
      \CCirc(300,0){2}{Black}{Black}\Text(302,-10)[]{$M_{Z}$}
      \Text(160,150)[l]{$A_{0},\, m_{1/2},\, Y^{GUT}$}
      \Text(210,105)[l]{$ A_{l},\,A_{\nu}$}
      \Text(215,95)[l]{$ A_{d},\,A_{u}$}
      \Text(250,60)[l]{$A_{l},\, A_{d},\,A_{u}$}
      \Text(300,10)[l]{$\rightarrow \delta_{LR}^{ij}$}
      \Line(240,120)(270,90)\Line(270,95)(270,90)\Line(265,90)(270,90)
      \CCirc(250,150){2}{Black}{Black}\Text(250,160)[]{$M_\text{GUT}$}
      \Line(250,150)(400,0)
      \Text(320,170)[]{\textbf{Sfermion and Higgs masses}}
      \CCirc(325,75){2}{Black}{Black}
      \CCirc(297.5,102.5){2}{Black}{Black}
      \CCirc(270,130){2}{Black}{Black}
      \CCirc(400,0){2}{Black}{Black}\Text(402,-10)[]{$M_{Z}$}
      \Text(260,150)[l]{$m_{0}$, universality}
      \Line(340,120)(370,90)\Line(370,95)(370,90)\Line(365,90)(370,90)
      \Text(290,125)[l]{$m_{L}^{2}$}
      \Text(300,115)[l]{$m_{\nu}^{2}$}
      \Text(310,105)[l]{$m_{e}^{2}$}
      \Text(320,95)[l]{$m_{Q}^{2}$}
      \Text(330,85)[l]{$m_{u}^{2}$}
      \Text(340,75)[l]{$m_{d}^{2}$}
      \Text(350,65)[l]{$m_{H_{u}}^{2}$}
      \Text(360,55)[l]{$m_{H_{d}}^{2}$}
      \Text(380,35)[l]{$\rightarrow \delta_{LL}^{ij}$}
      \Text(390,25)[l]{$ \delta_{RR}^{ij}$}
    \end{picture}
  \end{small}
  \caption{Illustration of the procedure used to solve the RGE}
  \label{fig:RGESchema}
\end{figure}
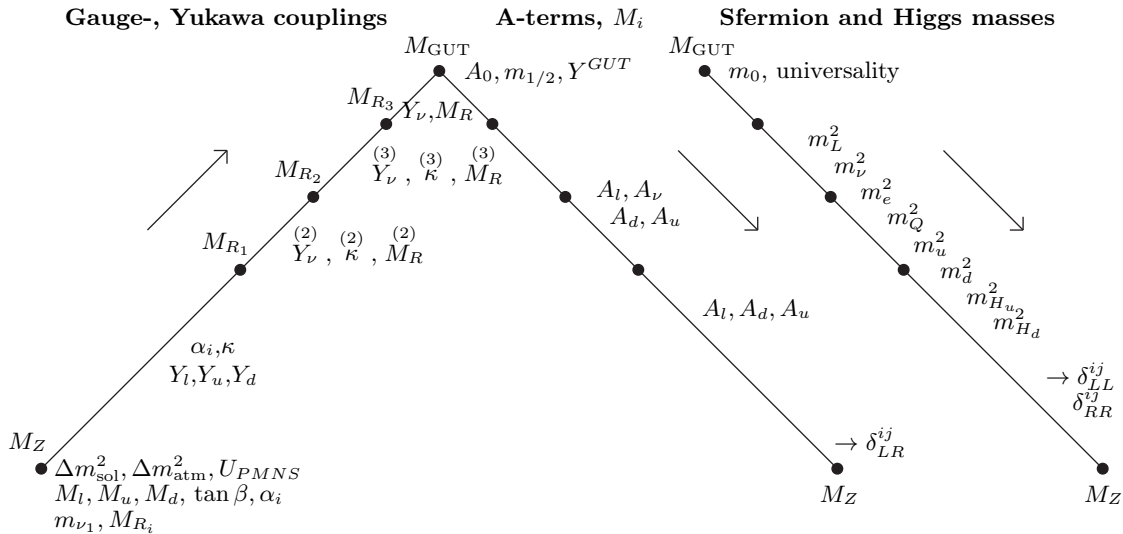

The procedure of solving the RGE is schematically depicted in
Figure~\ref{fig:RGESchema}.  With the experimental values of the
indicated parameters at the scale $M_{Z}$, we evaluate the gauge and
Yukawa couplings at the various mass scales.  The three heavy neutrinos
are included step by step.  At the GUT scale, $M_\text{GUT} = 2\cdot
10^{16}$~GeV, we assume universality of the supersymmetry breaking soft
parameters,
\begin{align}
  m_{\tilde{Q}}^{2} & = m_{\tilde{u}}^{2} = m_{\tilde{d}}^{2} =
  m_{\tilde{L}}^{2} = m_{\tilde{e}}^{2} = m_{0}^{2}\, \mathds{1} \,,
  \mspace{48mu} m_{H_{u}}^{2} = m_{H_{d}}^{2} = m_{0}^{2} \;, \nonumber
  \\[2pt]
  M_{i}^\text{GUT} & = m_{1/2}, \mspace{70mu} i = 1,\,2,\,3 \;,
  \nonumber
  \\[2pt]
  A^\text{GUT}_{f} & = A_{0} Y^\text{GUT}_f , \qquad f = u,\, d,\, l,\,
  \nu \;.
  \label{equ:AGUT}
\end{align}
Solving the RGE in leading order, one gets for the LFV off-diagonal
elements \cite{Casas:2001sr, PhysRevLett.57.961}\footnote{As we assume
  universality of the SUSY breaking parameters at $M_\text{GUT}$, we do
  not receive additional terms due to the coupling to coloured Higgs
  fields as in, e.g., Refs.~\cite{Hisano:1998fj,Ellis:2008st}.}
\begin{align}
  \label{equ:mmLapprox}
  \left(\Delta m_{\tilde{L}}^{2}\right)_{ij} & \simeq
  -\frac{3m_{0}^{2}+A_{0}^{2}}{8\pi^{2}} \left(Y_{\nu}^{\dagger}
    \ln\left(\frac{M_\text{GUT}}{M_{R_{i}}}\right)Y_{\nu}\right)_{ij},
  \quad i\neq j \,,
  \\
  \label{equ:mmeapprox} 
  \left(\Delta m_{\tilde{e}}^{2}\right)_{ij} & \simeq 0,\quad i\neq j
  \,,
  \\
  \label{equ:Alapprox}
  \left(A_{l}\right)_{ij} & \simeq -\frac{3}{8 \pi^{2}}
  A_{0}Y_{l_{i}}\left(Y_{\nu}^{\dagger}
    \ln\left(\frac{M_\text{GUT}}{M_{R_{i}}}\right)Y_{\nu}\right)_{ij},
  \quad i\neq j \,.
\end{align}
(Recall that $m_{\tilde{L}}^2$ and $m_{\tilde{e}}^{2}$ contribute to
$(m^2_L)_{LL}$ and $(m^2_L)_{RR}$ as shown in Eqs.~(\ref{equ:mLLLsoft})
and (\ref{equ:mLRRsoft}).)
The size of LFV depends essentially on the structure and magnitude of
the neutrino Yukawa coupling, the scale of the right handed neutrinos as
well as on the SUSY breaking parameters $m_{0}$ and $A_{0}$.  The only
source of flavour violation stems from $Y_{\nu}$.  According to
Eq.~(\ref{equ:mmeapprox}), off-diagonal RR elements $\delta_{RR}^{ij}$
are not generated at leading order.  The off-diagonal elements
$\left(A_{l}\right)_{i\neq j}$ are related to $\delta_{LR}^{ij}$, as
can be read off from the slepton mass matrix listed in Appendix
\ref{appendix:massesmixingfeynmanrules}.  They are proportional to
$A_{0}$ and suppressed by a Yukawa coupling.  So in general, the
generated $\delta_{LR}^{ij}$ at the weak scale are negligibly small
compared to the generated $\delta_{LL}^{ij}$ elements (see, however,
Section~\ref{sec:dim5}).

In view of the slepton mass matrix, the RGE are usually solved by
integrating out the right-handed neutrinos at one scale $M_{R}\simeq
\mathcal{O}(10^{13}-10^{14})$ GeV.  In most GUT models, however, the
heavy neutrinos are strongly hierarchical so that these degrees of
freedom should be integrated out successively.  As a result, we have a
number of effective field theories below the GUT scale; the details are
listed in Appendix \ref{appendix:RGE}.  The running of the mixing
angle can change significantly with three non-degenerate heavy neutrinos
and cannot be reproduced if all heavy neutrinos are integrated out at a
common scale $M_\text{int}$ \cite{Antusch:2005gp,Antusch:2002rr}.

Our input values are the gauge couplings, the masses of the leptons and
quarks at the electroweak scale, the neutrino mass differences $\Delta
m_\text{atm}^{2}$ and $\Delta m_\text{sol}^{2}$ as well as the PMNS
matrix.  We will assume the normal hierarchy for the masses of the light
neutrinos,
\begin{align}
  m_{\nu_{1}},\qquad m_{\nu_{2}} = \sqrt{m_{\nu_{1}}^{2} + \Delta
    m_\text{sol}^{2}},\qquad m_{\nu_{3}} = \sqrt{m_{\nu_{1}}^{2} + \Delta
    m_\text{atm}^{2}} \,,
\end{align}
so we are left with the mass of the lightest neutrino $m_{\nu_{1}}$, the
three masses of the Majorana neutrinos $M_{R_{i}}$, the factor
$\tan\beta$ and the uncertainties in the mixing angles of the PMNS
matrix as free parameters.  For numerical results, we will choose
$m_{\nu_{1}}\approx\mathcal{O}(10^{-3})$ eV.  Note that the heavy
neutrino masses $M_R$ are already fixed by Eqs.~(\ref{equ:MRCKMFall})
and (\ref{equ:MRmnsFall}),
\begin{align}
  \left( M_{R_{1}}, \, M_{R_{2}}, \, M_{R_{3}} \right) & =
  \begin{cases}
    \left( 4.0\cdot 10^9~\text{GeV}, ~4.0\cdot 10^9~\text{GeV},
      \mspace{13mu} 5.9\cdot 10^{14}~\text{GeV} \right) , & \text{PMNS
      case} ; \cr \left( 2.0\cdot 10^{6}~\text{GeV}, ~3.9\cdot
      10^{11}~\text{GeV}, ~7.4 \cdot 10^{15}~\text{GeV} \right) , &
    \text{CKM case} .
  \end{cases}
\end{align}
In addition, the soft SUSY breaking terms $A_{0}$, $m_{0}$, and
$m_{1/2}$ as well as $\tan\beta$ are free parameters.

\subsection{Numerical Results}
\label{subsec:numericalresults}

We start by considering the $\Delta m_{\tilde{L}}^2$ entries in
Eq.~(\ref{equ:YnuYu}) for the two scenarios.  In the PMNS case, we get
from Eq.~(\ref{equ:mmLapprox})
\begin{align}
  \left(\Delta m_{\tilde{L}}^2\right)_{12} & \approx -\frac{3m_{0}^{2} +
    A_{0}^{2}}{8\pi^{2}} \left( y_{t}^{2} U_{e3} U_{\mu 3}
    \ln\frac{M_\text{GUT}}{M_{R_{3}}} + y_{c}^{2} U_{e2} U_{\mu 2}
    \ln\frac{M_\text{GUT}}{M_{R_{2}}} \right) + \mathcal{O}(y_{u}^{2}),
  \nonumber
  \\
  \left(\Delta m_{\tilde{L}}^2\right)_{13} & \approx -\frac{3m_{0}^{2} +
    A_{0}^{2}}{8\pi^{2}} \left( y_{t}^{2} U_{e3} U_{\tau 3}
    \ln\frac{M_\text{GUT}}{M_{R_{3}}} + y_{c}^{2} U_{e2} U_{\tau 2}
    \ln\frac{M_\text{GUT}}{M_{R_{2}}} \right) + \mathcal{O}(y_{u}^{2}),
  \nonumber
  \\
  \left(\Delta m_{\tilde{L}}^2\right)_{23} & \approx -\frac{3m_{0}^{2} +
    A_{0}^{2}}{8\pi^{2}} \, y_{t}^{2} U_{\mu3} U_{\tau 3}
  \ln\frac{M_\text{GUT}}{M_{R_{3}}} + \mathcal{O}(y_{c}^{2}).
\end{align}
In these equations, we replaced $U^{(0)}$ by $U$ since we already know
that SUSY corrections do not spoil a possible symmetry at the high scale
(Sec.~\ref{se:pmns}). Hence, we can neglect the small difference
between $U^\text{phys}$ and $U^{(0)}$.  In the following, we will
distinguish between two different input values at $M_\text{GUT}$,
$\theta_{13} = 0^{\circ}$ and $\theta_{13} = 3^{\circ}$.

In the 12 and 13 elements, the large top Yukawa coupling compensates the
suppression by $U_{e3}$, i.e., $\theta_{13}$.  For $\theta_{13} =
3^{\circ}$, which is the sensitivity of the DOUBLE CHOOZ experiment, the
top contribution dominates,
\begin{align}
  \frac{y_{t}^{2} U_{e3} U_{\mu 3}
    \ln\frac{M_\text{GUT}}{M_{R_{3}}}}{y_{c}^{2} U_{e2} U_{\mu 2}
    \ln\frac{M_\text{GUT}}{M_{R_{2}}}} \approx 13000\cdot U_{e3} & \approx
  650 \,, & \frac{y_{t}^{2} U_{e3} U_{\tau 3}
    \ln\frac{M_\text{GUT}}{M_{R_{3}}}}{y_{c}^{2} U_{e2} U_{\tau 2}
    \ln\frac{M_\text{GUT}}{M_{R_{2}}}} \approx 12000 \cdot U_{e3} & \approx
  600 \,;
\end{align}
however, for much smaller angles, the contribution of the second
generation needs to be taken into account 
and leading-order RGE are not a good approximation anymore.
 Note that the dominant
contribution to $\delta_{LL}^{23}$ is independent of the unknown mixing
angle $\theta_{13}$.

We can perform the analogous analysis for the CKM case.  Then the
generated $\delta_{LL}^{ij}$ are one or two orders of magnitudes smaller
for this case, simply because the small CKM mixing angles replace the
large PMNS ones in $Y_\nu$ (see Eq.~(\ref{equ:YnuCKM})).

In the following, we will analyse how large the LFV off-diagonal
elements $\delta_{LL}^{ij}$ can get at the electroweak scale due to
renormalisation group running and study their sensitivity on
$\theta_{13}$.

\paragraph{PMNS case.}
The main contribution stems from the running between $M_\text{GUT}$ and
$M_{R_{3}}$, where the dominant entry of $Y_{\nu}$ is of the same order
as the top Yukawa coupling.  Below $M_{R_{3}}$ the entries of the
remaining neutrino Yukawa coupling is much smaller such that the result
is only weakly dependent on the two lighter Majorana masses.

Table \ref{tab:RGEdeltaLLbeide} lists the results for two different
neutrino mixing angles; we generally obtain
$\left|\delta_{LL}^{12}\right| \lesssim \left|\delta_{LL}^{13}\right|
\leq \left|\delta_{LL}^{23}\right|$.  As expected from
  the leading-order RGE the sizes of the (1,2) and (1,3) elements
  increase by two or three orders of magnitude for a sizable
  $U_{e3}\left(M_{Z}\right)=0.05$ element, compared to the case with $U_{e3}=0$. For
  $\theta_{13} = 0$ the estimate from the leading-order solution does
  no longer coincide with the exact numerical solution. While for
  $\theta_{13} = 3^\circ$ the relation $\left|\left(\Delta
      m_{\tilde{L}}^2\right)_{12}/\left(\Delta
      m_{\tilde{L}}^2\right)_{13}\right|\approx 1$ holds, the latter estimate 
  is no longer valid for $\theta_{13} = 0$.

\begin{table}
  \centering
  \begin{tabular}{lc|lllllll}
    \hline
    \rule[-3pt]{0pt}{13pt}
    & SPS & \multicolumn{1}{c}{1a} & \multicolumn{1}{c}{1b} &
    \multicolumn{1}{c}{2} & \multicolumn{1}{c}{3} &
    \multicolumn{1}{c}{4} & \multicolumn{1}{c}{A} &
    \multicolumn{1}{c}{B} \\
    \hline
    \hline
    \rule[-6pt]{0pt}{16pt}
    \boldmath{$\theta_{13}=0^{\circ}$}
    & $\left|\delta_{LL}^{12}\right|$ & 0.0000055 & 0.000044 &
    0.000006 & 0.0000007 & 0.000280 & 0.000237 & 
    0.0000039 
    \\
    \cline{2-9}
    \rule[-6pt]{0pt}{16pt}
	\boldmath{$U_{e3}=0$}
    & $\left|\delta_{LL}^{13}\right|$ & 0.000011 & 0.000049 & 
    0.000018 & 0.0000024 & 0.000314 & 0.000262 &
    0.000012 
    \\
    \cline{2-9}
    \rule[-6pt]{0pt}{16pt}
    & $\left|\delta_{LL}^{23}\right|$ & 0.0462 & 0.0244 & 0.0699 &
    0.0089 & 0.0647 & 0.0538 & 0.0458  \\
    \hline
    \hline
    \rule[-6pt]{0pt}{16pt}
    \boldmath{$\theta_{13}=3^{\circ}$} &
    $\left|\delta_{LL}^{12}\right|$ & 0.00332 & 0.00169 & 0.00497 &
    0.00065 & 0.00400 & 0.00338 & 0.00328
    \\
    \cline{2-9}
    \rule[-6pt]{0pt}{16pt}
	\boldmath{$U_{e3}=0.05$}
    & $\left|\delta_{LL}^{13}\right|$ & 0.00333 & 0.00173 &
0.00498 &
    0.00065 & 0.00433 & 0.00362 & 0.00329
    \\
    \cline{2-9}
    \rule[-6pt]{0pt}{16pt}
    & $\left|\delta_{LL}^{23}\right|$ & 0.0460 & 0.0243 & 0.0697 &
    0.0089 & 0.0646 & 0.0537 & 0.0455
    \\ 
    \hline
  \end{tabular}
  \caption{Results for the generated off-diagonal elements
    $\delta_{LL}^{ij}$ at $M_Z$ in the PMNS case for the different
    mSUGRA scenarios. We assume $\theta_{12}=33^{\circ}$ and
    $\theta_{23}=45^{\circ}$ according to the tri-bimaximal scenario
    and consider the two cases $\theta_{13}=0^{\circ}$ (top) and
    $\theta_{13}=3^{\circ}$ at $M_{Z}$ (bottom).}
  \label{tab:RGEdeltaLLbeide} 
\end{table}

Now we use $\delta^{12}_{LL}$ in order to derive an upper bound on
$\theta_{13}$ for the different mSUGRA scenarios and obtain
\begin{align}
  \left|\theta_{13}\right| & \leq \left(0.25^{\circ},\, 0.42^{\circ},\,
    1.1^{\circ},\, 2.2^{\circ},\, 0.30^{\circ},\, 0.5^{\circ},\,
    1.2^{\circ}\right)
\end{align}
for the respective scenarios.  The element $\delta_{LL}^{13}$ is far
less sensitive to $\theta_{13}$: even for of $\theta_{13}=3^{\circ}$, it
is at least one order of magnitude below the current experimental
bounds.
As discussed above, $\delta_{LL}^{23}$ is not sensitive to
$\theta_{13}$.  In the SPS1a and SPS4 scenarios, however, it is above
the experimental bound, whereas it is well below the limit in SPS2, SPS3
and B.  Hence, some region of the parameter space can be excluded by the
element $\delta_{LL}^{23}$.

\paragraph{CKM case.}
The neutrino Yukawa matrix contains an $\mathcal{O}(1)$-entry only above
the scale $M_{R_{3}}$.  Thus non-vanishing $\delta_{LL}^{ij}$ are
basically generated in the interval
$\left[M_{R_{3}},M_\text{GUT}\right]$.

\begin{table}
  \centering
  \begin{tabular}{c|ddddddd}
    \hline
    \rule[-3pt]{0pt}{13pt}
    SPS & \multicolumn{1}{c}{1a} & \multicolumn{1}{c}{1b} &
    \multicolumn{1}{c}{2} & \multicolumn{1}{c}{3} &
    \multicolumn{1}{c}{4} & \multicolumn{1}{c}{A} &
    \multicolumn{1}{c}{B} \\
    \hline
    \hline
    \rule[-6pt]{0pt}{16pt}
    $\left|\delta_{LL}^{12}\right|$ & 0.0000101 & 0.0000025 & 0.0000071
    & 0.0000006 & 0.0000091 & 0.0000076 & 0.0000046
    \\ 
    \hline
    \rule[-6pt]{0pt}{16pt}
    $\left|\delta_{LL}^{13}\right|$ & 0.000234 & 0.000051 & 0.000133 &
    0.000133 & 0.000200 & 0.000163 & 0.000087
    \\ 
    \hline
    \rule[-6pt]{0pt}{16pt}
    $\left|\delta_{LL}^{23}\right|$ & 0.00119 & 0.00026 & 0.00067 &
    0.00007 & 0.00099 & 0.00083 & 0.00044
    \\ 
    \hline
  \end{tabular}
  \caption{RGE induced off-diagonal elements at $M_Z$ $\delta_{LL}^{ij}$
    in the CKM case for the different mSUGRA scenarios.}
  \label{tab:deltaLLCKMFall} 
\end{table}

The results are shown in Table~\ref{tab:deltaLLCKMFall}.  As expected,
the values for the various $\delta_{LL}^{ij}$ are small, due to the
small CKM mixing angles.  They are well below the experimental bound so
that we cannot exclude parts of the parameter space in this case at all.

\smallskip

\begin{figure}
  \psfrag{deltaLL12}{\scalefont{0.9}\hspace{-0.0cm}$|\delta_{LL}^{12}|$} 
  \psfrag{deltaLL13}{\scalefont{0.9}\hspace{-0.0cm}$|\delta_{LL}^{13}|$} 
  \psfrag{deltaLL23}{\scalefont{0.9}\hspace{-0.0cm}$|\delta_{LL}^{23}|$} 
  \psfrag{Szenario}{\scalefont{0.9}\hspace{-0.1cm} Scenario} 
  \psfrag{1}{\scalefont{0.7}\hspace{-0.2cm} 1a}
  \psfrag{2}{\scalefont{0.7}\hspace{-0.2cm} 1b}
  \psfrag{3}{\scalefont{0.7}\hspace{-0.0cm} 2}
  \psfrag{4}{\scalefont{0.7}\hspace{-0.0cm} 3}
  \psfrag{5}{\scalefont{0.7}\hspace{-0.0cm} 4}
  \psfrag{6}{\scalefont{0.7}\hspace{-0.0cm} A}
  \psfrag{7}{\scalefont{0.7}\hspace{-0.0cm} B}
  \psfrag{0}{\scalefont{0.7}\hspace{-0.0cm} }
  \includegraphics[width=.45\linewidth]{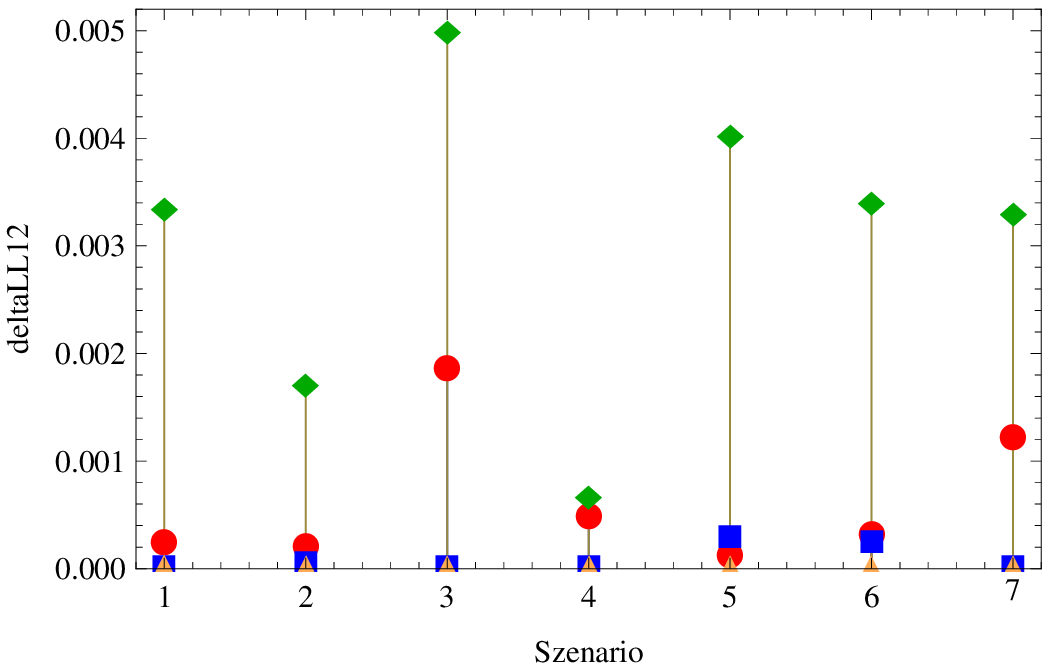}
  \includegraphics[width=.45\linewidth]{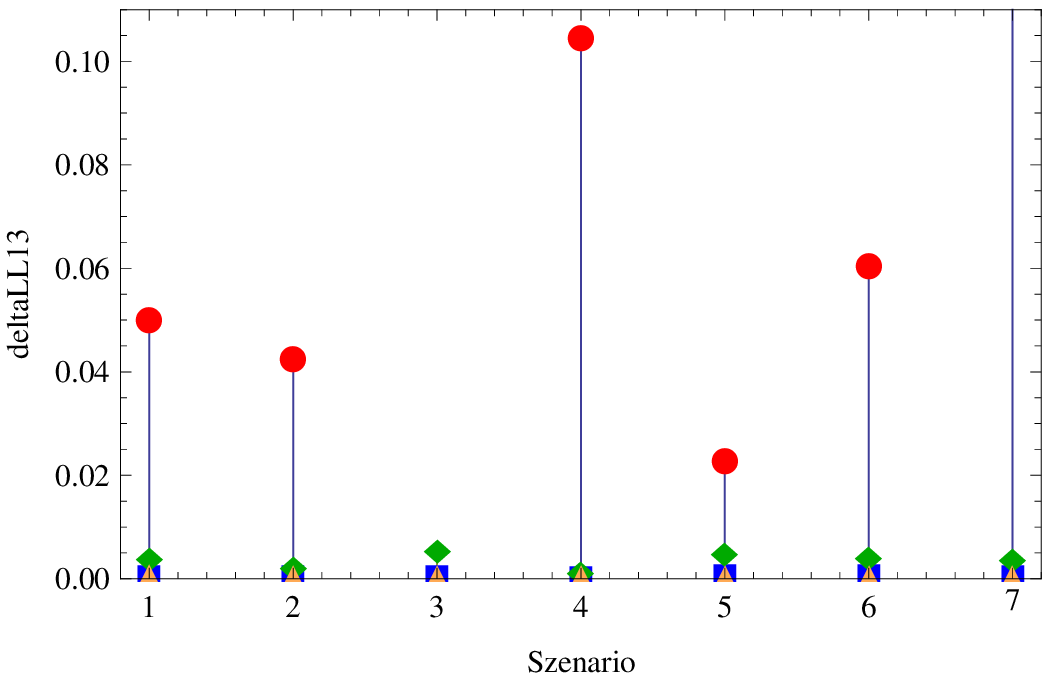}
  \includegraphics[width=.45\linewidth]{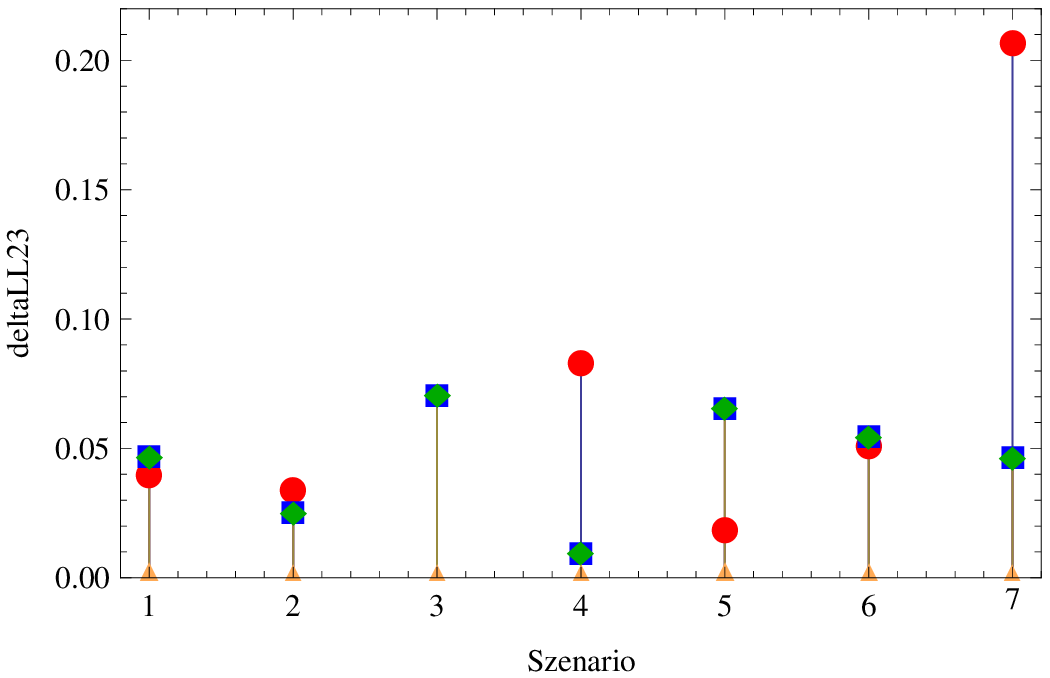}
  \caption{Summary of the results, as shown in
    Tables~\ref{tab:deltaLLSchranken}, \ref{tab:RGEdeltaLLbeide} and
    \ref{tab:deltaLLCKMFall}.  The red circle denotes the experimental
    bound, the blue square the PMNS case with $\theta_{13} = 0^{\circ}$,
    the green diamond the PMNS case with $\theta_{13} = 3^{\circ}$, and
    the orange triangle the CKM case.}
  \label{fig:Sektor121323} 
\end{figure}

In summary, we have seen that LFV processes offer a window to look at
the structure of SUSY GUT scenarios.  The results of the various cases
considered in this paper (see Tab.~\ref{tab:deltaLLSchranken},
\ref{tab:RGEdeltaLLbeide} and \ref{tab:deltaLLCKMFall}) are compared in
Figure~\ref{fig:Sektor121323}.
The decay $\tau\rightarrow \mu \gamma$ can exclude the PMNS case through
$\delta_{LL}^{23}$ for some SUSY mass spectra, irrespective of $U_{e3}$.
In addition, $\theta_{13}$ is bounded by $\delta_{LL}^{12}$; the PMNS case
allows for small values of $\theta_{13}$ only.  A more precise
measurement of $U_{e3}$ will make it possible to disfavour or even to
exclude models.  In the CKM case, $\mu \to e \gamma$ and $\tau \to \mu
\gamma$ should be observable in the near future \cite{Mori:2007zza}.
Here, the $\tan\beta$-enhanced corrections should also be included.

Most of these conclusions hold irrespective of the GUT correction terms
which will be discussed in the following subsection.  These do not
affect the LL sector; however the LR sector will be modified
significantly.
Furthermore, note that any observation of $\tau \to e \gamma$ calls for
additional sources of LFV; as discussed above, the needed
$\delta_{LL}^{13}$ cannot be generated if we start with universal
boundary conditions at $M_\text{GUT}$.

\subsection{Effects from Fermion Mass Corrections}
\label{sec:dim5}

In grand-unified models, the Yukawa couplings arise from few basic
couplings, relating the couplings of the SM fields.  In particular,
minimal GUT models predict the unification of the down quark and charged
lepton masses.  While those of the bottom quark and the tau are in
remarkable agreement at $M_\text{GUT}$, the relation is violated for the
first and second generation.
The failure for the lighter generations, however, is naturally explained
by the presence of higher dimensional operators due to physics at the
Planck scale that induces corrections of the order
$M_\text{GUT}/M_\text{Pl}$ \cite{Ellis:1979fg}.

These nonrenormalisable operators do not only help to
  get realistic fermion mass relations, they also cure another problem
  of SUSY SU(5) models: the too large proton decay rate stemming from
  couplings of the color triplet Higgs field 
  \cite{Bajc:2002bv,EmmanuelCosta:2003pu,Berezhiani:1998hg,Bajc:2002pg,Borzumati:2007bn}.
  The consequences of the higher-dimensional operators on flavour
  physics observables, however, were for a long time neglected. First
  studies were carried out in
  Refs.~\cite{ArkaniHamed:1995fs,Hisano:1998cx} with vanishing neutrino
  masses. In Ref.~\cite{Hisano:1998cx} the RGE of the RR slepton sector
  involves the CKM matrix together with an additional rotation matrix.
  Ref.~\cite{ArkaniHamed:1995fs} discusses an SO(10) SUSY GUT model with
  nonminimal Yukawa interaction. Massive neutrinos via the seesaw
  mechanism were first analysed within SUSY SU(5) in
  Ref.~\cite{Baek:2001kh}. The authors include one higher-dimensional
  operator whose main effect is described by two mixing angles that
  parametrise rotations between the down-type quarks and charged leptons
  in the first and second generations. In Ref.~\cite{Ko:2008zu} the
  correlation between the flavour-violating mass insertions of the
  squark and slepton sectors in a SUSY SU(5) scenario are studied,
  including the corrections to fix the quark-lepton mass relations.  The
  authors of Ref.~\cite{Trine:2009ns} study an SO(10) SUSY GUT model in
  which the large atmospheric mixing angle can induce $b_R-s_R$
  transitions. They parametrise the effect of higher-dimensional
  operators on flavour transitions in terms of a mixing angle and a
  (CP-violating) phase and place tight constraints on these parameters
  from a simultaneous study of \kkm, \bbmd\ and \bbms.  A very detailed
  theoretical analysis generalising the approach of
  Ref.~\cite{Baek:2001kh} has recently been performed in
  Ref.~\cite{Borzumati:2009hu}, for a compact summary see
  Ref.~\cite{Borzumati:2009wh}. These papers contain a complete list of
  RGEs for SUSY SU(5) including nonrenormalisable operators for all
  three types of the seesaw mechanism. This setup drastically increases the
  number of free parameters which show up in several diagonalisation
  matrices.  Even with flavour-blind and field-type-independent
  mediation of SUSY breaking, the higher-dimensional operators give rise
  to tree-level flavour-violating entries in the sfermion mass matrices.
  Their effective trilinear couplings are no longer aligned with the
  effective Yukawa couplings. Since the $A$-terms contribute to the
  slepton mass matrices already at tree level, these misalignments are
  potentially very dangerous.  The authors of
  Ref.~\cite{Borzumati:2009hu} study special types of K\"ahler
  potentials and superpotentials in which such terms can be
  avoided. With some approximations they recover the parametrisation
  with mixing angles between the first and second generation adopted in 
  Ref.~\cite{Baek:2001kh}. We will adopt a similar approach explained
  below. A comprehensive phenomenological analysis with the RGE of
  Ref.~\cite{Borzumati:2009hu} and its very general diagonalisation
  matrices has not been done yet. We will make a simplified ansatz:
  Instead of using the most general setup, we concentrate on the lepton
  sector and parametrise the effect of higher dimensional operators as a
  rotation between the first and second generation without the inclusion
  of any phases (which are not probed by current experiments). Further
  we only use the RGE of the MSSM and focus on the effect in the
  trilinear terms.

If we denote the renormalisable and the higher-dimensional couplings as
$Y_\text{GUT}$ and $Y_\sigma$, respectively, we can express the Yukawa
couplings of down quarks and charged leptons at $M_\text{GUT}$
as\footnote{Here, we neglect the higher-dimensional operators which
  contribute equally to $Y_d$ and $Y_l$.  In this discussion, we can
  absorb them in $Y_\text{GUT}$; however, they become important for $B$
  and $L$ violating processes \cite{Bajc:2002bv,EmmanuelCosta:2003pu}.}
\begin{align}
  Y_d & = Y_\text{GUT} + k_d\, \frac{\sigma}{M_\text{Pl}} Y_\sigma \;, &
  Y_l^\top & = Y_\text{GUT} + k_e\, \frac{\sigma}{M_\text{Pl}} Y_\sigma
  \;,
\end{align}
where $\sigma=\mathcal{O}\left(M_\text{GUT}\right)$.  The coefficients
$k_d$ and $k_e$ are determined by the direction of the GUT breaking
vevs.  The relative transposition between $Y_d$ and $Y_l$ is due to
their embedding in SU(5) multiplets.
Even though we can calculate the masses of the fermions at
$M_\text{GUT}$ with fairly good precision, we cannot fix the various
couplings.  The reason is simply that the observed mixing matrices
diagonalise the products or combinations of Yukawa matrices.  In the
simplest case, where all matrices but $Y_d$ and $Y_l$ are diagonal, the
quark mixing matrix diagonalises
\begin{align}
  Y_d Y_d^\dagger = Y_\text{GUT} Y_\text{GUT}^\dagger + k_d
  \frac{\sigma}{M_\text{Pl}} \left( Y_\text{GUT} Y_\sigma^\dagger +
    Y_\sigma Y_\text{GUT}^\dagger \right) + \left( k_d
    \frac{\sigma}{M_\text{Pl}} \right)^2 Y_\sigma Y_\sigma^\dagger \,,
\end{align}
while the leptonic mixing matrix diagonalises
\begin{align}
  Y_l Y_l^\dagger = Y_\text{GUT}^\top Y_\text{GUT}^\ast + k_e
  \frac{\sigma}{M_\text{Pl}} \left( Y_\text{GUT}^\top Y_\sigma^\ast +
    Y_\sigma^\top Y_\text{GUT}^\ast \right) + \left( k_e
    \frac{\sigma}{M_\text{Pl}} \right)^2 Y_\sigma^\top Y_\sigma^\ast \,.
\end{align}
(Again, these relations hold at $M_\text{GUT}$.)  In addition, as
indicated by the factors $k$, the matrices are model-dependent.

Since we do not want to restrict ourselves to a special version of a
particular model, we proceed as follows.
In the basis of diagonal charged lepton Yukawa coupling one gets
\begin{align}
  \label{eq:dim5-lepton}
  \mathcal{D}_{l} = U_{1}^{\dagger} \mathcal{D}_{d} U_{2} +
  \frac{\sigma}{M_\text{Pl}} Y_\sigma^\prime
\end{align}
with unitary matrices $U_1$ and $U_2$.  The starting point for universal
$A$ terms (cf.~Eq.~(\ref{equ:AGUT})) is the renormalisable Yukawa
coupling,%
\begin{align}
  \label{eq:a_dim5-lepton}
  A_l & = A_d = A_{0}\, Y_\text{GUT} = A_{0} \left( Y_{l} - k_e
    \frac{\sigma}{M_\text{Pl}} Y_\sigma \right)
\end{align}
Now we know that the entries of $Y_\text{GUT}$ are generally of the
right order of magnitude.  Since the contributions from $Y_\sigma$ are
suppressed by a factor $M_\text{GUT}/M_\text{Pl}$ and bottom-tau
unification works well, they do not change the third generation's
entries significantly.  Then we can approximate the effect of the
higher-dimensional operators with an additional rotation in the
12-sector, parametrised by one single mixing angle $\theta$,
\begin{align}
  A_l & \simeq A_{0}
  \begin{pmatrix}
    \cos\theta & -\sin\theta & 0 \cr \sin\theta & \cos\theta & 0 \cr 0 &
    0 & 1
  \end{pmatrix}
  Y_{l} \;.
  \label{eq:a_dim5-appr}
\end{align}

This parametrisation is similar to the one of 
Ref.~\cite{Baek:2001kh}. We restrict ourselves 
to the lepton sector and place our boundary conditions at the 
GUT scale rather than the Planck scale, since we aim at constraints on
the GUT parameters derived from low-energy data on LFV.
In Section~\ref{subsec:numericalresults}, we saw that the the LR
off-diagonal elements of the slepton mass matrix are negligibly small.
These elements are expected to become sizable now, due to inclusion of
the additional mixing, parametrised by $\theta$.  
In order to be as model-independent as possible, we will continue to
assume universality of the soft SUSY terms at $M_\text{GUT}$.  Then the
mixing does not affect the derived results for $\delta_{LL}^{ij}$
because the LL elements are not sensitive to $\theta$.  In a given GUT
model, it may be more natural to assume universality at $M_\text{Pl}$,
as is the case in Refs.~\cite{Hisano:1998cx,Ellis:2008st}.  Naturally,
their results are model-specific.

\subsubsection*{Numerical Results for the LR Sector}

\begin{figure}
  \centering
  \psfrag{BRmalzehnhochelf}{\scalefont{0.8}\hspace{0.0cm}{\slshape
      BR}$\left(\mu\rightarrow e\gamma\right)\times 10^{11}$} 
  \psfrag{deltaLR12}{\scalefont{0.8}\hspace{0.5cm}\raisebox{0.2cm}{
      $|\delta_{LR}^{12}|$}}
  \psfrag{exp}{\scalefont{0.7}\hspace{0.0cm}$\text{exp.}$}
  \psfrag{Grenze}{\scalefont{0.7}\hspace{0.0cm}$\text{ bound}$}
  \psfrag{SPS1a}{\scalefont{0.7}$\text{SPS1a}$}
  \psfrag{SPS1b}{\scalefont{0.7}$\text{SPS1b}$}
  \psfrag{SPS2}{\scalefont{0.7}$\text{SPS2}$}
  \psfrag{SPS3}{\scalefont{0.7}$\text{SPS3}$}
  \psfrag{SPS4}{\scalefont{0.7}$\text{SPS4}$}
\psfrag{A}{\scalefont{0.7}$\text{A}$}
  \psfrag{B}{\scalefont{0.7}$\text{B}$}

  \includegraphics[width=.6\linewidth]{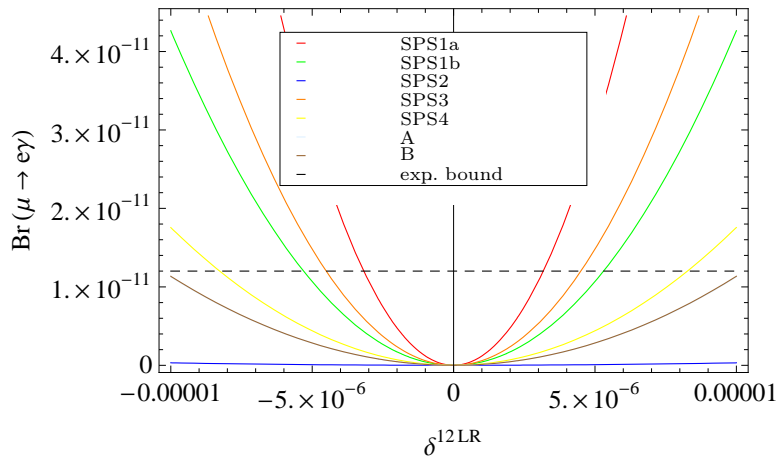}
  \caption{{\slshape BR}$\left(\mu\rightarrow e\gamma\right)\times
    10^{11}$ as a function of $\delta_{LR}^{12}$ for the different
    mSUGRA scenarios: From top to bottom: SPS1a: red, SPS3: orange,
    SPS1b: green, SPS4: yellow, A: b light blue, B: brown, SPS2: blue,
    experimental upper bound black dashed.}
  \label{fig:plotLR12}
\end{figure}

In order to derive upper bounds on $\left|\delta_{LR}^{ij}\right|$, we
assume all other off-diagonal elements are zero.  The results for the
different scenarios are listed in Table~\ref{tab:deltaLRSchranken}.  We
show the relation between the branching ratio {\slshape BR}$\left(\mu\to
  e\gamma\right)$ and $\delta_{LR}^{12}$ in Figure~\ref{fig:plotLR12}.
The main contribution comes from a bino exchange which is independent of
$\tan\beta$, contrary to the LL elements.

\begin{table}
  \centering
  \begin{tabular}{l|ddddddd}
    \hline
    \rule[-3pt]{0pt}{13pt}
    SPS & \multicolumn{1}{c}{1a} & \multicolumn{1}{c}{1b} &
    \multicolumn{1}{c}{2} & \multicolumn{1}{c}{3} &
    \multicolumn{1}{c}{4} & \multicolumn{1}{c}{A} &
    \multicolumn{1}{c}{B} \\
    \hline
    \hline
    \rule[-6pt]{0pt}{16pt}
    $\left|\delta_{LR}^{12}\right|\leq$ & 0.0000032 & 0.0000053
    & 0.000062 & 0.0000045 & 0.0000082 & 0.0000103 & 0.0000103 
    \\ 
    \hline
    \rule[-6pt]{0pt}{16pt}
    $ |\delta_{LR}^{13}|\leq$ & 0.012 & 0.019 & 0.232 & 0.017 & 
    0.028 & 0.036 & 0.039 
    \\ 
    \hline
    \rule[-6pt]{0pt}{16pt}
    $ |\delta_{LR}^{23}|\leq$ &  0.009 & 0.015 & 0.182 & 0.013 &
    0.022 & 0.028 & 0.030 \\ 
    \hline
  \end{tabular}
  \caption{Upper bounds on $\left|\delta_{LR}^{12}\right|$,
    $\left|\delta_{LR}^{13}\right|$ and $\left|\delta_{LR}^{23}\right|$
    for the mSUGRA scenarios from $l_{j} \to l_{i}\gamma$.}
  \label{tab:deltaLRSchranken} 
\end{table}

Assuming diagonal slepton mass matrix at the GUT scale, the generated
$\delta_{LR}^{ij}$ at the weak scale depend basically on $A_{0}$ and
$\theta$.  The bounds on $\delta_{LR}^{13}$ and $\delta_{LR}^{23}$ are
too loose and the generated off-diagonal elements stay far below them.
Only the 12 element can reach the experimental sensitivity.  As long as
one chooses $A_{0}=0$ , $\delta_{LR}^{12}$ is negligible small even for
a large mixing angle $\theta$.  

\begin{figure}[t]
  \centering
  \psfrag{theta}{\scalefont{1}\hspace{-0.0cm}$\theta$ in $^{\circ}$}
  \psfrag{deltaLR12}{\scalefont{1}\hspace{0.2cm}$|\delta_{LR}^{12}|$} 
  \includegraphics[width=.45\linewidth]{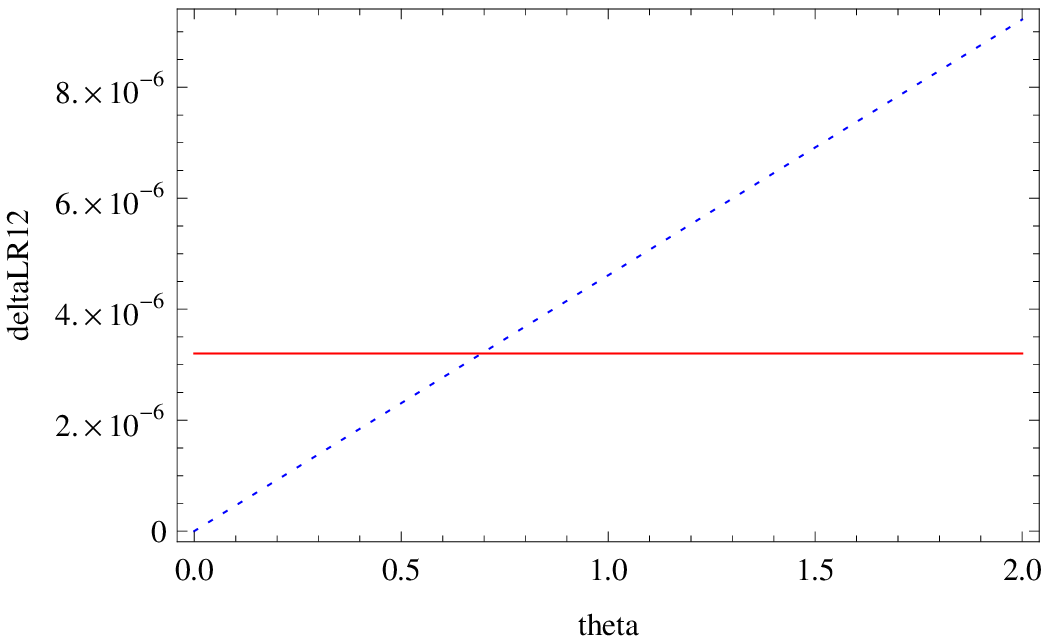}
  \hspace{.05\linewidth}
  \includegraphics[width=.45\linewidth]{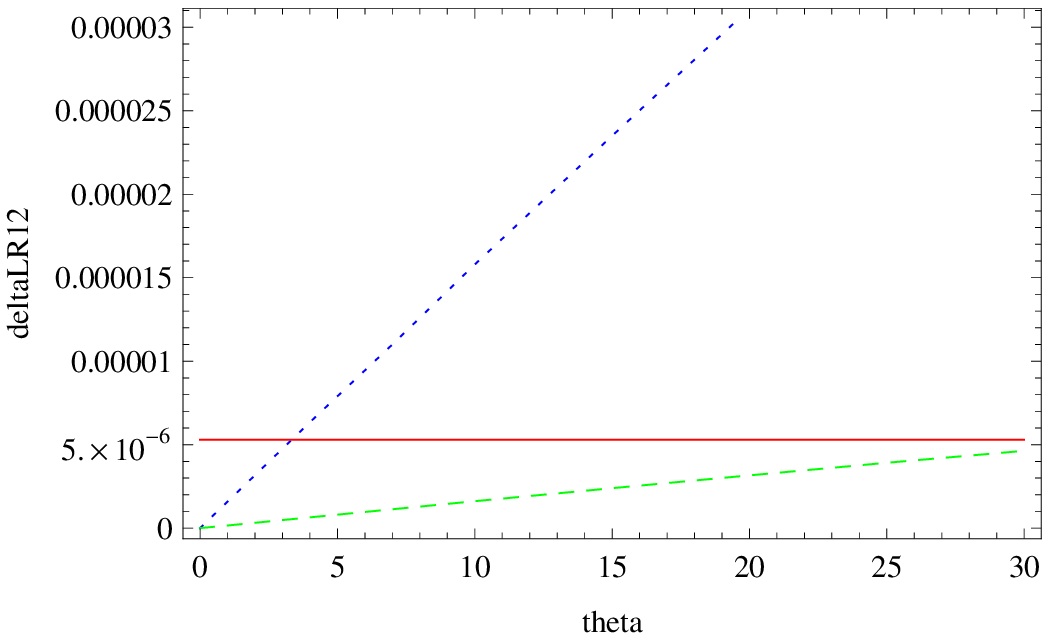}
  \caption{$\delta_{LR}^{12}$ as a function of $\theta$ and the
    experimental bounds (red).  Left hand side: SPS1a with $A_{0} =
    -100$~GeV (blue dotted).  Right hand side: SPS1b with $A_{0} = -100$~GeV
    (blue dotted) and $A_{0} = -10$~GeV (green dashed).}
  \label{fig:thetadelta}
\end{figure}

\begin{figure}
  \centering
  \psfrag{thetamax}{\scalefont{1}\hspace{0.0cm}$\theta_{max}$ }
  \psfrag{ao}{\scalefont{1}\hspace{0.0cm}\raisebox{0.0cm}{$A_{0}$}} 
  \includegraphics[width=8cm]{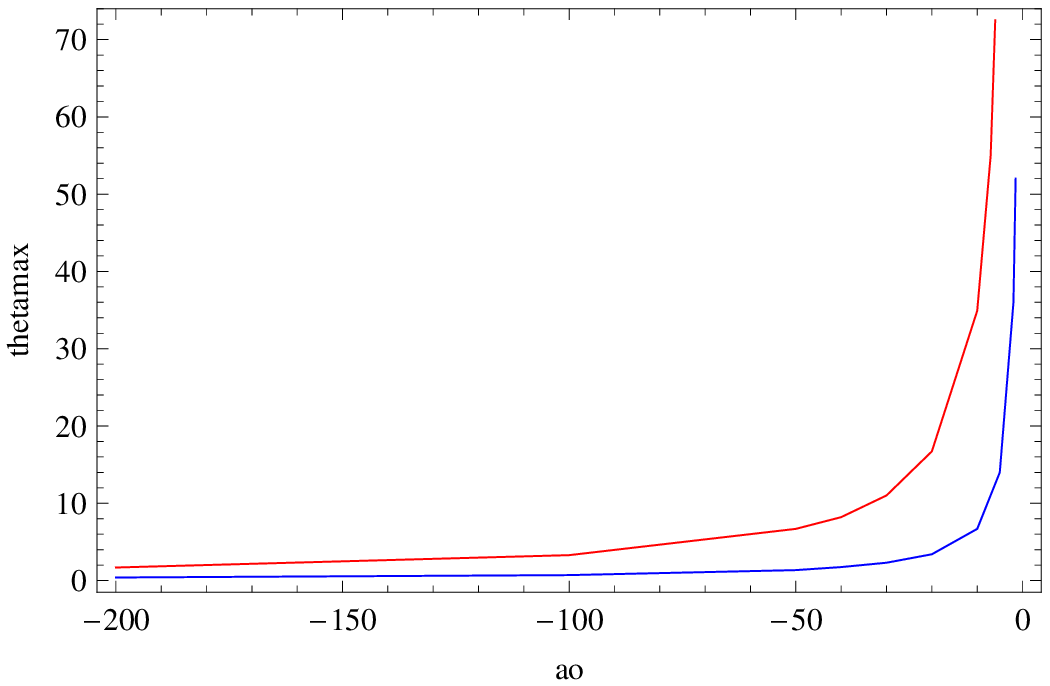}
  \caption{Upper bound for $\theta$ as a function of $A_{0}$ for SPS1a
    (down blue) and SPS1b (top red) with varying $A_{0}$ (in GeV).  The allowed
    $\left(A_{0}, \,\theta\right)$ region lies below the curve
    respectively.}
  \label{fig:A0theta}
\end{figure}

Let us now vary $A_{0}$.  This variation slightly modifies the mass
spectrum at the electroweak scale via the RGE but the upper bounds on
$\delta_{LR}^{12}$ do not change significantly.  For instance, in a
modified SPS1a scenario with $A_{0}$ varying from $-200$ to $0$, the
bound lies within $(3.22 - 3.34) \cdot 10^{-6}$.
The generated $\delta_{LR}^{12}$ element, however, can quickly exceed
the experimental bounds, even for small values of $\theta$
(Fig.~\ref{fig:thetadelta}).  Then we can derive a relation between
$A_{0}$ and the maximal allowed value for $\theta$;
Figure~\ref{fig:A0theta} shows the maximally allowed
  value for $\theta$ as a function of $A_0$ for the SPS1a and 1b
  scenarios.  The additional rotation reflects the different flavour
  structure of the down and charged lepton Yukawa couplings (see
  Eq.~(\ref{eq:a_dim5-lepton})).  Given the relation
  (\ref{eq:dim5-lepton}), we conclude that for sizable $A_0$, the
  higher-dimensional operators respect the flavour structure of the
  tree-level couplings.

In contrast to the LL sector, the results for the PMNS and CKM cases do
not differ much for the LR sector.  In both cases, the generated
$\delta_{LR}^{12}$ are negligible for vanishing mixing, $\theta =
0^{\circ}$.  We can easily understand this behaviour as we read off from
Eq.~(\ref{equ:Alapprox}) that $\delta_{LR}^{12}$ contains the
mixing from both $Y_{\nu}$ and $Y_{l}$.  Any mixing coming from
$\theta\neq 0$ contributes to $Y_{l}$ and dominates over the mixing in
$Y_{\nu}$.  Hence, there is no significant difference between CKM and
PMNS cases. Ref.~\cite{Borzumati:2009hu} has arrived at
 a similar conclusion, stating 
 that intrinsic, arbitrary flavour violations in the slepton mass
 parameters at the high scale, even if relatively
small, can completely obscure the loop effects induced by the seesaw mechanism.

\section{Conclusions}
\label{sec:summary}

Apart from neutrino oscillations, lepton flavour violating (LFV)
processes have not been observed up to now and the individual lepton
numbers have succeeded as good quantum numbers in charged lepton decays.
Weak-scale supersymmetry, however, generically introduces an additional
source of flavour violation.  Hence, these rare processes enable us to
study the supersymmetry breaking sector.  In this paper we have
performed a comprehensive study of the quantities $\delta_{XY}^{ij}$
parametrising the flavour structure of the leptonic soft
supersymmetry-breaking terms in the MSSM. Novel features of our analysis
are the consideration of mass and anomalous magnetic moment of the
electron and the (finite) renormalisation of the PMNS matrix by
supersymmetric loops with soft terms. Further, we include
$\tan\beta$-enhanced two-loop corrections to the LFV decays $l_j\to
l_i\gamma$ in a diagrammatic approach. Unlike previous analyses our
method a priori does not involve any expansion in $v^2/M_{\rm SUSY}^2$,
which becomes questionable in the case of large slepton mixing.  We have
subsequently expanded the exact result in $v^2/M_{\rm SUSY}^2$ and have
checked the accuracy of the expanded results.  Our analysis of the PMNS
matrix and the radiative decays follows the line of
Refs.~\cite{Crivellin:2008mq,Hofer:2009xb,Crivellin:2009ar}, which have
addressed similar problems in the quark sector.  We finally analyse the
effect of dimension-5 Yukawa couplings in the context of SO(10) GUT
scenarios.

  Studying the one-loop renormalisation of lepton masses and PMNS
  elements at large $\tan\beta$ we have found potentially large finite
loop contributions to
  the electron mass $m_e$ while corrections to the PMNS matrix stay
rather small. Applying a
  standard naturalness criterion to $m_e$ leads to the requirement that
  the loop contributions must not exceed the measured value. As a result
  we find $|\delta^{13}_{LL} \delta^{13}_{RR}|\lsim 0.1$ which involves
  the otherwise poorly constrained quantity $\delta^{13}_{RR}$. The same
  parameter combination is constrained by the anomalous magnetic moment
  of the electron, $a_e$, for which the MSSM contributions decouple if
  the corresponding mass scale $M_{\text{SUSY}}$ becomes heavy. $a_e$
  gives the same constraint on $|\delta^{13}_{LL} \delta^{13}_{RR}|$ as
  $m_e$ for $M_{\text{SUSY}}=500 \, \text{GeV}$.  Further we have
  pointed out that the flavour-changing counterterms renormalising the
  PMNS elements generically appear in the charged-Higgs couplings, even
  if the latter are expressed in terms of weak neutrino eigenstates. The
  corresponding loop-corrected vertices are summarised in
  Eqs.~(\ref{effHiggsenu})--(\ref{effHiggstaunu}).  Our  two-loop
  corrections to the LFV radiative decays change the decay rates by up
  to 10\% for large values of $\tan\beta$. Results on the inferred
  bounds on $|\delta^{ij}_{LL}|$ for selected MSSM parameter points can
  be found in Table~\ref{tab:deltaLLSchranken}. Assuming reasonable
SUSY masses $\leq 1$~TeV
  we find that $BR(\tau\to e \gamma)$ severely limits the size of the
  loop correction $\delta U_{e3}$ to the PMNS element $U_{e3}$: 
   For $M_{\rm SUSY}\leq 500\,$GeV we find
  $|\delta U_{e3}|<10^{-3}$ corresponding to a correction to the mixing
  angle $\theta_{13}$ of at most 0.06$^\circ$. Therefore SUSY loop
  corrections cannot fake a deviation from $U_{e3}=0$ implied by
  tri-bimaximal neutrino mixing, if this PMNS element is probed with the
  precision of the DOUBLE CHOOZ experiment. Stated differently, DOUBLE
  CHOOZ will probe the Yukawa sector and not the soft SUSY-breaking
  sector.

The bounds on $\delta^{ij}_{XY}$ are known to be severe, motivating the
assumption that the SUSY breaking terms respect the SM flavour
structure.  As the symmetries and the particle content of the standard
model point towards grand unification, one frequently assumes that these
terms are universal at the scale $M_\text{GUT}$, where the SM gauge
couplings converge.  Then the RGE running generates non-vanishing
$\delta^{ij}_{XY}$ at the weak scale.  In Sec.~4 of this paper, we
have considered the Yukawa structure of two simple GUT scenarios.  We
calculated the size of the generated $\delta^{ij}_{XY}$ for various SUSY
spectra, using the RGE for the MSSM extended with singlet neutrinos.
The comparison with our bounds obtained from $l_j\rightarrow
l_i\gamma$ allows to constrain or even to exclude particular scenarios.

In our RGE analysis we include the effect of higher-dimensional Yukawa
operators (of dimension 5 or higher) which are needed to reconcile
Yukawa unification with the experimental values of the fermion masses of
the first two generations. If SUSY-breaking occurs above the GUT scale,
flavour universality will naturally align the trilinear breaking terms
with the dimension-4 Yukawa couplings, leaving the higher-dimensional
terms as potential new sources of flavour violation. We have
parametrised this effect by a new mixing angle $\theta$ in
\eq{eq:a_dim5-appr}.  For typical values of the universal trilinear term
$A_0$ one finds very stringent bounds on $\theta$, as depicted in
Fig.~\ref{fig:A0theta}.  As a consequence, the flavour structure of
down-quark and charged-lepton Yukawa couplings must be similar for
sizable $A_0$.  This result hints at flavour symmetries which are
respected by the higher-dimensional Yukawa operators.  Note that it also
applies to renormalisable couplings with a higher-dimensional Higgs
representation, which couple differently to down quarks and charged
fermions. In addition, the higher-dimensional Yukawa operators are
generally consistent with all symmetries, hence appear naturally and
yield significant corrections to the light generations' masses.  The
same qualitative result, aligned flavour structures of dimension-4 and
higher-dimensional Yukawa couplings, has been found in a complementary
analysis of the quark sector \cite{Trine:2009ns}.

  With the upcoming LHC experiments we will explore whether weak-scale
  supersymmetry is realised in nature.  In addition, new flavour
  experiments like MEG will probe lepton number violation.  Our
    analysis stresses once more the importance of lepton flavour physics
    to map out the parameter space of the MSSM. Our GUT analysis
    exemplifies the well-known potential of lepton flavour physics to
    probe theories valid at very high energies.

\section*{Acknowledgements}
The authors thank Lars Hofer, Paride Paradisi and Dominik Scherer for
useful discussions. The presented work is supported by project C6 of the
DFG Research Unit SFB--TR 9 \emph{Computergest\"utzte Theoretische
  Teilchenphysik} and by the EU Contract No.~MRTN-CT-2006-035482, \lq\lq
FLAVIAnet''. J.G.\ and S.M.\ acknowledge the financial support by
\emph{Studienstiftung des deutschen Volkes}\ and the DFG Graduate
College \emph{High Energy Physics and Particle Astrophysics},
respectively.

\begin{appendix}

  \section{Loop integrals}\label{appendix:loopintegrals}
  
  We list the loop integrals, used in Section~\ref{se:lfv}:
  {\allowdisplaybreaks
    \begin{align}
      B_{0}(x,y) & = -\Delta - \frac{x}{x-y} \ln{\frac{x}{\mu^{2}}} -
      \frac{y}{y-x} \ln{\frac{y}{\mu^{2}}} \qquad \textrm{with}\quad
      \Delta = \frac{1}{\epsilon}-\gamma_{E} + \ln 4\pi
      \\[3pt]
      f_{1}(x,y,z) & = \frac{xy\ln{\frac{x}{y}} + xz\ln{\frac{z}{x}} +
        yz\ln{\frac{y}{z}}}{(x-y)(x-z)(y-z)}  
      \label{i-def}
      \\[3pt]
      f_{2}(x,y,z,w) & = \frac{f_1(x,y,z)-f_1(x,y,w)}{z-w}
      \\[3pt]
      & = \frac{w\ln{\frac{w}{z}}}{(w-x)(w-y)(w-z)} +
      \frac{y\ln{\frac{y}{x}}}{(y-w)(y-x)(y-z)} +
      \frac{z\ln{\frac{z}{x}}}{(z-w)(z-x)(z-y)}\nonumber
      \\[3pt]
      F_{0}(x,y,z,v,w) & = -\frac{f_2(x,y,z,v)-f_2(x,y,z,w)}{v-w}
      \\[3pt]
     C_0(x,y,z) & = - \frac{xy\ln{\frac{x}{y}} + xz\ln{\frac{z}{x}} +
        yz\ln{\frac{y}{z}}}{(x-y)(x-z)(y-z)}\\[3pt]
	C_2(x,y,z) & =-\Delta - \ln\frac{x}{\mu^2} - \frac{y^2\ln\frac{y}{x}}{(x-y)(z-y)} - \frac{z^2\ln\frac{z}{x}}{(x-z)(y-z)},
      \\[3pt]
      D_2(x,y,z,w) & = - \frac{y^2\ln\frac{y}{x}}{(y-x)(y-z)(y-w)}-\frac{z^2\ln\frac{z}{x}}{(z-x)(z-y)(z-w)} - \frac{w^2\ln\frac{w}{x}}{(w-x)(w-y)(w-z)},
      \\[3pt]
      F_1^N(x) & = \frac{2}{(1-x)^4}\left[1-6x+3x^2-6x^2\log x\right] ,
      \nonumber
      \\
      F_2^N(x) & = \frac{3}{(1-x)^3}\left[1-x^2+2x\log x\right] ,
      \nonumber
      \\
      F_1^C(x) & = \frac{2}{(1-x)^4}\left[2+3x-6x^2+x^3+6x\log x\right]
      , \nonumber
      \\
      F_2^C(x) & = \frac{3}{2(1-x)^3}\left[-3+4x-x^2-2\log x\right] ,
      \label{Schleifenfunktionen}
    \end{align}
  }%
  \section{Interaction of gauginos, sfermions and fermions}
  \label{appendix:massesmixingfeynmanrules}

  The convention and notation of \cite{Rosiek} is used with some little
  modification.  The factors $\sqrt{2}$ associated with the vacuum
  expectation value is omitted, such that
  \begin{equation}
    v = \sqrt{v_{u}^{2} + v_{d}^{2}}=174\textrm{ GeV}
  \end{equation}
  and the ratio between the vacuum expectation values are denoted by
  $\tan\beta = v_{u}/v_{d}$.  The Yukawa couplings are defined
  as follows
  \begin{align}
    m_{u} & = v_{u} Y_{u} \ , &  m_{d} & = -v_{d} Y_{d} \ , & 
    m_{l} & = -v_{d} Y_{l} \ .
  \end{align}

  \subsubsection*{Neutralinos \boldmath{$\tilde{\chi}^{0}_{i}$}}

  \begin{align}
    \Psi^{0} & = \left(\tilde{B},\tilde{W},\tilde{H}^{0}_{d},
      \tilde{H}^{0}_{u}\right) , & \mathcal{L}_{\tilde{\chi}^{0}_{mass}}
    & = -\frac{1}{2}(\Psi^{0})^\top M_{N}\Psi^{0} + \text{h.c.}
    \nonumber
    \\
    \label{Neutralinomassenmatrix}
    & & M_{N} & =
    \begin{pmatrix}
      M_{1} & 0 & -\frac{g_{1}v_{d}}{\sqrt{2}} &
      \frac{g_{1}v_{u}}{\sqrt{2}}
      \\
      0 & M_{2} & \frac{g_{2}v_{d}}{\sqrt{2}} &
      -\frac{g_{2}v_{u}}{\sqrt{2}}
      \\
      -\frac{g_{1}v_{d}}{\sqrt{2}} & \frac{g_{2}v_{d}}{\sqrt{2}}& 0&
      -\mu
      \\
      \frac{g_{1}v_{u}}{\sqrt{2}} & -\frac{g_{2}v_{u}}{\sqrt{2}}& -\mu&
      0
    \end{pmatrix}
  \end{align}
  $M_{N}$ can be diagonalised with an unitary transformation such that
  the eigenvalues are real and positive.
  \begin{equation}
    \label{Neutralinodiag}
    Z_{N}^\top M_{N}Z_{N}=M_{N}^{D} = 
    \begin{pmatrix}
      m_{\tilde{\chi}_{1}^{0}} & &0 \cr &\ddots & \cr 0 & &
      m_{\tilde{\chi}_{4}^{0}}
    \end{pmatrix}
  \end{equation}
  For that purpose, $Z_{N}^{\dagger} M_{N}^{\dagger} M_{N} Z_{N} =
  (M_{N}^{D})^{2}$ can be used.  $Z_{N}$ consists of the eigenvectors of
  the hermitian matrix $M_{N}^{\dagger} M_{N}$.  Then the columns can be
  multiplied with phases $e^{i\phi}$, such that
  $Z_{N}^\top M_{N}Z_{N}=M_{N}^{D}$ has positive and real diagonal
  elements.

  \subsubsection*{Charginos \boldmath{$\tilde{\chi}^{\pm}_{i}$}}

  \begin{align}
    \Psi^{\pm} & = \left( \tilde{W}^+,\, \tilde{H}_{u}^+,\,
      \tilde{W}^{-},\, \tilde{H}_{d}^{-} \right), &
    \mathcal{L}_{\tilde{\chi}^{\pm}_{mass}} & = -\frac{1}{2}
    \left(\Psi^{\pm}\right)^\top M_{C} \Psi^{\pm} + \text{h.c.}
    \nonumber
    \\
    & & M_{C} & =
    \begin{pmatrix}
      0 & X^\top \cr X & 0
    \end{pmatrix}
    , \qquad X = 
    \begin{pmatrix}
      M_{2} & g_{2} v_{u} \cr g_{2} v_{d} & \mu
    \end{pmatrix}
    \label{Charginomassenmatrix}
  \end{align}
  The rotation matrices for the positive and negative charged fermions
  differ, such that
  \begin{equation}
    Z_{-}^\top  X Z_{+} = 
    \begin{pmatrix}
      m_{\tilde{\chi}_{1}} & 0 \cr 0 & m_{\tilde{\chi}_{2}}
    \end{pmatrix}
    .
  \end{equation}
  
  \subsubsection*{Sleptons and sneutrinos} 
  
  There are two ways of arranging the sleptons in a vector, either by
  family or by chiralities.  The latter approach is adapted for the most
  general case and is used in Ref.~\cite{Rosiek}, whereas the former is
  convenient if LFC is assumed or for small off-diagonal elements
  treated as a perturbation.
  In this latter case, we have
  {\allowdisplaybreaks
    \begin{subequations}
      \label{Sleptonmassenmatrix}
      \begin{align}
        \mathcal{L}_{m} & = -\frac{1}{2} \left(
          \tilde{e}_{L}^{\dagger}\, \tilde{e}_{R}^{\dagger}\,
          \tilde{\mu}_{L}^{\dagger}\, \tilde{\mu}_{R}^{\dagger}\,
          \tilde{\tau}_{L}^{\dagger}\, \tilde{\tau}_{R}^{\dagger}
        \right) M_{\tilde{l}}^{2} \left( \tilde{e}_{L}\, \tilde{e}_{R}\,
          \tilde{\mu}_{L}\, \tilde{\mu}_{R}\, \tilde{\tau}_{L}\,
          \tilde{\tau}_{R} \right)^\top \nonumber
        \\
        M_{\tilde{l}}^{2} & =
        \begin{pmatrix}
          m_{e_{L}}^{2}& m_{e_{LR}}^{2}&\Delta m_{LL}^{e\mu}&\Delta
          m_{LR}^{e\mu}&\Delta m_{LL}^{e\tau}&\Delta m_{LR}^{e\tau}
          \\
          m_{e_{RL}}^{2}&m_{e_{R}}^{2}&\Delta m_{RL}^{e\mu}&\Delta
          m_{RR}^{e\mu}&\Delta m_{RL}^{e\tau}&\Delta m_{RR}^{e\tau}
          \\
          \Delta m_{LL}^{\mu e}&\Delta m_{LR}^{\mu
            e}&m_{\mu_{L}}^{2}&m_{\mu_{LR}}^{2}&\Delta
          m_{LL}^{\nu\tau}&\Delta m_{LR}^{\mu\tau}
          \\
          \Delta m_{RL}^{\mu e}&\Delta m_{RR}^{\mu
            e}&m_{\mu_{RL}}^{2}&m_{\mu_{R}}^{2}&\Delta
          m_{RL}^{\nu\tau}&\Delta m_{RR}^{\mu\tau}
          \\
          \Delta m_{LL}^{\tau e}&\Delta m_{LR}^{\tau e}& \Delta
          m_{LL}^{\tau\mu}&\Delta m_{LR}^{\tau\mu}&
          m_{\tau_{L}}^{2}&m_{\tau_{LR}}^{2}
          \\
          \Delta m_{RL}^{\tau e}&\Delta m_{RR}^{\tau e}&\Delta
          m_{RL}^{\tau\mu}&\Delta m_{RR}^{\tau\mu}&
          m_{\tau_{RL}}^{2}&m_{\tau_LR}^{2}
        \end{pmatrix}
      \end{align}
      with
      \begin{align}
        m_{e_{RL}}^{2} & = m_{e_{LR}}^{2*}, \quad \Delta
        m_{LR}^{e\mu}=\Delta m_{RL}^{\mu e *},\quad\dots \nonumber
        \\[2pt]
        \left(m_{L}^{2}\right)_{LL}^{ij} & = \frac{e^{2}
          \left(v_{d}^{2}-v_{u}^{2}\right) \left(1-2
            c_{W}^{2}\right)}{4s_{W}^{2} c_{W}^{2}} \delta_{ij} +
        v_{d}^{2}Y_{l_{i}}^{2}\delta_{ij} + (m_{\tilde{L}}^{2})_{ji}
        \label{equ:mLLLsoft}
        \\
        \left(m_{L}^{2}\right)_{RR}^{ij} & = -\frac{e^{2}
          \left(v_{d}^{2}-v_{u}^{2}\right)}{2 c_{W}^{2}}\delta_{ij} +
        v_{d}^{2}Y_{l_{i}}^{2}\delta_{ij} +
        m_{\tilde{\overline{e}}ij}^{2}
        \label{equ:mLRRsoft}
        \\
        \left(m_{L}^{2}\right)_{LR}^{ij} & = v_{u}\mu Y_{l}^{ij*} +
        v_{d}A_{l}^{ij*}
      \end{align}
    \end{subequations}
  }%
  The rotation matrix $Z$ is defined as
  \begin{align}
    Z^{\dagger}M_{\tilde{l}}^{2}Z = \diag\left( m_{1}^{2}, \ldots,
      m_{6}^{2} \right) .
  \end{align}
  This matrix can be spit up into three parts
  \begin{equation}
    Z = 
    \begin{pmatrix}
      Z_{e} \cr Z_{\mu} \cr Z _{\tau}
    \end{pmatrix}
    .
  \end{equation}
  In general, these $Z_{l}$ are $2\times6$ matrices, which reduce to
  $2\times2$ matrices, respectively, with zeros in the remaining entries
  in case of vanishing LFV.  Then for every generation one can write:
  \begin{equation}
    Z_{l}^{\dagger}
    \begin{pmatrix}
      \left(m_{l}^{2}\right)_{LL}& \left(m_{l}^{2}\right)_{LR} \cr
      \left(m_{l}^{2}\right)_{RL}^{*}& \left(m_{l}^{2}\right)_{RR}
    \end{pmatrix}
    Z_{l}=
    \begin{pmatrix}
      m_{l_{1}}^{2}& 0 \cr 0 & m_{l_{2}}^{2}
    \end{pmatrix}
    , \qquad l = \tilde{e},\tilde{\mu},\tilde{\tau}
  \end{equation}
  
  Alternatively, in the former case, one defines $\tilde{L}_{2}^{I} :=
  \tilde{l}_{L}^{I}$ and $\tilde{R}^{I} := \tilde{e}_{R}^{\dagger I}$,
  which mix to six charged mass eigenstates
  $\tilde{L}_{i}\,\,i=1\dots6$,
  \begin{align}
    \tilde{L}_{2}^{I} & = Z_{L}^{Ii*}\tilde{L}_{i}^{-} , \quad
    \tilde{R}^{I} = Z_{L}^{(I + 3)i}\tilde{L}_{i}^{ + } , \quad
    Z_{L}^{\dagger}
    \begin{pmatrix}
      (m_{L}^{2})_{LL}& (m_{L}^{2})_{LR} \cr
      (m_{L}^{2})_{RL}^{\dagger}&(m_{L}^{2})_{RR}
    \end{pmatrix}
    Z_{L}= \diag\left(m_{L_{1}}^{2}, \ldots, m_{L_{6}}^{2} \right) .
  \end{align}

  These two approaches can be translated into each other by the
  following substitutions (with $i = 1,2$ for LFC and $i=1,\dots 6$ for
  LFV):
  \begin{align}
    Z_{L}^{1i} & = Z_{e}^{1i} & Z_{L}^{2i} & = Z_{\mu}^{1i} & Z_{L}^{3i}
    & = Z_{\tau}^{1i} & Z_{L}^{4i} & = Z_{e}^{2i} & Z_{L}^{5i} & =
    Z_{\mu}^{2i} & Z_{L}^{6i} & = Z_{\tau}^{2i} \ .
  \end{align}

  The sneutrinos and the left-handed sleptons have a common SUSY
  breaking soft mass.  The weak eigenstates $\tilde{L}_{1} =
  \tilde{\nu}_{l}$ can be rotated in the sneutrino mass eigenstates
  $\tilde{\nu}_{j}$ via $Z_{\nu}$,
  \begin{align}
    \tilde{L}_{1}^{i} & = Z_{\nu}^{ij}\tilde{\nu}_{j} \ , &
    Z_{\nu}^{\dagger} \mathcal{M}_{\nu}^{2} Z_{\nu} & = \diag\left(
      m_{\tilde{\nu}_{1}}^{2}, m_{\tilde{\nu}_{2}}^{2} ,
      m_{\tilde{\nu}_{3}}^{2} \right) , & \mathcal{M}_{\nu}^{2} & =
    \frac{e^{2}\left(v_{d}^{2}-v_{u}^{2}\right)}{8
      s_{W}^{2}c_{W}^{2}}\mathds{1} + m_{\tilde{L}}^{2} \ .
    \label{equ:Mnusoft}
  \end{align}

  \subsubsection*{Lepton-slepton-neutralino}

  \begin{itemize}
  \item incoming lepton $l$, outgoing neutralino and slepton
    $\tilde{l}_{i}$:
    \begin{align}
      i\Gamma_{l}^{\tilde{\chi}_{j}^{0}\tilde{l}_{i}} =
      i\underbrace{\left(\frac{Z_{l}^{1i}}{\sqrt{2}}
          \left(g_{1}Z_{N}^{1j} + g_{2}Z_{N}^{2j}\right) +
          Y_{l}Z_{l}^{2i}Z_{N}^{3j}\right)}_{=
        \Gamma_{l_{L}}^{\tilde{\chi}_{j}^{0}\tilde{l}_{i}}}P_{L} +
      i\underbrace{\left(-g_{1}\sqrt{2}Z_{l}^{2i}Z_{N}^{1j*} +
          Y_{l}Z_{l}^{1i}Z_{N}^{3j*}\right)}_{=
        \Gamma_{l_{R}}^{\tilde{\chi}_{j}^{0}\tilde{l}_{i}}}P_{R}
      \nonumber
    \end{align}
  \item outgoing lepton $l$, incoming neutralino and slepton
    $\tilde{l}_{i}$:
    \begin{align}
      i\left(\Gamma_{l}^{\tilde{\chi}_{j}^{0} \tilde{l}_{i}}\right)^{*}
      = i\underbrace{\left(-g_{1}\sqrt{2}Z_{l}^{2i*}Z_{N}^{1j} +
          Y_{l}Z_{l}^{1i*}Z_{N}^{3j}\right)}_{= \left(
          \Gamma_{l_{R}}^{\tilde{\chi}_{j}^{0}\tilde{l}_{i}}
        \right)^{*}}P_{L}
      + i\underbrace{\left(\frac{Z_{l}^{1i*}}{\sqrt{2}}\left(
            g_{1}Z_{N}^{1j*} + g_{2}Z_{N}^{2j*}\right) +
          Y_{l}Z_{l}^{2i*}Z_{N}^{3j*}\right)}_{= \left(
          \Gamma_{l_{L}}^{\tilde{\chi}_{j}^{0}\tilde{l}_{i}}
        \right)^{*}}P_{R}
      \nonumber
    \end{align}
  \end{itemize}

  \subsubsection*{Lepton-sneutrino-chargino}
  
  \begin{itemize}
  \item incoming lepton, outgoing sneutrino and chargino:
    \begin{equation}
      i\Gamma_{l_{i}}^{\tilde{\nu}_{j}\tilde{\chi}^{\pm}_{k}} =
      -i\left(g_{2}Z_{ + }^{1k} P_{L}  +  Y_{l_{i}}Z_{-}^{2k*}
        P_{R}\right)Z_{\nu}^{ij*} \nonumber
    \end{equation}
  \item outgoing lepton, incoming sneutrino and chargino:
    \begin{equation}
      i\left(\Gamma_{l_{i}}^{\tilde{\nu}_{j}\tilde{\chi}^{\pm}_{k}}
      \right)^{*} = -i\left(Y_{l_{i}}Z_{-}^{2k}P_{L}  +   g_{2}Z_{ +
        }^{1k*} P_{R}\right)Z_{\nu}^{ij} \nonumber
    \end{equation}
  \end{itemize}
  
  \section{Renormalisation group equations}\label{appendix:RGE}

  In the following $\mu$ denotes the energy scale (and not the $\mu$
  parameter of the superpotential) and $t=\ln(\mu)$.  For $g_{1}$ the
  GUT normalisation is used ($g_{1}^{GUT}
  =\sqrt{\frac{5}{3}}g_{1}^{SM}$).

  At one loop order the gauge coupling in the MSSM evolve according to
  \begin{align}
    \frac{d}{dt}\alpha_{1}(t)& = \frac{1}{4
      \pi}\frac{66}{5}\alpha_{1}^{2}(t), \nonumber
    \\
    \frac{d}{dt}\alpha_{2}(t) &= \frac{1}{4 \pi}2\alpha_{2}^{2}(t),
    \nonumber
    \\
    \frac{d}{dt}\alpha_{3}(t) & = - \frac{1}{4 \pi}6\alpha_{2}^{2}(t).
    \label{equ:RGEalpha3}
  \end{align}
  For the running of the gaugino masses, we use that
  $g_{i}^{2}(t)/M_{i}(t)$ is independent of the scale $t$ at one loop
  order.  Defining $k= g_{3}^{2}(t_\text{GUT})/m_{1/2}$ and assuming
  universal gaugino masses $m_{1/2}$ at the GUT scale, you can use
  $M_{i}(t)=g_{i}^{2}(t)/k$ in the RGE.

  The running of the Yukawa couplings and the Majorana mass matrix
  between $M_\text{GUT}$ and $M_{R}$ at one loop level is given by the
  following differential equations:
  \begin{subequations}
    \begin{align}
      \label{equ:RGEYu}
      \frac{d}{dt}Y_{u} & = \frac{1}{16 \pi^{2}}Y_{u}\left[\left( \tr
          \left(3 Y_{u}Y_{u}^{\dagger} + Y_{\nu}Y_{\nu}^{\dagger}\right)
          - \frac{16}{3}g_{3}^{2} - 3 g_{2}^{2} -
          \frac{13}{15}g_{1}^{2}\right) \mathds{1} + 3
        Y_{u}^{\dagger}Y_{u} + Y_{d}^{\dagger}Y_{d} \right] ,
      \\
      \label{equ:RGEYd}
      \frac{d}{dt}Y_{d} & = \frac{1}{16 \pi^{2}} Y_{d} \left[ \left( \tr
          \left(3 Y_{d}Y_{d}^{\dagger} + Y_{l}Y_{l}^{\dagger} \right) -
          \frac{16}{3}g_{3}^{2} - 3 g_{2}^{2} - \frac{7}{15}g_{1}^{2}
        \right) \mathds{1} + 3 Y_{d}^{\dagger}Y_{d} + Y_{u}^{\dagger}
        Y_{u} \right] ,
      \\
      \label{equ:RGEYnu}
      \frac{d}{dt}Y_{\nu} & = \frac{1}{16 \pi^{2}} Y_{\nu} \left[ \left(
          \tr \left(3 Y_{u}Y_{u}^{\dagger} +
            Y_{\nu}Y_{\nu}^{\dagger}\right) - 3 g_{2}^{2} -
          \frac{3}{5}g_{1}^{2} \right) \mathds{1} + 3
        Y_{\nu}^{\dagger}Y_{\nu} + Y_{l}^{\dagger} Y_{l}\right] ,
      \\
      \label{equ:RGEYl}
      \frac{d}{dt}Y_{l} & = \frac{1}{16 \pi^{2}} Y_{l} \left[ \left( \tr
          \left(3 Y_{d}Y_{d}^{\dagger} + Y_{l}Y_{l}^{\dagger}\right) - 3
          g_{2}^{2} - \frac{9}{5}g_{1}^{2} \right) \mathds{1} + 3
        Y_{l}^{\dagger}Y_{l} + Y_{\nu}^{\dagger} Y_{\nu} \right] ,
      \\
      \label{equ:RGEMR}
      \frac{d}{dt}M_{R} & =\frac{1}{8 \pi} \left[ M_{R}
        \left(Y_{\nu}Y_{\nu}^{\dagger}\right)^\top  + Y_{\nu}
        Y_{\nu}^{\dagger} M_{R} \right] .
    \end{align}
  \end{subequations}

  The running of the $A$-terms is given by 
  {\allowdisplaybreaks
    \begin{subequations}
      \label{eq:a-terms}
      \begin{align}
        \frac{d}{dt}A_{u} & = \frac{1}{16 \pi^{2}}\left[A_{u}\left( \tr
            (3 Y_{u}Y_{u}^{\dagger} + Y_{\nu}Y_{\nu}^{\dagger}) -
            \frac{16}{3}g_{3}^{2} -
            3g_{2}^{2}-\frac{13}{15}g_{1}^{2}\right)\right.\nonumber
        \\
        & \mspace{75mu} + Y_{u}\left( \tr (6 Y_{u}^{\dagger} A_{u} + 2
          Y_{\nu}^{\dagger}A_{\nu}) + \frac{32}{3}g_{3}^{2}M_{3} + 6
          g_{2}^{2}M_{2} + \frac{26}{15}g_{1}^{2}M_{1} \right)\nonumber
        \\
        & \mspace{75mu} \left.  + 4 Y_{u}Y_{u}^{\dagger}A_{u} + 5
          A_{u}Y_{u}^{\dagger}Y_{u} + 2 Y_{u}Y_{d}^{\dagger} A_{d} +
          A_{u}Y_{d}^{\dagger}Y_{d}\rule{0cm}{0.5cm}\right],
        \\
        \frac{d}{dt}A_{d} & = \frac{1}{16 \pi^{2}}\left[A_{d}\left( \tr
            (3 Y_{d}Y_{d}^{\dagger} + Y_{l}Y_{l}^{\dagger}) -
            \frac{16}{3}g_{3}^{2} - 3g_{2}^{2} -
            \frac{7}{15}g_{1}^{2}\right)\right.\nonumber
        \\
        & \mspace{75mu} + Y_{d}\left( \tr (6 Y_{d}^{\dagger} A_{d} + 2
          Y_{l}^{\dagger}A_{l}) + \frac{32}{3}g_{3}^{2}M_{3} + 6
          g_{2}^{2}M_{2} + \frac{14}{15}g_{1}^{2}M_{1} \right)\nonumber
        \\
        & \mspace{75mu} \left.  + 4 Y_{d}Y_{d}^{\dagger}A_{d} + 5
          A_{d}Y_{d}^{\dagger}Y_{d} + 2 Y_{d}Y_{u}^{\dagger} A_{u} +
          A_{d}Y_{u}^{\dagger}Y_{u}\rule{0cm}{0.5cm}\right],
        \\
        \frac{d}{dt}A_{\nu} & = \frac{1}{16 \pi^{2}}\left[A_{\nu}\left(
            \tr (3 Y_{u}Y_{u}^{\dagger} + Y_{\nu}Y_{\nu}^{\dagger}) -
            3g_{2}^{2} - \frac{3}{5}g_{1}^{2}\right)\right.\nonumber
        \\
        & \mspace{75mu} + Y_{\nu}\left( \tr (6 Y_{u}^{\dagger} A_{u} + 2
          Y_{\nu}^{\dagger}A_{\nu}) + 6 g_{2}^{2}M_{2} +
          \frac{6}{5}g_{1}^{2}M_{1} \right)\nonumber
        \\
        & \mspace{75mu} \left.  + 4 Y_{\nu}Y_{\nu}^{\dagger}A_{\nu} + 5
          A_{\nu}Y_{\nu}^{\dagger}Y_{\nu} + 2 Y_{\nu}Y_{l}^{\dagger}
          A_{l} + A_{\nu}Y_{l}^{\dagger}Y_{l}\rule{0cm}{0.5cm}\right],
        \\
        \label{equ:RGEAl}
        \frac{d}{dt}A_{l} & = \frac{1}{16 \pi^{2}}\left[A_{l}\left( \tr
            (3 Y_{d}Y_{d}^{\dagger} + Y_{l}Y_{l}^{\dagger}) - 3g_{2}^{2}
            - \frac{9}{5}g_{1}^{2}\right)\right.\nonumber
        \\
        & \mspace{75mu} + Y_{l}\left( \tr (6 Y_{d}^{\dagger} A_{d} + 2
          Y_{l}^{\dagger}A_{l}) + 6 g_{2}^{2}M_{2} +
          \frac{18}{5}g_{1}^{2}M_{1} \right)\nonumber
        \\
        & \mspace{75mu} \left.  + 4 Y_{l}Y_{l}^{\dagger}A_{l} + 5
          A_{l}Y_{l}^{\dagger}Y_{l} + 2 Y_{l}Y_{\nu}^{\dagger} A_{\nu} +
          A_{l}Y_{\nu}^{\dagger}Y_{\nu}\rule{0cm}{0.5cm}\right].
      \end{align}
    \end{subequations}
  }%

  Sfermion- and Higgs masses (Notation: $m_{\tilde{Q}}^{2}=m_{Q}^{2}$,
  $m_{\tilde{\overline{u}}}^{2}=m_{u}^{2},$
  $m_{\tilde{\overline{d}}}^{2}=m_{d}^{2},$
  $m_{\tilde{L}}^{2}=m_{L}^{2},$
  $m_{\tilde{\overline{e}}}^{2}=m_{e}^{2}$):
  \begin{subequations}
    \label{eq:soft-masses}
    {\allowdisplaybreaks
      \begin{align}
        \frac{d}{dt}m_{Q}^{2} & = \frac{1}{16
          \pi^{2}}\left[m_{Q}^{2}Y_{u}^{\dagger}Y_{u} +
          Y_{u}^{\dagger}Y_{u}m_{Q}^{2} + m_{Q}^{2}Y_{d}^{\dagger}Y_{d}
          +
          Y_{d}^{\dagger}Y_{d}m_{Q}^{2}\rule{0cm}{0.5cm}\right.\nonumber
        \\
        & \mspace{75mu} + 2\left(Y_{d}^{\dagger}m_{d}^{2}Y_{d} +
          m_{H_{d}}^{2}Y_{d}^{\dagger}Y_{d} +
          A_{d}^{\dagger}A_{d}\right) +
        2\left(Y_{u}^{\dagger}m_{u}^{2}Y_{u} +
          m_{H_{u}}^{2}Y_{u}^{\dagger}Y_{u} +
          A_{u}^{\dagger}A_{u}\right)\nonumber
        \\
        & \mspace{75mu} \left. +
          \left(-\frac{2}{15}g_{1}^{2}|M_{1}|^{2}- 6
            g_{2}^{2}|M_{2}|^{2} - \frac{32}{3}g_{3}^{2}|M_{3}|^{2} +
            \frac{1}{5}g_{1}^{2} S\right)\mathds{1} \right],
        \\
        \frac{d}{dt}m_{u}^{2} & = \frac{1}{16
          \pi^{2}}\left[2\left(m_{u}^{2}Y_{u}Y_{u}^{\dagger} +
            Y_{u}Y_{u}^{\dagger}m_{u}^{2}\right) +
          4\left(Y_{u}m_{Q}^{2}Y_{u}^{\dagger} +
            m_{H_{u}}^{2}Y_{u}Y_{u}^{\dagger} +
            A_{u}A_{u}^{\dagger}\right)\right.\nonumber
        \\
        & \mspace{75mu} \left. +
          \left(-\frac{32}{15}g_{1}^{2}|M_{1}|^{2} -
            \frac{32}{3}g_{3}^{2}|M_{3}|^{2} - \frac{4}{5}g_{1}^{2}
            S\right)\mathds{1}\right],
        \\
        \frac{d}{dt}m_{d}^{2} & = \frac{1}{16
          \pi^{2}}\left[2\left(m_{d}^{2}Y_{d}Y_{d}^{\dagger} +
            Y_{d}Y_{d}^{\dagger}m_{d}^{2}\right) +
          4\left(Y_{d}m_{Q}^{2}Y_{d}^{\dagger} +
            m_{H_{d}}^{2}Y_{d}Y_{d}^{\dagger} +
            A_{d}A_{d}^{\dagger}\right)\right.\nonumber
        \\
        & \mspace{75mu} \left. + \left(-\frac{8}{15}g_{1}^{2}|M_{1}|^{2}
            - \frac{32}{3}g_{3}^{2}|M_{3}|^{2} + \frac{2}{5}g_{1}^{2}
            S\right)\mathds{1} \right],
        \\
        \label{equ:RGEmmL}
        \frac{d}{dt}m_{L}^{2} & = \frac{1}{16
          \pi^{2}}\left[m_{L}^{2}Y_{l}^{\dagger}Y_{l} +
          Y_{l}^{\dagger}Y_{l}m_{L}^{2} +
          m_{L}^{2}Y_{\nu}^{\dagger}Y_{\nu} +
          Y_{\nu}^{\dagger}Y_{\nu}m_{L}^{2}\rule{0cm}{0.5cm}\right.\nonumber
        \\
        & \mspace{75mu} + 2\left(Y_{e}^{\dagger}m_{e}^{2}Y_{e} +
          m_{H_{d}}^{2}Y_{e}^{\dagger}Y_{e} +
          A_{e}^{\dagger}A_{e}\right) +
        2\left(Y_{\nu}^{\dagger}m_{\nu}^{2}Y_{\nu} +
          m_{H_{u}}^{2}Y_{\nu}^{\dagger}Y_{\nu} +
          A_{\nu}^{\dagger}A_{\nu}\right)\nonumber
        \\
        & \mspace{75mu} \left.-\left(\frac{6}{5}g_{1}^{2}|M_{1}|^{2} + 6
            g_{2}^{2}|M_{2}|^{2}-\frac{3}{5}g_{1}^{2} S\right)\mathds{1}
        \right],
        \\
        \label{equ:RGEmme}
        \frac{d}{dt}m_{e}^{2} & = \frac{1}{16
          \pi^{2}}\left[2\left(m_{e}^{2}Y_{l}Y_{l}^{\dagger} +
            Y_{l}Y_{l}^{\dagger}m_{e}^{2}\right) +
          4\left(Y_{e}m_{L}^{2}Y_{e}^{\dagger} +
            m_{H_{d}}^{2}Y_{e}Y_{e}^{\dagger} +
            A_{e}A_{e}^{\dagger}\right)\right.\nonumber
        \\
        & \mspace{75mu}  \left. + \left(-\frac{24}{5}g_{1}^{2}|M_{1}|^{2} +
            \frac{6}{5}g_{1}^{2} S\right)\mathds{1} \right],
        \\
        \frac{d}{dt}m_{\nu}^{2} & = \frac{1}{16
          \pi^{2}}\left[2\left(m_{\nu}^{2}Y_{\nu}Y_{\nu}^{\dagger} +
            Y_{\nu}Y_{\nu}^{\dagger}m_{\nu}^{2}\right) +
          4\left(Y_{\nu}m_{L}^{2}Y_{\nu}^{\dagger} +
            m_{H_{u}}^{2}Y_{\nu}Y_{\nu}^{\dagger} +
            A_{\nu}A_{\nu}^{\dagger}\right)\right],
        \\
        \label{equ:RGEmmhu}
        \frac{d}{dt}m_{H_{u}}^{2} & = \frac{1}{16 \pi^{2}}\left[6 \tr
          \left(Y_{u}^{\dagger}(m_{Q}^{2} + m_{u}^{2} +
            m_{H_{u}}^{2}\mathds{1})Y_{u} +
            A_{u}^{\dagger}A_{u}\right)\right.
        \\
        \label{equ:RGEmmhd}
        & \mspace{75mu} \left. + 2 \tr \left(Y_{\nu}^{\dagger}(m_{L}^{2}
            + m_{\nu}^{2} + m_{H_{u}}^{2}\mathds{1})Y_{\nu} +
            A_{\nu}^{\dagger}A_{\nu}\right)-\frac{6}{5}g_{1}^{2}|M_{1}|^{2}-6
          g_{2}^{2}|M_{2}|^{2} + \frac{3}{5}g_{1}^{2} S\right],\nonumber
        \\
        \frac{d}{dt}m_{H_{d}}^{2} & = \frac{1}{16 \pi^{2}}\left[6 \tr
          \left(Y_{d}^{\dagger}(m_{Q}^{2} + m_{d}^{2} +
            m_{H_{d}}^{2}\mathds{1})Y_{d} +
            A_{d}^{\dagger}A_{d}\right)\right.
        \\
        & \mspace{75mu} \left. + 2 \tr \left(Y_{l}^{\dagger}(m_{L}^{2} +
            m_{e}^{2} + m_{H_{d}}^{2}\mathds{1})Y_{l} +
            A_{l}^{\dagger}A_{l}\right)-\frac{6}{5}g_{1}^{2}|M_{1}|^{2}-6
          g_{2}^{2}|M_{2}|^{2} - \frac{3}{5}g_{1}^{2} S\right],\nonumber
      \end{align}
    }%
    with
    \begin{align}
      S = \tr \left(m_{Q}^{2} + m_{d}^{2}-2 m_{u}^{2}-m_{L}^{2} +
        m_{e}^{2}\right)-m_{H_{d}}^{2} + m_{H_{u}}^{2}.
    \end{align}
  \end{subequations}
  
  The neutrino Yukawa coupling $Y_{\nu}$ decouples from the RGE below
  the Majorana mass scale and thus disappears from the equations.  Some
  peculiarities occur if you integrate out the right handed neutrinos
  separately, as we do.

\end{appendix}

\bibliography{LiteraturPaper}
\bibliographystyle{JHEP}
\end{document}